\documentclass[useAMS,usenatbib]{mn2e}
\usepackage{url}
\usepackage[pdftex]{graphicx}
\usepackage{hyperref}
\usepackage{stmaryrd}

\usepackage{threeparttable}

\def\aj{AJ}%
\def\apj{ApJ}%
\def\apjl{ApJ}%
\def\apjs{ApJS}%
\def\aaps{A\&AS}%
\def\mnras{MNRAS}%
\def\pasp{PASP}%
\def\nat{Nature}%
\def\hi{H\,{\sc i}}

\def\m20{$M_{20}$}
\def\gm{$G_{M}$}

\def\whisp{{\em WHISP}}

\def\fuv{{\sc fuv}}

\def\xuv{{\sc xuv}}

\def\uv{{UV}}

\begin{document}

\title[\hi\ Morphology of Dwarf Galaxies]{Quantified \hi\ Morphology VII:\\ 
star-formation and tidal influence on local dwarf \hi\ morphology}

\author[Holwerda et al.]{
B. W. Holwerda$^{1}$\thanks{E-mail: benne.holwerda@esa.int},
N. Pirzkal,$^{2}$,
W.J.G. de Blok$^{3}$, and
S--L. Blyth$^{4}$ \\
$^{1}$ European Space Agency, ESTEC, Keplerlaan 1, 2200 AG, Noordwijk, the Netherlands\\
$^{2}$ Space Telescope Science Institute, Baltimore, MD 21218, USA\\
$^{3}$ Netherlands Institute for Radio Astronomy, Schattenberg 1, 9433 TA Zwiggelte, The Netherlands\\
$^{4}$ Astrophysics, Cosmology and Gravity Centre, Department of Astronomy, \\
University of Cape Town, Private Bag X3, Rondebosch 7701, Republic of South Africa
}
\date{Accepted 1988 December 15. Received 1988 December 14; in original form 1988 October 11}

\pagerange{\pageref{firstpage}--\pageref{lastpage}} \pubyear{2002}

\maketitle

\label{firstpage}


\begin{abstract}

Scale-invariant morphology parameters applied to atomic hydrogen maps (\hi) of galaxies
can be used to quantify the effects of tidal interaction or star-formation on the ISM. Here we apply these 
parameters, Concentration, Asymmetry, Smoothness, Gini, \m20, and the \gm\ parameter, to two public 
surveys of nearby dwarf galaxies, the VLA-ANGST and LITTLE-THINGS survey, to explore whether 
tidal interaction or the ongoing or past star-formation is a dominant force shaping the \hi\ disk of 
these dwarfs.

Previously, \hi\ morphological criteria were identified for ongoing spiral-spiral interactions. 
When we apply these to the Irregular dwarf population, they either select almost all or none of the 
population. We find that only the Asymmetry-based criteria can be used to identify very isolated 
dwarfs (i.e., these have a low tidal indication). Otherwise, there is little or no relation between the 
level of tidal interaction and the \hi\ morphology.

We compare the \hi\ morphology to three star-formation rates based on either H$\alpha$, \fuv\ or 
the resolved stellar population, probing different star-formation time-scales. The \hi\ morphology 
parameters that trace the inequality of the distribution, the Gini, \gm, and \m20\ parameters, 
correlate weakly with all these star-formation rates. This is in line with the picture that local 
physics dominates the ISM appearance and not tidal effects. 

Finally, we compare the SDSS measures of star-formation and stellar mass to the \hi\ morphological 
parameters for all four \hi\ surveys. In the two lower-resolution \hi\ surveys (12"), there is no relation 
between star-formation measures and \hi\ morphology. The morphology of the two high-resolution 
\hi\ surveys (6"), the Asymmetry, Smoothness, Gini, \m20, and \gm, do show a link to the total 
star-formation, but a weak one.
\end{abstract}

\begin{keywords}
(galaxies:) Local Group 
galaxies: dwarf 
galaxies: ISM 
galaxies: structure 
ISM: structure 
radio lines: ISM 
\end{keywords}

\section{\label{s:intro}Introduction}

It has recently become clear that the ongoing star formation in smaller galactic systems strongly 
influences the structure of the dwarf system's interstellar matter (ISM), and vice versa 
\citep[e.g.,][]{Weisz09a}.  The low-density and -metallicity environment as well as strong effects of feedback make local dwarfs an outstanding 
laboratory to understand the physics of star-formation. In addition, 
$\Lambda$CDM predicts the dynamics to be dominated by their dark matter content but this is 
observationally still debated \citep[e.g.,][]{Oh11, Swaters11}.

Dwarf galaxy morphology is related to their environment, as evident from the relation of stellar morphology 
with tidal index \citep{Weisz11b}, as is their gas content \citep{Grcevich09}. Both strongly point to the gas 
content as the main driver of dwarf morphology. Similarly, \cite{Geha12} find that all field and central dwarf galaxies have a low fraction 
of quenched star-formation, i.e. they all have a substantial gas reservoir. This gas is at significantly sub-solar metallicities 
\citep{Berg12} and there is strong evidence for metal loss from supernova \citep{Tremonti04, Dalcanton07, Bouche07, Kirby11}. 

For these reasons, the local sample of low-mass galaxies has been studied extensively using ultra-violet, optical and near-infrared tracer of star-formation \citep[e.g.,][]{Hunter06,McQuinn12a}, ISM \citep{littlethings, Ott12}, and 
resolved stellar populations \citep{angst, Weisz11b, Weisz11a}. 
These studies have been made possible by space observatories which allow for observations 
of stars and dusty ISM with surface brightnesses, and large programs on the Karl. G. Jansky Very Large Array (VLA) 
which observe the 21cm line of neutral atomic hydrogen (\hi).

A key HST program, the ANGST survey \citep{angst}, has observed a large sample of nearby galaxies, 
mostly dwarfs, uniformly and in unprecedented depth with the Advanced Camera for Surveys (ACS) 
on the Hubble Space Telescope (HST). To accompany the HST observations, a large VLA program, 
the VLA-ANGST survey has observed the neutral ISM in great detail \citep{Ott12}.
The star-formation history from the resolved stellar populations' colours and luminosities has already 
revealed that star-bursts in these galaxies occur stochastically in both time and location \citep{McQuinn12b} 
over the last several hundred Myrs \citep[see also][]{McQuinn09,McQuinn10a,McQuinn10b}
A second program, LITTLE-THINGS, has observed a different set of nearby dwarfs with Herschel, 
Spitzer, GALEX and a large program on the VLA. 

The present consensus from these programs is that the processes related to star-formation are all 
inefficient in dwarf galaxies: the star-formation efficiency, the quenching of star-formation, and the 
interactions between the star-formation and the ISM dynamics \citep[][]{Skillman12}.

In this series of papers, we have explored the quantified morphology of available \hi\ maps with the 
common parameters for observed optical or ultra-violet morphology: 
concentration-asymmetry-smoothness \citep{CAS}, Gini and \m20 \citep{Lotz04} and $G_M$ 
\citep{Holwerdapra09,Holwerda11c}. Recent 
interest in these morphology parameters has shifted from high-mass spirals and major interaction 
to more unequal mass interactions \citep{Lotz10a}, more gas-rich interactions \citep{Lotz10b}, 
both of which typically involve dwarf galaxies, and the visibility times of mergers in this parameter 
space \citep{Lotz11b}.
In \cite{Holwerda11a}, we compare the \hi\  morphology to those at other wavelengths for the THINGS 
sample, noting that the \hi\ and ultraviolet morphologies are closely related, which would make 
quantified \hi\ morphology a reasonable tracer for interactions.
In the next papers of the series, we use the \hi\ morphology to identify mergers \citep{Holwerda11b}, 
their visibility time \citep{Holwerda11c}, and subsequently infer a merger rate from the \whisp\ survey 
\citep{Holwerda11d}, as well as identify phenomena unique to cluster members \citep{Holwerda11e}
and the those \hi\ disks hosting an extended ultraviolet disks \citep[\xuv,][]{Holwerda12c}. 

In this paper, we explore the \hi\ morphology of low-mass local dwarf galaxies. These have recently 
been observed in 21cm radio emission (\hi) by the VLA-ANGST and LITTLE-THINGS surveys. A third 
survey is underway to observe the lower mass galaxies in the local volume (the ÒSurvey of H I in 
Extremely Low-mass DwarfsÓ \citep[SHIELD,][]{shield} but we do not include it due to its low spatial 
resolution ($\sim 20"$ beam). The combined VLA-ANGST and LITTLE-THINGS span a representative selection of the 
smallest members of the local volume (60 \hi\ maps). The general picture that emerges from these 
surveys of low-mass galaxies in the local Universe is that the appearance of the \hi\ becomes 
amorphous with lower masses: there is a progression from disks with spiral structure to mostly 
featureless rotating disks to a collection of clouds supported by both rotation and dispersion. 
Our motivation for this study was to explore how much information there still is in the morphology 
of the \hi\ in these systems. These morphological parameterizations are used to high redshift with 
HST imaging of distant galaxies, which equally appear less structured beyond $z\sim2$, i.e., more as a collection of star-forming regions rather than organized in disks with spiral pattern and bulges. 
We shall compare the \hi\ morphological parameters to indicators of tidal disturbance and star-formation. 

While there are many outstanding questions as to the nature of dwarf galaxies and their ISM and 
star-formation \citep[see][]{Skillman10, Skillman12}, we focus here on two: What is the impact of 
the star formation on the structure of their ISM? Does star formation induce or quench further star formation?, 
i.e., does star formation propagate through the host galaxy or is it stochastic?.

The paper is organized as follows: 
Section \ref{s:data} describes the data products and sample from the two surveys used for this paper, 
Section \ref{s:qm} briefly describes the six morphological parameters,
Section \ref{s:results} presents the results, 
Section \ref{s:sdss} compares the \hi\ morphology of all our catalogs to SDSS estimates of star-formation and mass,
Section \ref{s:disc} briefly discusses them, and 
Section \ref{s:concl} lists our conclusions.

\section{Data}
\label{s:data}

The ``Local Irregulars That Trace Luminosity Extremes" \citep[LITTLE-THINGS,][]{littlethings}  and 
the VLA-ANGST \citep{Ott12} surveys, are close in observational setup to the ÒThe H I Nearby Galaxy SurveyÓ 
\citep[THINGS,][]{Walter08}, the sample for our first paper \citep{Holwerda11a}. Both surveys were 
conducted while the VLA transitioned to the Karl G. Jansky Very Large Array.
For this paper, we use the robustly-weighted \hi\ surface density maps (RO). These maps are the 
highest resolution, contain the most small detail, essential for quantified morphology measurements, 
at the expense of some large-scale faint structure. This trade-off is essential for quantified morphological 
measurements which are the most sensitive when sampling at sub-kiloparsec physical scales 
\citep{Lotz04}, at which point the diffuse large-scale \hi\ emission barely contributes signal in most 
parameters \citep[see the comparison in][]{Holwerda11a}.

\subsection{LITTLE-THINGS}
\label{s:little-things}

The LITTLE-THINGS sample \citep{littlethings} is made up of 42 dwarf irregular (dIm) and Blue Compact 
Dwarf (BCD) galaxies. The \hi\ observations are a mix of new (21 galaxies) and archival observations. 
Some galaxies were dropped from the sample due to issues with individual observations. 
The LITTLE-THINGS sample was drawn from a larger multi-wavelength effort \citep{Hunter04, Hunter06a} 
and there are extensive ancillary data available for the full sample.
Observational setup was kept identical to the THINGS survey \citep{Walter08} and data is public at 
 \url{https://science.nrao.edu/science/surveys/littlethings/}. We converted the the 
LITTLE-THINGS moment 0 maps into column density maps using the expression in \cite{Walter08}, their equation 5, 
and the major and minor axes from \cite{littlethings} to conform to the VLA-ANGST data products.
Typical resolution is slightly lower than VLA-ANGST ($\sim$6-10"), depending on the observational 
configuration.

\subsection{VLA-ANGST}
\label{s:vla-angst}

The VLA-ANGST sample is based on the volume-limited ANGST survey \citep{angst}. The galaxies in 
these surveys drawn from the local volume compilation from \cite{Karachentsev04}. This catalog lists 
relevant parameters and, of specific interest, the tidal index $\Theta$ (see \S \ref{s:theta}). 
The ANGST survey targets the local volume, mostly less than 3.5 Mpc away with a limiting distance of 4 Mpc, in order 
to resolve low-level star-formation from resolved stellar populations. From the 89 ANGST galaxies, 
VLA-ANGST is a subset of 29 detected galaxies, excluding southern objects and those with no 
single-dish \hi\ detections or low star-formation. All VLA-ANGST galaxies were observed in both high 
spatial ($\sim$6", corresponding to $\sim$100 pc) and spectral (corresponding to 0.65-2.6 km/s in 
velocity) resolution in the VLA B, C, and D array configurations. For this study, the high velocity 
resolution is not pertinent but the spatial resolution and depth comparable to the THINGS survey are.
Data are available at \url{https://science.nrao.edu/science/surveys/vla-angst/} and described in detail in 
\cite{Ott12} and \cite{Warren12} presents the \hi\ profiles for these galaxies.

\subsection{Final Sample}

Some of the \hi\ observations for LITTLE-THINGS are not yet archived and there is some overlap 
with the VLA-ANGST and LITTLE-THINGS surveys with galaxies included under a different name. 
Omitted galaxies are NGC1156, NGC6822, DDO6, KDG63, HS117, NGC4190, DDO113, DDO125
DDO181 and DDO183. The galaxies CVnIdwA (UGCA292), GR 8 (DDO155) and UGC 8508  are in 
both the LITTLE-THINGS and VLA-ANGST surveys. The final tally of \hi\ maps is 60 galaxies.

\begin{figure*}
\begin{center}
	\includegraphics[width=0.49\textwidth]{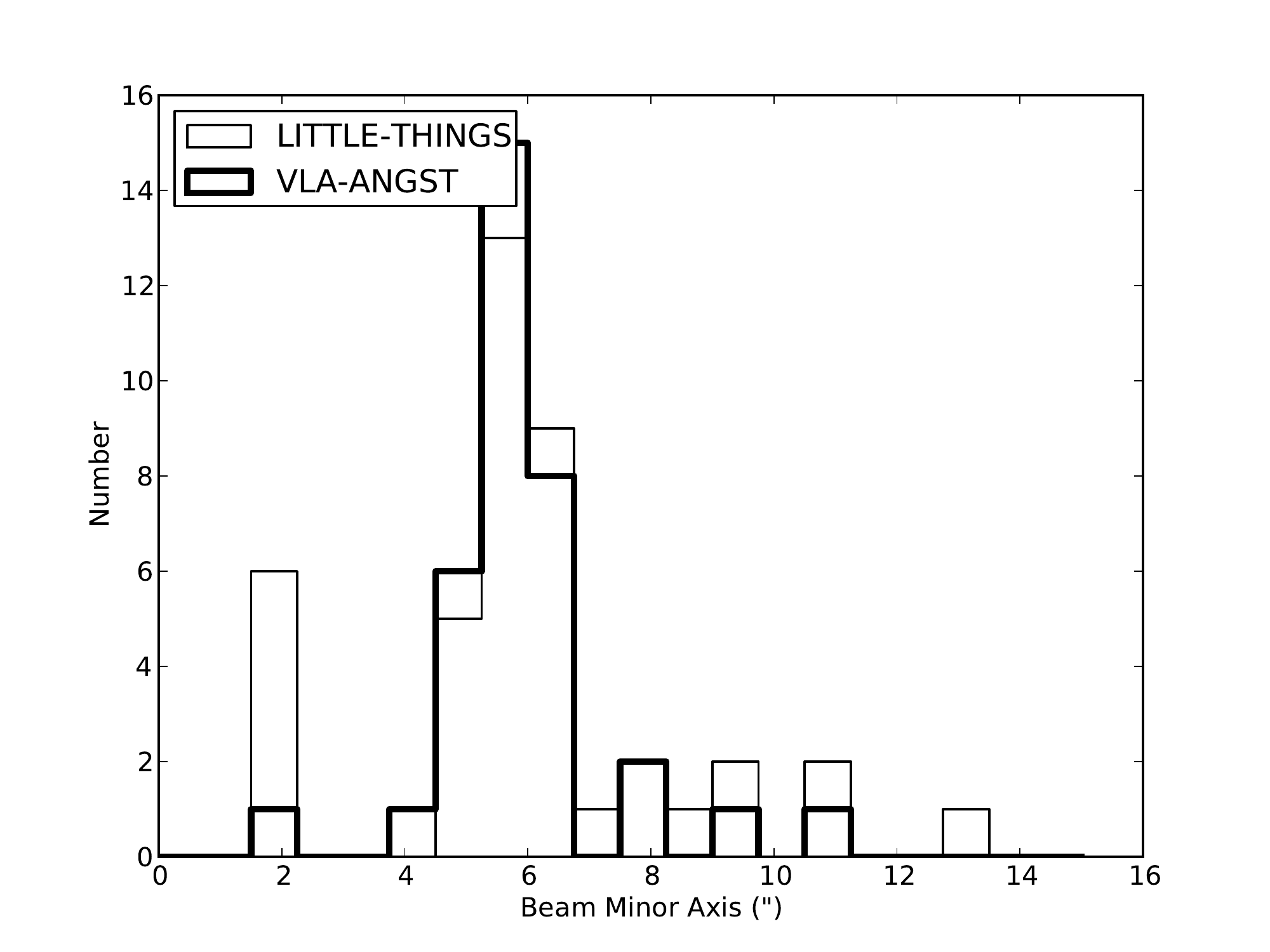}
	\includegraphics[width=0.49\textwidth]{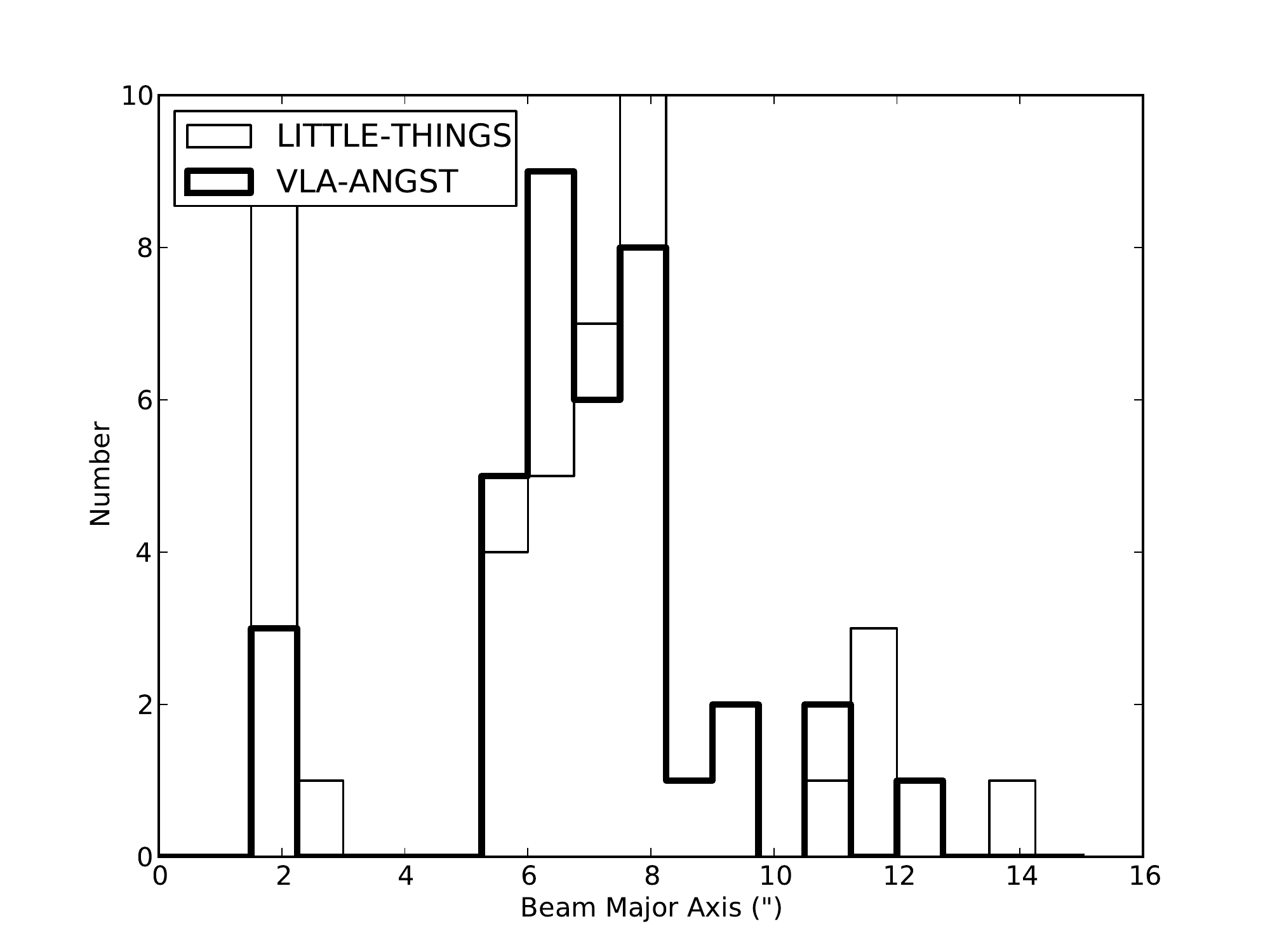}
	\caption{\label{f:B} The distribution of minor and major axes of the LITTLE-THINGS and VLA-ANGST \hi\ observations.}
\end{center}
\end{figure*}

As noted, the sampling of the two surveys is slightly different: the mean LITTLE-THINGS beam is $7\farcs2 \times 8\farcs8$ and the VLA-ANGST one $6\farcs1 \times 7\farcs4$ but these resolutions are comparable in the sampling of the \hi\ disk (Figure \ref{f:B}) and the two surveys can be treated as a single data-set.

\begin{table}
\caption{\label{t:beam} The spatial resolution for the Robustly Weighted column density maps from the LITTLE-THINGS and VLA-ANGST surveys. Information from \protect\cite{littlethings} and \protect\cite{Ott12}.}
\begin{center}
\begin{tabular}{l l l l}

Galaxy 	& Beam Size &  Survey \\
		& Minor	& Major	& \\
\hline
\hline
CVnIdwA & 10.5	& 10.9	& LITTLE-THINGS \\
DDO 43	& 6.0		& 8.1		& LITTLE-THINGS \\
DDO 46	& 5.2 	& 6.3		& LITTLE-THINGS \\
DDO 47	& 9.0 	& 10.4	& LITTLE-THINGS \\
DDO 50	& 6.1 	& 7.0		& LITTLE-THINGS \\
DDO 52	& 5.2		& 6.8		& LITTLE-THINGS \\
DDO 53	& 5.7		& 6.3		& LITTLE-THINGS \\
DDO 63 	& 6.0		& 7.8		& LITTLE-THINGS \\
DDO 69	& 5.4		& 5.8		& LITTLE-THINGS \\
DDO 70	& 13.2	& 13.8	& LITTLE-THINGS \\
DDO 75	& 6.5		& 7.5		& LITTLE-THINGS \\
DDO 87	& 6.2		& 7.6		& LITTLE-THINGS \\
DDO 101	& 7.0		& 8.3		& LITTLE-THINGS \\
DDO 126	& 5.6		& 6.9		& LITTLE-THINGS \\
DDO 133	& 10.8	& 12.4	& LITTLE-THINGS \\
DDO 154	& 6.3		& 7.9		& LITTLE-THINGS \\
DDO 155	& 10.1	& 11.3	& LITTLE-THINGS \\
DDO 165	& 7.6		& 10.0	& LITTLE-THINGS \\
DDO 167	& 5.3		& 7.3		& LITTLE-THINGS \\
DDO 168	& 5.8		& 7.7		& LITTLE-THINGS \\
DDO 187	& 5.5		& 6.2		& LITTLE-THINGS \\
DDO 210	& 8.6		& 11.7	& LITTLE-THINGS \\
DDO 216	& 15.4	& 16.2	& LITTLE-THINGS \\
IC 10	& 5.5		& 5.9		& LITTLE-THINGS \\
IC 1613	& 6.5		& 7.7		& LITTLE-THINGS \\
LGC 3	& 9.3		& 11.8	& LITTLE-THINGS \\
M81dwA	& 6.3		& 7.8		& LITTLE-THINGS \\
NGC 1569	& 5.2 & 5.9	& LITTLE-THINGS \\
NGC 2366	& 5.9	 & 6.9	& LITTLE-THINGS \\
NGC 3738	& 5.5 & 6.3	& LITTLE-THINGS \\
NGC 4163	& 5.9	 & 9.7	& LITTLE-THINGS \\
NGC 4214	& 6.4	 & 7.6 	& LITTLE-THINGS \\
SagDIG		& 16.9 & 28.2	& LITTLE-THINGS \\
UGC 8508	& 4.9	  & 5.9	& LITTLE-THINGS \\
WLM 		& 5.1	 & 7.6	& LITTLE-THINGS \\
Haro 29		& 5.6	 & 6.8	& LITTLE-THINGS \\
Haro 36		& 5.8	 & 7.0	& LITTLE-THINGS \\
Mrk 178		& 5.5	 & 6.2	& LITTLE-THINGS \\
VIIZw 403		& 7.7 & 9.4	& LITTLE-THINGS \\
\hline
\end{tabular}
\end{center}
\end{table}%

\begin{table}
\begin{center}
\begin{tabular}{l l l l}

Galaxy 	& Beam Size &  Survey \\
		& Minor	& Major	& \\
\hline
\hline

NGC 247		& 6.2	 & 9.0	& VLA-ANGST \\
DDO 6		& 6.3 & 7.2	& VLA-ANGST \\
NGC 404		& 6.1	 & 7.1	& VLA-ANGST \\
KKH37		& 5.8 & 6.5	& VLA-ANGST \\
UGC 4483	& 5.7	 & 7.6	& VLA-ANGST \\
KK 77		& 5.8	 & 6.1	& VLA-ANGST \\
BK3N		& 5.8 & 6.3	& VLA-ANGST \\
AO0952+69	& 5.9 & 6.4	& VLA-ANGST \\
Sextans B		& 7.5	 & 9.5	& VLA-ANGST \\
NGC 3109	& 5.0 & 7.6	& VLA-ANGST \\
Antlia		& 9.6 & 10.5	& VLA-ANGST \\
KDG 63		& 6.0 & 6.2	& VLA-ANGST \\
Sextans A		& 6.0 & 7.3	& VLA-ANGST \\
HS 117		& 6.1 & 8.6	& VLA-ANGST \\
DDO 82		& 5.7 & 5.8	& VLA-ANGST \\
KDG 73		& 5.6 & 6.9	& VLA-ANGST \\
NGC 3741	& 4.8 & 5.5	& VLA-ANGST \\
DDO 99		& 5.2 & 7.7	& VLA-ANGST \\
NGC 4163	& 5.4	 & 7.6	& VLA-ANGST \\
NGC 4190	& 5.3 & 6.1	& VLA-ANGST \\
DDO 113		& 7.7 & 9.9 	& VLA-ANGST \\
MCG +09-20-131 & 5.3 & 6.1 	& VLA-ANGST \\
DDO 125		& 5.4 & 6.3	& VLA-ANGST \\
UGCA 292	& 5.0	 & 7.0	& VLA-ANGST \\
GR 8 		& 5.4 & 5.8	& VLA-ANGST \\
UGC 8508	& 6.4 & 8.2	& VLA-ANGST \\
DDO 181		& 5.5	 & 7.6	& VLA-ANGST \\
DDO 183		& 6.2 & 7.6 	& VLA-ANGST \\
KKH 86 		& 5.8 & 7.5	& VLA-ANGST \\
UGC 8833	& 11.2 & 12.4	& VLA-ANGST \\
KKH 230 		& 5.2	 & 5.9 	& VLA-ANGST \\
DDO187		& 5.7 & 7.1	& VLA-ANGST \\
DDO 190		& 9.9 & 10.8	& VLA-ANGST \\
KKR 25		& 4.4 & 5.5 	& VLA-ANGST \\
KKH 98		& 5.2 & 6.2	& VLA-ANGST \\
\hline
\end{tabular}
\end{center}
\end{table}%

\section{Quantifying Morphology}
\label{s:qm}

We use the Concentration-Asymmetry-Smoothness parameters 
\citep[CAS,][]{CAS}, combined with the Gini-$M_{20}$ parameters from \cite{Lotz04}, and our own 
$G_M$. We have discussed the definitions of these parameters in the previous papers, as well 
as how we estimate uncertainties for each. Here, we will give a brief overview but for details we 
refer the reader to \cite{Holwerda11a,Holwerda11b} or Holwerda et al. {\em submitted}.

We select pixels in an image as belonging to the galaxy based on the outer \hi\ contour ($5. \times 10^{19}$ atoms/cm$^2$) and adopt 
the position from the respective survey catalogs as the central position of the galaxy 
\citep[as reported in][respectively]{littlethings,Ott12}. Given a set of $n$ pixels in each object, iterating over pixel 
$i$ with value $I_i$, pixel position $x_i,y_i$ with the centre of the object at $x_c,y_c$ these 
parameters are defined as:
\begin{equation}
C = 5 ~ \log (r_{80} /  r_{20}),
\label{eq:c}
\end{equation}
\noindent with $r_{f}$ as the radial aperture, centered on $x_c,y_c$ containing percentage $f$ of the 
light of the galaxy \citep[see definitions of $r_f$ in][]{se,seman}\footnote{We must note that the earlier 
version of our code contained an error, artificially inflating the concentration values. A check revealed 
this to be $C_{new} = 0.38 \times C_{old}$, and we adopt the new, correct values in this paper.}. 
This concentration index can be used to quickly discern between light profiles; a de Vaucouleurs 
profile ($I \propto R^{-4}$) has Concentration value of $C=5.2$, and a purely exponential one 
has a value of $C=2.7$. It also can be used to identify unique phenomena, for for example H\,I disk 
stripping \citep{Holwerda11e}. 
\begin{equation}
A = {\Sigma_{i} | I_i - I_{180} |  \over \Sigma_{i} | I(i) |  },
\label{eq:a}
\end{equation}
\noindent where $I_{180}$ is the pixel at position $i$ in the galaxy's image, after it was rotated $180^\circ$ 
around the centre of the galaxy. Fully symmetric galaxies  have very low values of Asymmetry. A regular 
spiral need not show a high value of Asymmetry, e.g., a grand-design spiral galaxy's spiral arms map 
onto each other with a 180$^\circ$ rotation \citep[the rotational symmetry of galaxies can be used to 
infer dust extinction in pairs of galaxies, see][]{kw92,kw00a,kw00b,kw01a,kw01b,Holwerda07c,Keel13, Holwerda13a, Holwerda13b}.
Flocculant spirals can be expected to be slightly more Asymmetric still. The highest values of Asymmetry 
can be found in galaxies with strong tidal disruptions, provided the tidal structures are included in the 
calculation, which they are in \hi.

\begin{equation}
S = {\Sigma_{i,j} | I(i,j) - I_{S}(i,j) | \over \Sigma_{i,j} | I(i,j) | },
\label{eq:s}
\end{equation}
\noindent where $I_{S}$ is pixel $i$ in a smoothed image. The type of smoothing (e.g., boxcar or 
Gaussian) has changed over the years. We chose a fixed 5" Gaussian smoothing kernel for simplicity. 
We note that we use the term "Smoothness" for historical reasons as this has become the de facto 
designation of this parameter (the CAS scheme), even though an increase in its value means a more clumpy 
appearance of the image (hence its original designation "clumpiness"). Very smooth galaxies have very low 
values of Smoothness but in other galaxies, the value of the Smoothness parameter depends on the size of 
the smoothing kernel used. If the kernel's size correspond to, for example, the width of spiral arms at the 
distance of the galaxy, then grand design spirals will have relatively high Smoothness values.

The Gini coefficient is defined as:
\begin{equation}
G = {1\over \bar{I} n (n-1)} \Sigma_i (2i - n - 1) I_i ,
\label{eq:g}
\end{equation}
\noindent where the list of $n$ pixels was first ordered according to value and $\bar{I}$ is the mean 
pixel value in the image. 

\cite{Lotz04} introduce the relative second-order moment (\m20) of an object.
The second-order moment of a pixel is: $M_i = I_i \times R_i = I_i \times [(x_i ? x_c)^2 + (y_i ? y_c)^2]$.
The total second-order moment of an image is defined as:
\begin{equation}
M_{tot} = \Sigma M_i = \Sigma I_i [(x_i - x_c)^2 + (y_i - y_c)^2]
\end{equation}
The relative second-order moment of the brightest 20\% of the flux:
\begin{equation}
M_{20} = \log \left( {\Sigma_i^k M_i  \over  M_{tot}}\right), ~ {\rm for ~ which} ~ \Sigma_i^k I_i < 0.2 ~ I_{tot} {\rm ~ is ~ true}.\\
\label{eq:m20}
\end{equation}
\noindent where pixel $k$ marks the top 20\% point in the flux-ordered pixel-list. 
The \m20\ parameter is a parameter that is sensitive to bright structure away from the center of the galaxy: 
flux is weighted in favor of the outer parts. It therefore is relatively sensitive to tidal structures. 

Instead of using the intensity of  pixel $i$, the Gini parameter can be defined using the second 
order moment:
\begin{equation}
G_M = {1\over \bar{M} n (n-1)} \Sigma_i (2i - n - 1) M_i ,
\label{eq:gm}
\end{equation}

These parameters trace different structural characteristics of a galaxy's image but these do not span 
an orthogonal parameter space \citep[see also the discussion in][Holwerda et al. {\em in preparation}]{Scarlata07}. Two crucial input 
parameters for the computation of the morphology are the central position ($x_c$ and $y_c$) and 
the threshold for including pixels into the calculations. We use the positions reported by \cite{Ott12} 
for the VLA-ANGST galaxies and those in the NED database for the LITTLE-THINGS galaxies for 
the central pixel position. To determine which pixels to include, we adopt a threshold of 
$5 \times 10^{19}$ atoms/cm$^2$, the practical limiting depth of both of these surveys. 

\cite{Holwerda11a} discuss the uncertainties in these parameters in detail. To estimate their errors, 
we both vary the input central position and compute the rms from the resulting spread in values. 
Secondly, we scramble the pixels (but keep the central position identical) to asses the effect of 
random noise. Thirdly, in the case of the Gini parameter, there is no dependence on the central 
position. In this case we compute the variance by sub-sampling the pixel collection.

\subsection{Spatial Sampling}
\label{s:samp}

Interferometric radio observations filter out large-scale faint emission, a unique feature with respect to the characterization of morphology.  To remedy this, the total-power information from short baseline observations are needed, i.e., a large single-dish telescope or a radio array more compact than the VLA-A configuration. 

\begin{figure*}
\begin{center}
	\includegraphics[width=0.49\textwidth]{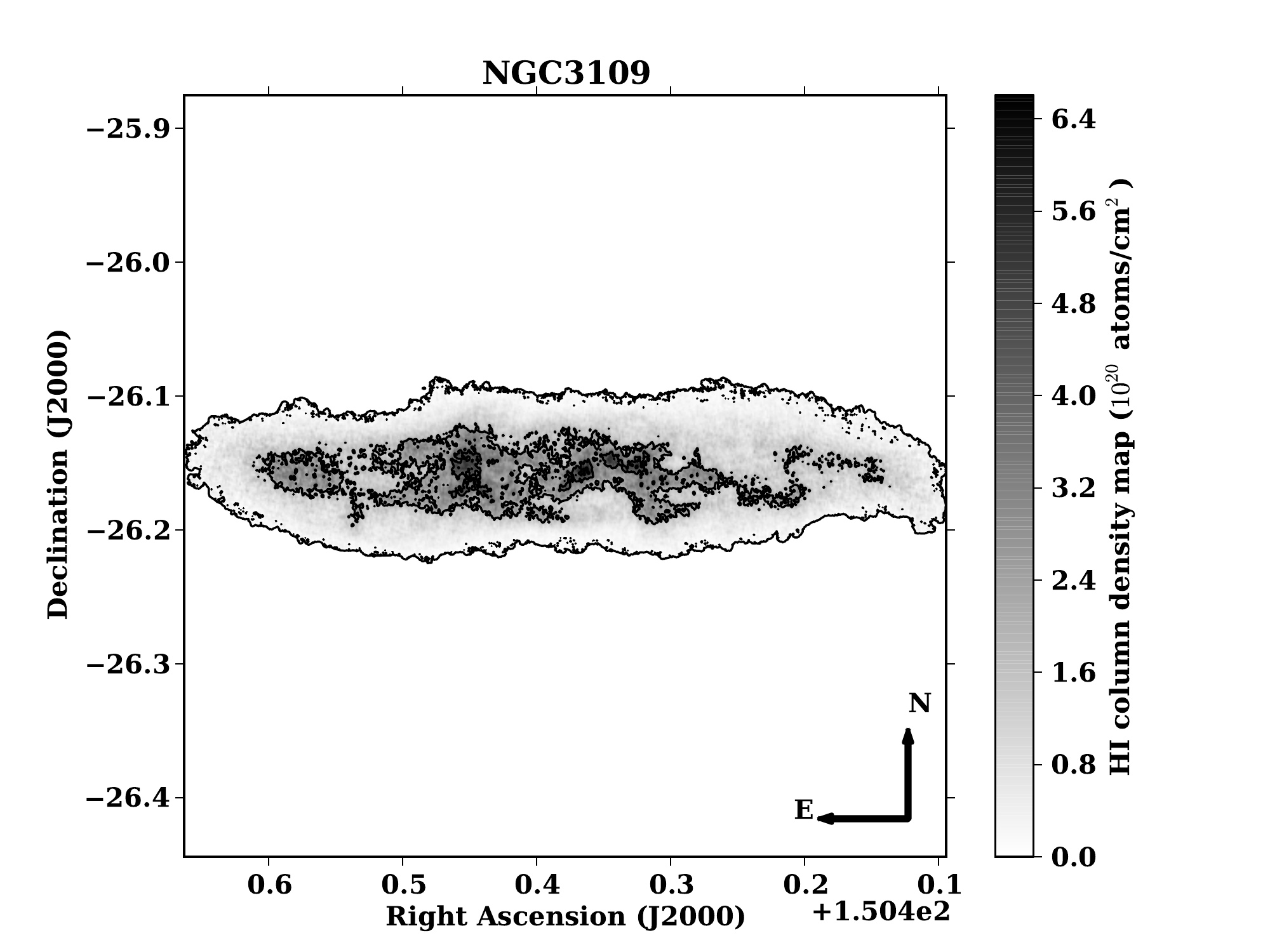}
	\includegraphics[width=0.49\textwidth]{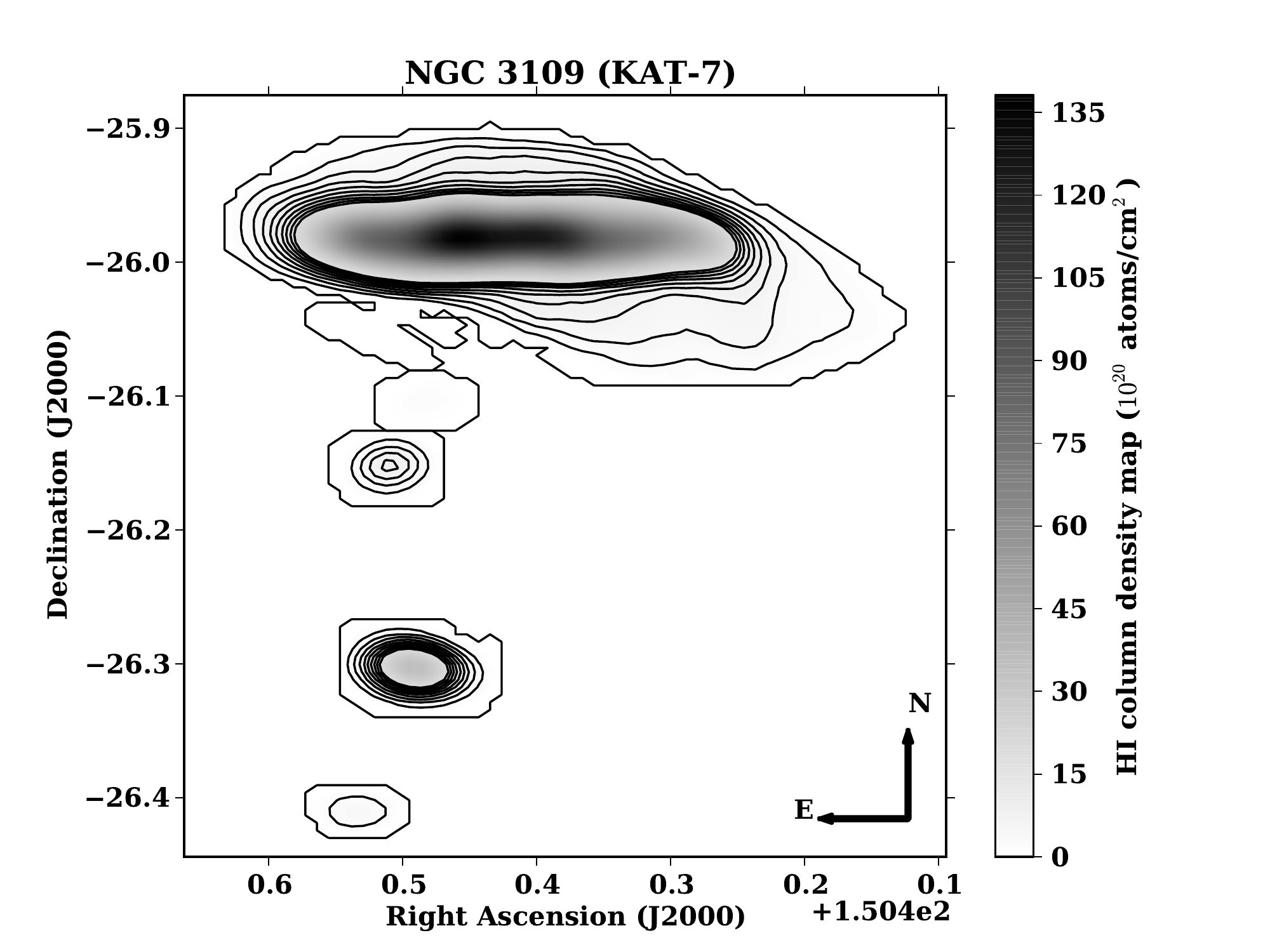}
	\caption{\label{f:n3109} The \hi\ map of NGC 3109 from the VLA-ANGST survey \protect\citep{Ott12} and the KAT-7 observations \protect\citep{Carignan13}, which include several smaller, unresolved galaxies. The outer VLA-contour corresponds to the inner one (32Jy/Beam) for the KAT-7 mosaic. While there is substantial information in the outer regions, their weight in flux is minimal.}
\end{center}
\end{figure*}

Fortunately, one of these galaxies, NGC 3109, was observed with the Karoo Array Telescope (KAT-7), a seven-dish precursor array to the MeerKAT telescope \citep{MeerKAT, meerkat1,meerkat2}. These observations and results are described in detail in \cite{Carignan13}. The resulting \hi\ map is sensitive to larger scale \hi\ features such as wide tidal tails or warps. Figure \ref{f:n3109} shows both \hi\ maps to illustrate the lack of large-scale, diffuse emission in VLA observations. 
For example. \cite{Carignan13} note that the {\em total} \hi\ mass estimated from the KAT-7 observations agrees well with single-dish observations which do not resolve out any structure.

Figure \ref{f:n3109} shows how the KAT-7 observations reveal a pronounced warp in the edge-on \hi\ disk while this is only visible as a slight dip in the VLA data. 
The question remains if the addition of an additional diffuse level will change the global morphology parameters or if their value is mostly determined by the morphological detail in the VLA data. 

We ran our morphological code on the KAT-7 image twice, delineated by different contours, one similar to the area covered by the VLA-ANGST outer contour and one defining the limit of the diffuse emission. Table \ref{t:kat} lists the resulting parameters. There are notable differences between the VLA and KAT-7 observations, to both the outer contour as well as an area corresponding to the VLA-ANGST outer contour.
 
The differences between the two KAT-7 contours are noticeable in S, G and \gm. The inclusion of a large number of low-intensity pixels will result in a completely different distribution and hence Gini and \gm parameters. The higher range in contrast results in a higher Smoothness --meaning a clumpier image-- compared to just the inner contour.

Comparing the inner contour in the KAT-7 observations and the VLA-ANGST observations (second and fourth column in Table \ref{t:kat}), we note differences in C, S, G, and to a lesser extend \m20 and \gm. 

The majority of morphological parameters are modified if we change spatial resolution, especially sampling over areas greater than a kpc. The addition of a large-scale structure only changes the measures of (in)equality in the distribution: Gini and \gm. Thus, while large-scale structure is missed by VLA interferometric surveys such as LITTLE-THINGS and VLA-ANGST, most of the morphological information is contained in the small-scale structures that are resolved by such observations.

\begin{table}
\caption{The different morphological parameters for NGC 3109. }
\begin{center}
\begin{tabular}{l l l l}
Parameter	& VLA-ANGST					& KAT-7			& inner contour\\
			&							&				& (32Jy/Beam) \\
\hline
\hline
C			& 0.0 $\pm$ 0.010				& $0.20\pm 0.02$ 	& $0.20 \pm 0.08$ \\	
A			& 1.0 $\pm$ 0.000				& $1.0 \pm 0.0$	& $1.0 0.0$\\
S			& 0.047 $\pm$ 0.031			& $0.22 \pm 0.09$ 	& $0.38 \pm 0.16$\\
G			& 0.445 $\pm$ 0.010 			& $ 0.68 \pm 0.01$ 	& $0.23 \pm 0.11$\\
\m20			& -0.741 $\pm$ 0.002			& $-0.71\pm 0.02$ 	& $-0.70 \pm 0.019$\\
\gm			& 0.429 $\pm$ 0.011			& $0.67 \pm 0.01$	& $0.22 \pm 0.11$ \\
\hline
\end{tabular}	
\end{center}
\label{t:kat}
\end{table}%

\section{Results}
\label{s:results}

To explore the relationships between the \hi\ morphological parameters and the tidal and star-formation
tracers, we show two plots, one where we compare \hi\ parameters against each other, 
colour-coded with a comparison parameter, if available. This is to identify possible sections of \hi\ 
morphology parameter space where special cases reside. 

Secondly, we plot the comparison parameter (e.g., a star-formation measure) against the six \hi\ 
morphological parameters directly and, thirdly, we calculate the Spearman ranking (-1 perfectly 
anti-correlated, 0 uncorrelated, and 1 fully correlated) between the comparison parameter and 
\hi\ morphological parameter.

The LITTLE-THINGS sample was drawn from \cite{Hunter04} and the VLA-ANGST from 
\cite{Karachentsev04}, meaning that the parameters on tidal effect or star-formation from the literature 
are not available for our full sample. 
We compare the \hi\ morphology to the tidal disturbance, and several parameterizations of the 
ongoing and past star-formation, to explore which of these are the dominant factor in the overall 
shape of the \hi\ in these dwarf galaxies.

\begin{figure*}
\begin{center}
     \begin{minipage}{0.65\linewidth}
	\includegraphics[width=\textwidth]{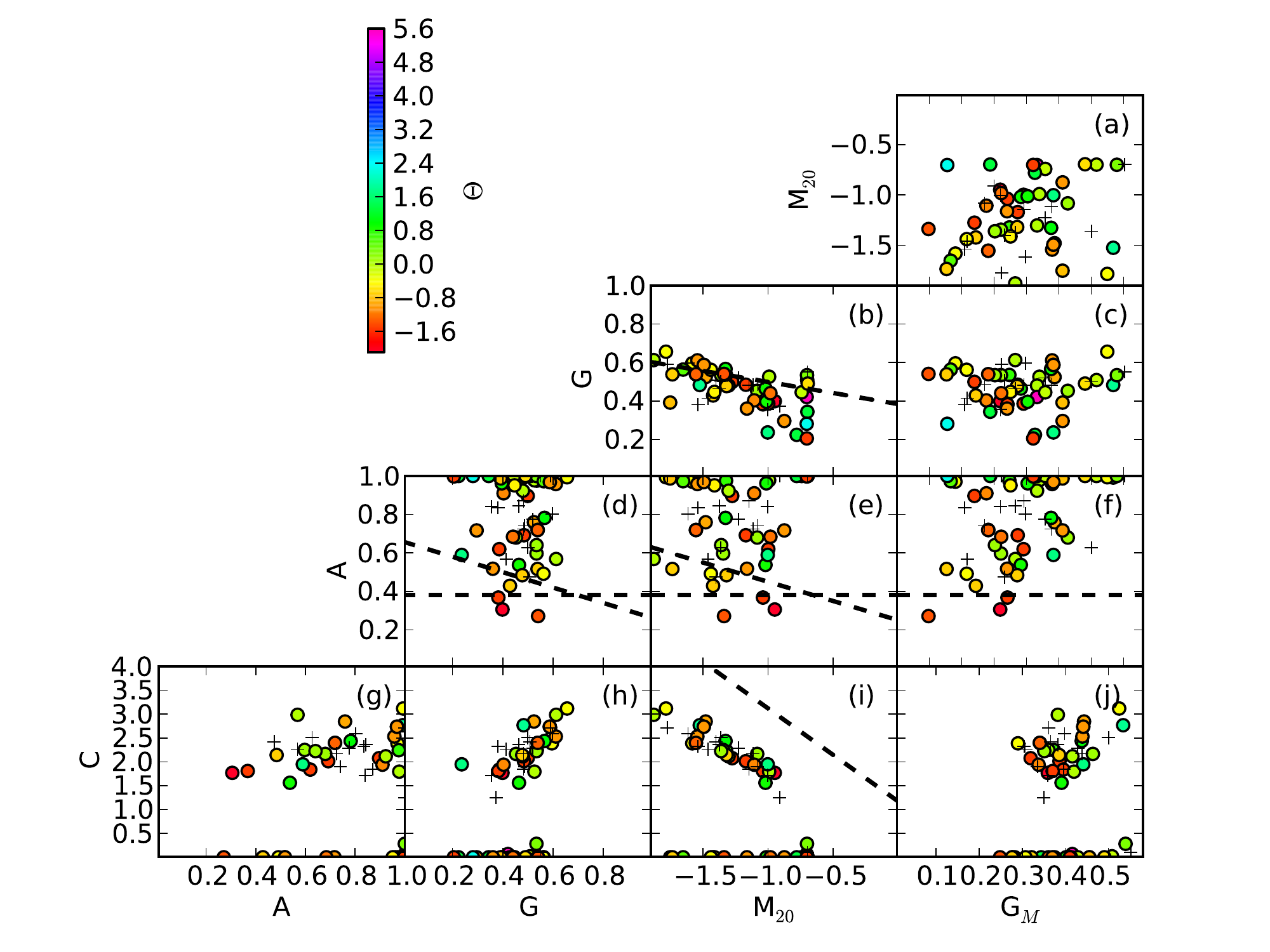}
    \end{minipage}\hfill
    \begin{minipage}{0.34\linewidth}
	\includegraphics[width=1.1\textwidth]{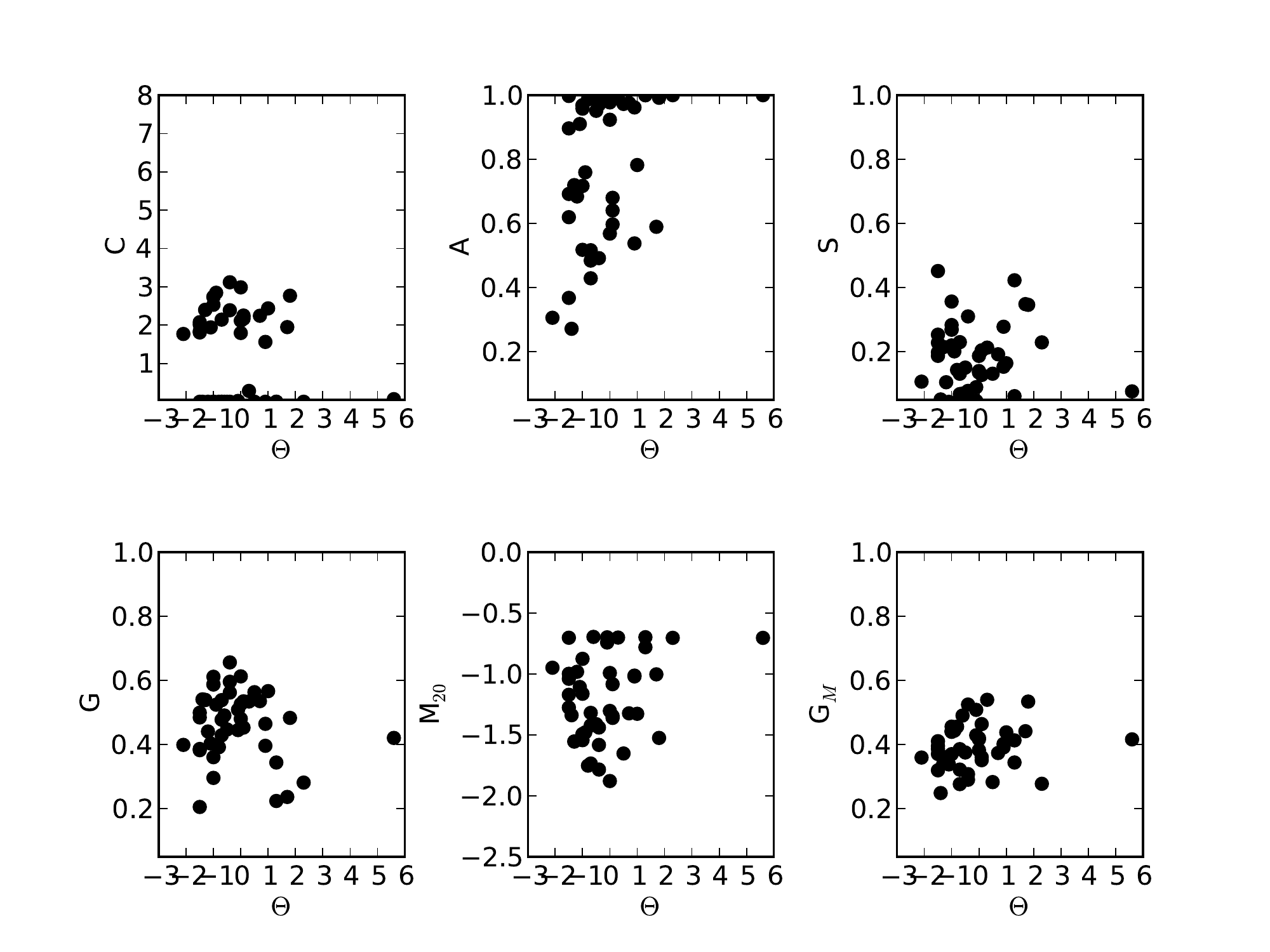}
	\caption{\label{f:Theta:6par}The six \hi\ morphology parameters as a function of the tidal parameter $\Theta$ from \protect\cite{Karachentsev04}. $\Theta=0$ is undisturbed, a negative value means residing in an under-density and a positive one means increased tidal influence by neighbouring systems.}
    \end{minipage}
    \begin{minipage}{\linewidth}
	\caption{\label{f:Theta} The distribution of the \hi\ morphological parameters with the tidal index $\Theta$ from \protect\cite{Karachentsev04}. Dashed lines are the criteria for major interaction from \protect\cite{Holwerda11c}, based on the WHISP sample or established morphological selections of mergers in optical data. Values of the tidal index appear poorly related to the position of the dwarfs in \hi\ morphology space. Only very isolated dwarfs ($\Theta < -1$) are also very symmetric in \hi\ ($A < 0.4$), the \hi\ parameter with the highest correlation with tidal index (Table \ref{t:spear}).}
    \end{minipage}
\end{center}
\end{figure*}

\subsection{Tidal Index}
\label{s:theta}

Figure \ref{f:Theta} shows the distribution of \hi\ column density map morphologies, coded by the 
tidal parameter ($\Theta$) from \cite{Karachentsev04}. Figure \ref{f:Theta:6par} shows the direct 
relation between the six \hi\ morphological parameters and the tidal parameter. Of all the parameters, 
only \hi\ Asymmetry is weakly related to $\Theta$ (see also Table \ref{t:spear}). 
The six morphological criteria for interaction of more massive galaxies are denoted with dashed lines 
in Figure \ref{f:Theta} and further.

These criteria are:
\begin{equation} 
A > 0.38 ~ {\rm and} ~S>A
\label{eq:AS}
\end{equation}
\noindent from \cite{CAS}. This is the straight dashed line in sub-panels (d), (e) and (f) in Figure \ref{f:Theta} etc.

\cite{Lotz04} added two different criteria, one using Gini and $M_{20}$:
\begin{equation} 
G > -0.115 \times M_{20} + 0.384
\label{eq:GM20}
\end{equation}
\noindent shown by the dashed line in sub-panel (b) in Figure \ref{f:Theta}.

\cite{Lotz04} also defined a interaction criterion based on Gini and Asymmetry: 
\begin{equation} 
G > -0.4 \times A + 0.66 ~ \rm or ~ A > 0.4.
\label{eq:GA}
\end{equation}
\noindent which is shown as an inclined dashed line in sub-panel (d) in Figure \ref{f:Theta}.
This latter criterion is a refinement of the Conselice et al. A-S criterion in equation \ref{eq:AS}.

\cite{Holwerda11b} defined three interaction criteria specifically for \hi\ data (typically lower spatial resolution, affected by spatial filtering (i.e., sensitivity to a specific angular scale), and smaller dynamical range than optical data). Ongoing spiral-spiral tidal interactions can be identified by:
\begin{equation}  
G_M > 0.6,
\label{eq:GM}
\end{equation}
\noindent which is not shown in Figure \ref{f:Theta} as the range of \gm values in the Dwarf galaxy \hi\ surveys does not extend this high.
However, it is shown as the vertical dashed line in sub-panels (a), (c), (f) and (j) in Figures \ref{f:sdss:sm}, \ref{f:sdss:sf} and \ref{f:sdss:ssf}.

Or their interaction can be identified based on Asymmetry and \m20:
\begin{equation} 
A > -0.2 \times M_{20} + 0.25,
\label{eq:AM20}
\end{equation}
\noindent or concentration and \m20, similar to the criteria from \cite{Lotz04} (equations \ref{eq:GM20} and \ref{eq:GA}), as:
\begin{equation} 
C_{82} > -5 \times M_{20} + 3. 
\label{eq:CM20}
\end{equation}

The first thing we note, is that the vast majority of dwarfs galaxies lie on one side of these criteria. 
The G-A and G-\m20\ criteria include almost all; the C-\m20\ and \gm\ criteria both completely 
exclude the dwarf from the tidally interacting. Only the Gini-\m20\ criterion bisects the dwarf sample. If we compare 
these criteria to the values of the tidal index $\Theta$, there is little correlation with the position in \hi\ 
morphology parameter space. The exception are three galaxies with low values of $\Theta$, i.e. very 
isolated, and a low asymmetry value ($A<0.4$). 

Figures  \ref{f:Theta:6par} and \ref{f:Theta} show that the \hi\ morphology is not primarily affected by the gravitational 
interaction. One can identify very isolated galaxies from the \hi\ morphology ($A<0.4$) but the 
majority of criteria that apply to spiral galaxies cannot be applied to dwarf \hi\ morphology to identify 
or even rank the level of interaction. We identify DDO47, DDO87, and UGC8833 as the most isolated 
dwarfs in our sample, based on their Asymmetry.

\subsection{Ongoing and Past Star-Formation}
\label{s:sfr}

The star-formation can be measured by a variety of techniques corresponding to different typical 
timescales: 
(a) H$\alpha$ emission which traces the currently forming massive stars still in their ionized birth clouds (tens of Myr), 
(b) far-ultraviolet (\fuv) emission which traces the population of massive young stars, after the surrounding gas has 
dissipated (hundreds of Myr), and 
(c) resolved stellar populations which trace the star-formation history to Gyr timescales.

Here we compare the \hi\ morphologies to these three star-formation tracers to explore which 
time-scale of star-formation informs the morphology of the atomic gas: current from H$\alpha$ 
emission, reported in \cite{Hunter06}, recent from \fuv\ fluxes, reported in \cite{Hunter10} and 
McQuinn et al. {\em in preparation}, or long-term star-formation history from HST resolved stellar 
populations, reported in \cite{Weisz11b} and \cite{McQuinn12a}.

\begin{figure*}
\begin{center}
     \begin{minipage}{0.65\linewidth}
	\includegraphics[width=\textwidth]{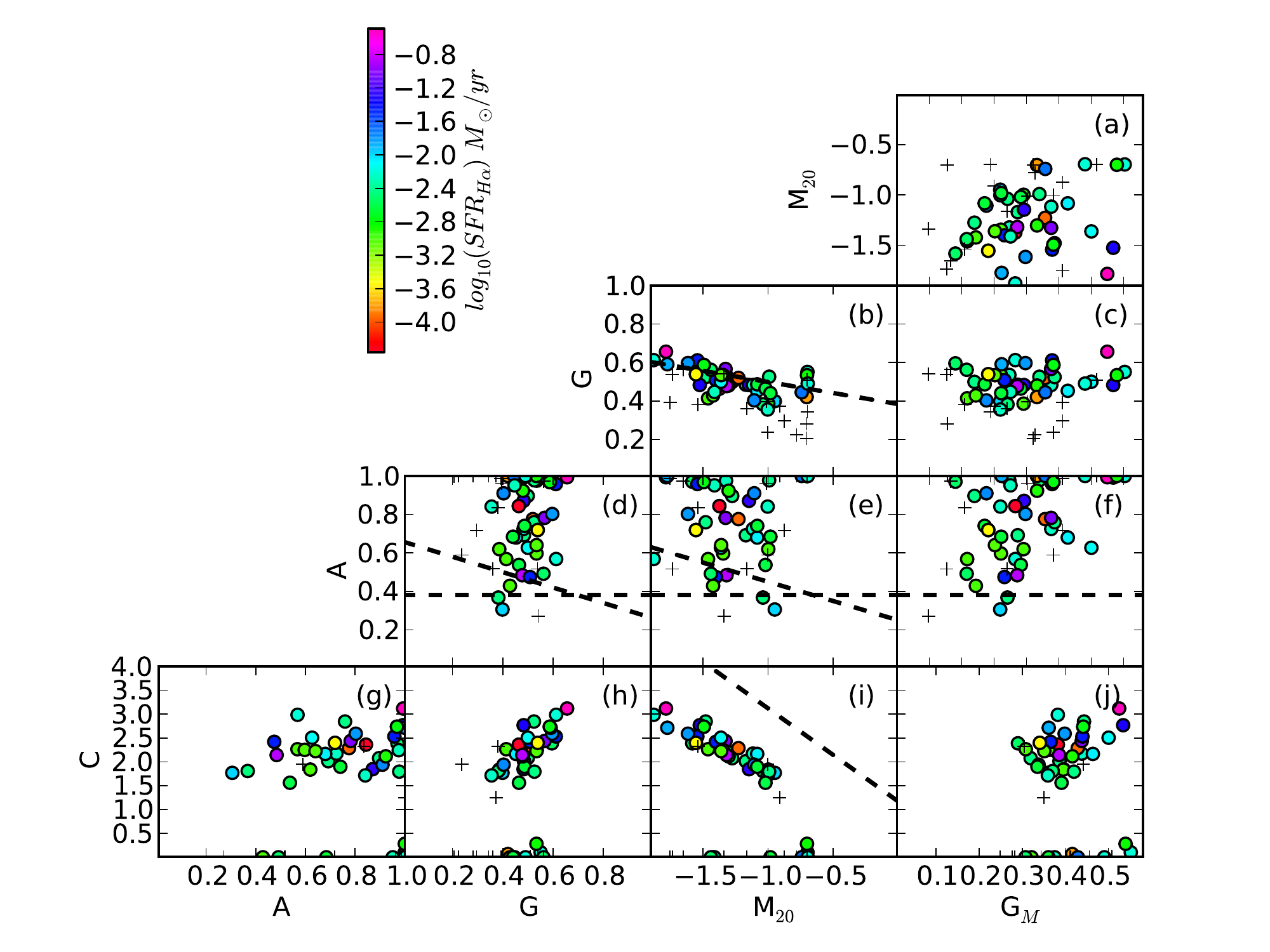}
    \end{minipage}\hfill
    \begin{minipage}{0.34\linewidth}
	\includegraphics[width=1.1\textwidth]{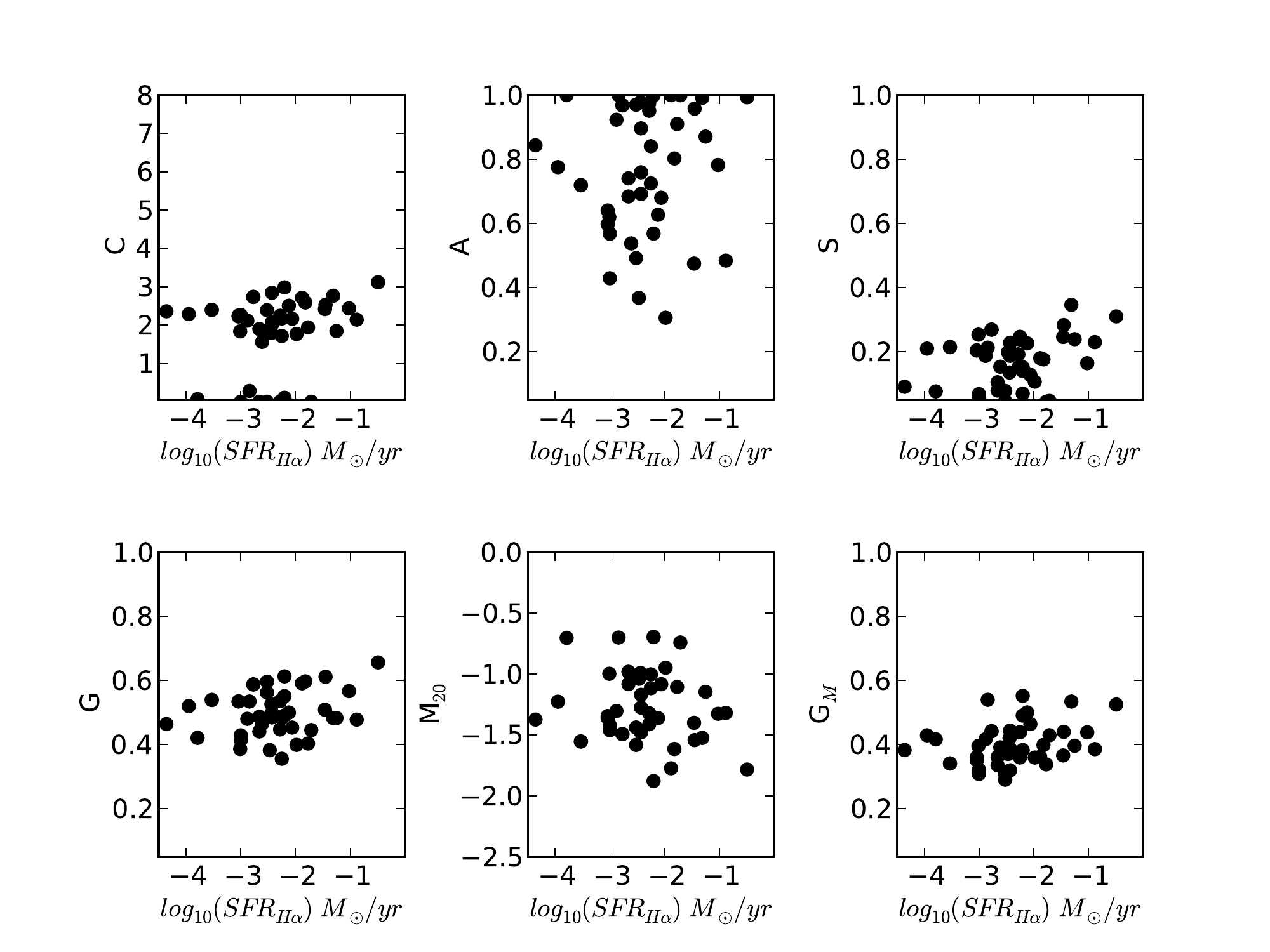}
	\caption{\label{f:sfr:6par} The relation between H$\alpha$-derived SFR from \protect\cite{Hunter06} and the six morphology parameters. }
    \end{minipage}
    \begin{minipage}{\linewidth}
	\caption{\label{f:sfr} The distribution of the \hi\ morphological parameters colour coded by the {\em total} star-formation rate inferred from H$\alpha$ flux from \protect\cite{Hunter04}. None of the \hi\ morphology parameter relate to the {\em total} star-formation, $log_{10}(SFR) ~ M_\odot yr^{-1}$ (Table \ref{t:spear}).}
    \end{minipage}
\end{center}
\end{figure*}

\begin{figure*}
\begin{center}
     \begin{minipage}{0.65\linewidth}
	\includegraphics[width=\textwidth]{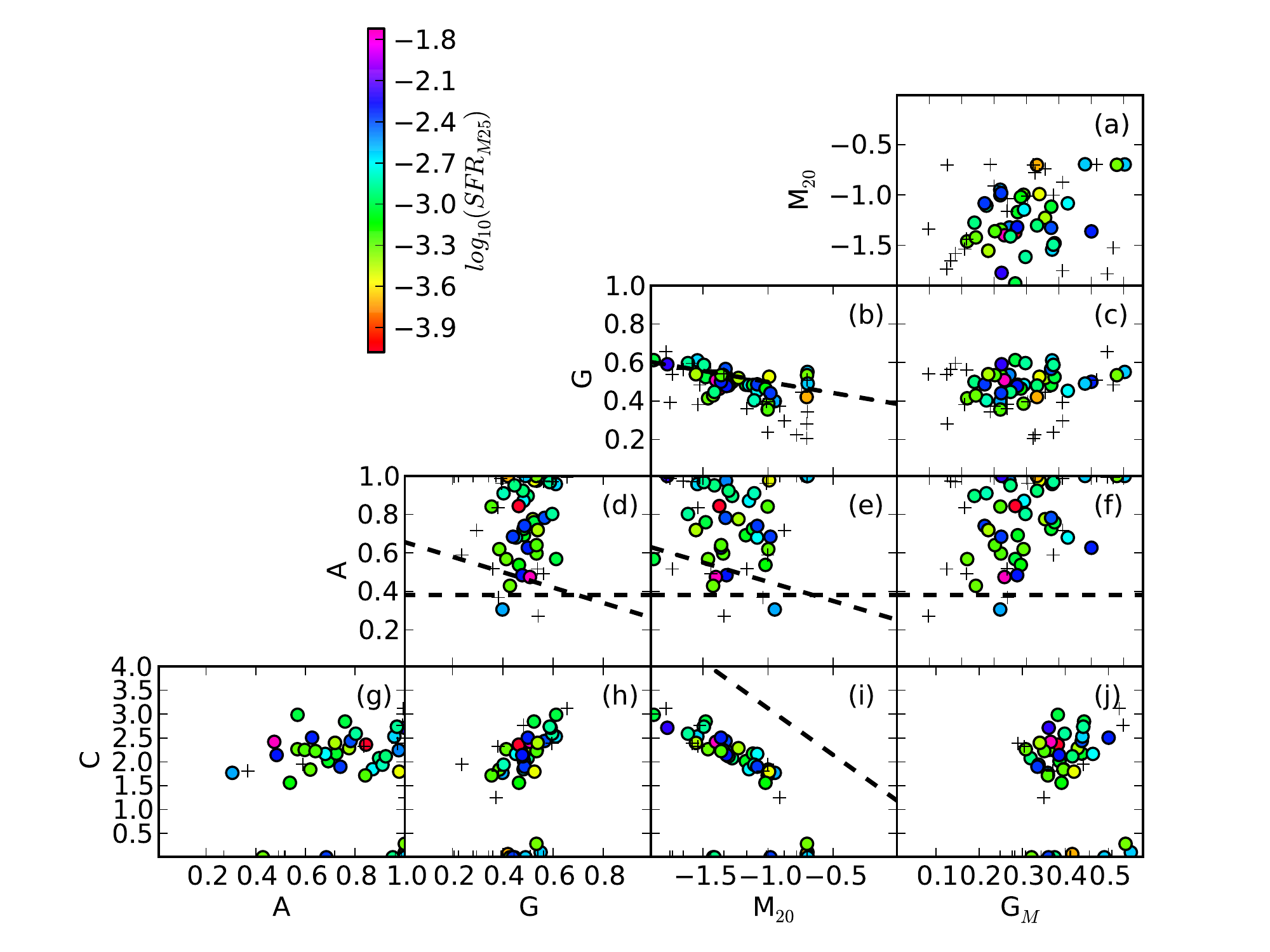}
    \end{minipage}\hfill
    \begin{minipage}{0.34\linewidth}
	\includegraphics[width=1.1\textwidth]{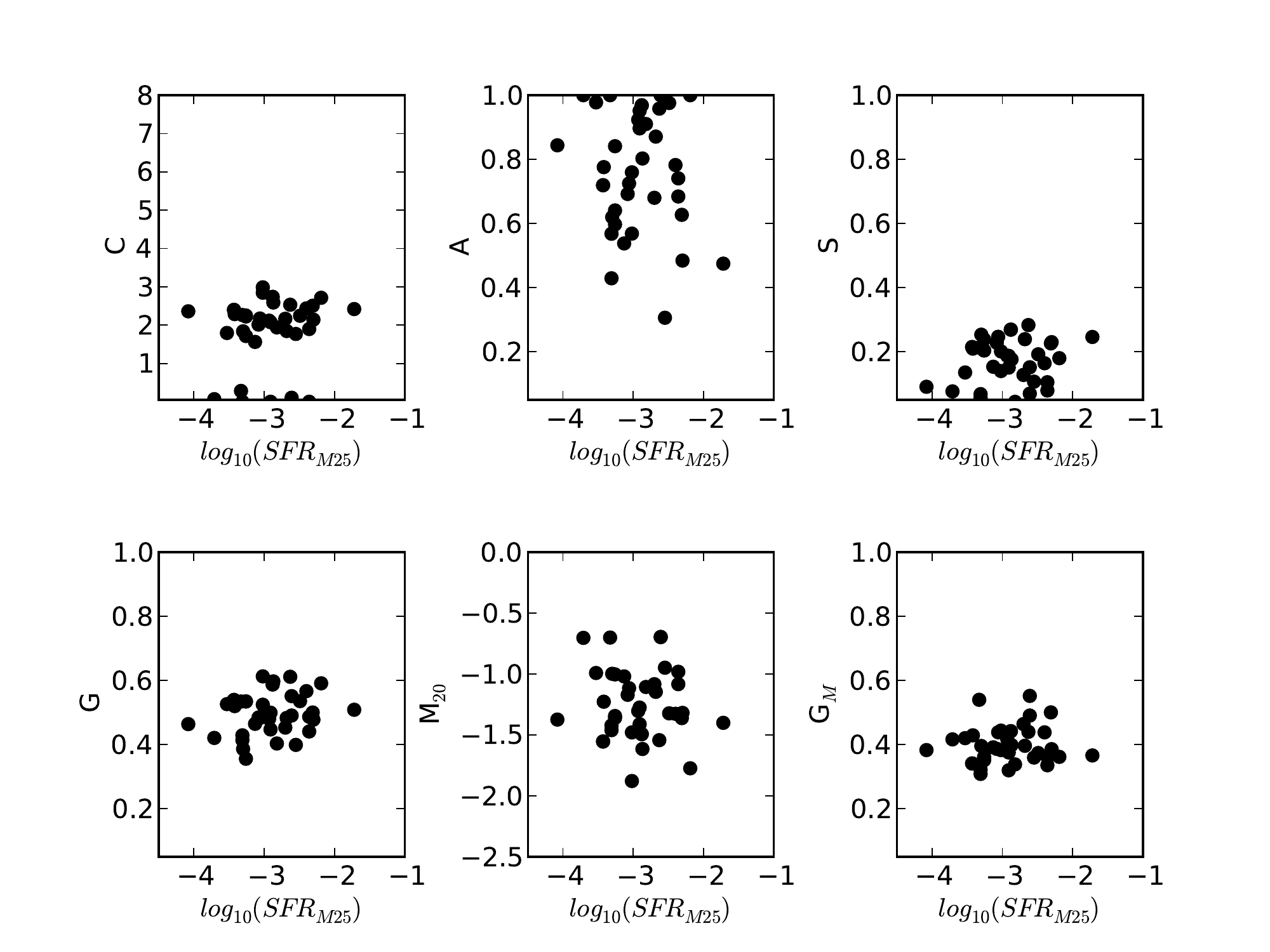}
	\caption{\label{f:sfd:6par} The star-formation surface density over the optical disk ($R_{25}$) from \protect\cite{Hunter04} compared to the \hi\ morphological parameters  None of the \hi\ morphology parameters are closely related to the H$\alpha$ surface brightness (Table \ref{t:spear}).}
    \end{minipage}
    \begin{minipage}{\linewidth}
	\caption{\label{f:sfd} The distribution of the \hi\ morphological parameters colour coded by the star-formation surface density over the optical disk ($R_{25}$), inferred from the H$\alpha$ flux, from \protect\cite{Hunter04}. None of the \hi\ morphology parameters are closely related to the H$\alpha$ surface brightness (Table \ref{t:spear}).}
    \end{minipage}
\end{center}
\end{figure*}

\begin{figure}
\begin{center}
\includegraphics[width=0.5\textwidth]{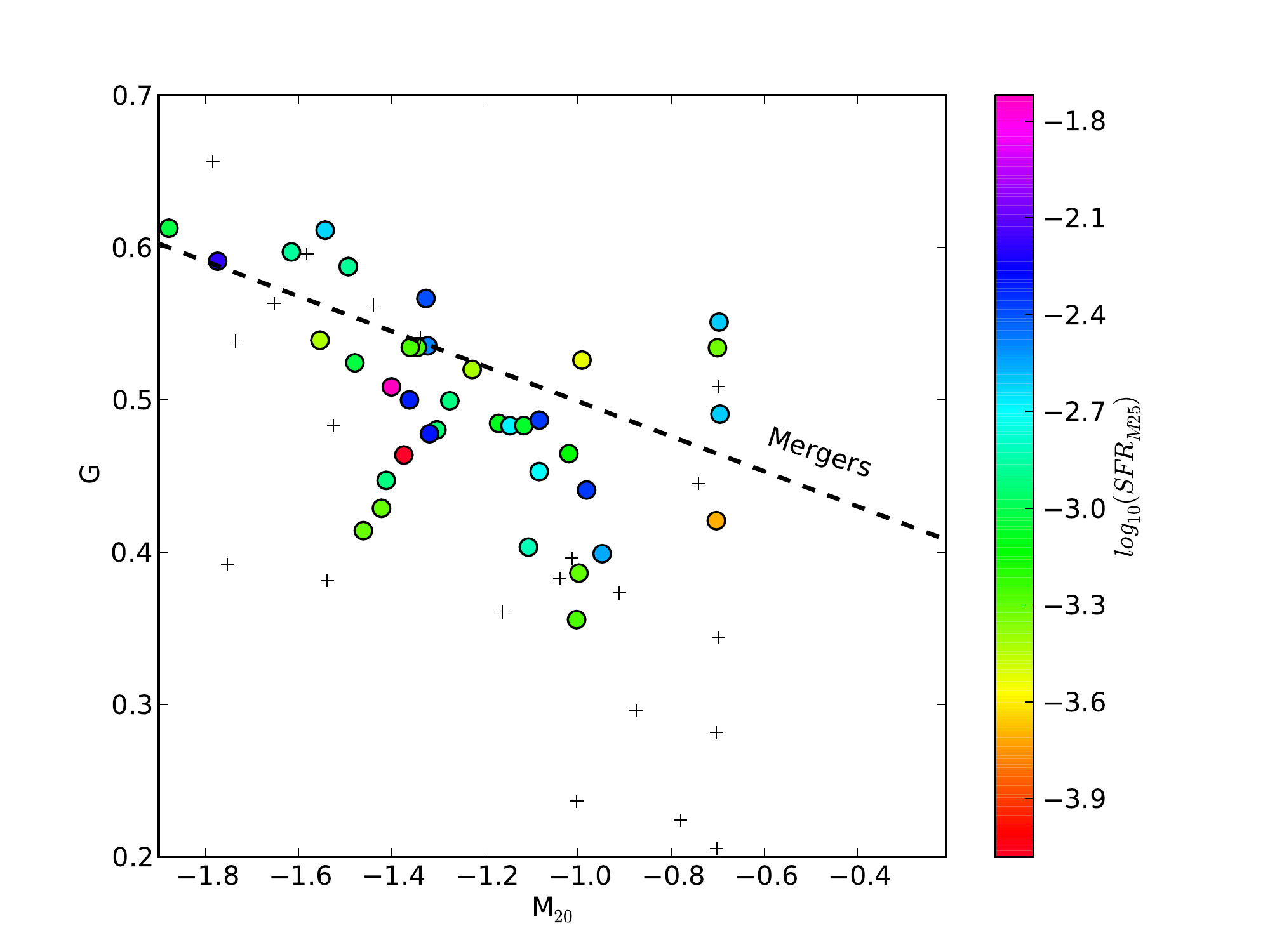}
\caption{The \m20\ and Gini parameters of the \hi\ maps colour coded by the star-formation surface density over the optical disk ($log_{10}(SFR_{M25})$) from \protect\cite{Hunter04}. The dashed line is a criterion for major interaction from \protect\cite{Holwerda11c, Lotz04}.  }
\label{f:GM20:sfrd}
\end{center}
\end{figure}

\subsubsection{Current Star-Formation: $SFR_{H\alpha}$}
\label{s:sfrHa}

Figure \ref{f:sfr} shows the distribution of \hi\ column density map morphologies, coded by their 
{\em total} star-formation, inferred from H$\alpha$ flux from \cite{Hunter04}, their Table 3. 
Typical low star-formation rates from H$\alpha$ flux values ($SFR_{H\alpha} \sim -2.8$) are found 
predominantly in low \hi\ Asymmetry galaxies. They also suggestively cluster elsewhere in \hi\ 
morphology space, e.g., Figure \ref{f:sfr}, sub-panels (a), (g) or (j). However, a direct comparison between 
current star-formation and the \hi\ morphology reveals little direct correlation between the \hi\ 
morphology parameters and the current star-formation (Figure \ref{f:sfr:6par} and Table \ref{t:spear}).

Figure \ref{f:sfr} shows the distribution of \hi\ column density map morphologies, coded by the 
star-formation surface density ($log_{10}(SFR_{M25})$) from \cite{Hunter04}, their Table 3, 
based on the H$\alpha$ luminosity over the 25 mag/arcsec$^2$ radius ($R_{25}$). A similar 
value can be obtained over the optical radius ($R_D$). The star-formation surface density stands 
out in \hi\ morphology space in the Gini parameter (when computed over $R_D$, see Table \ref{t:spear}). 

Because the dwarfs straddle the Gini-\m20\ criterion for interaction (equation \ref{eq:GM20}), 
we explore these parameters with current star-formation surface density in detail in Figure \ref{f:GM20:sfrd}.
Lower star-formation surface densities ($log_{10}(SFR_{M25}) < -3$) tend to lie below this interaction criterion while
those above it have star-formation surface densities above it. That is not to say that those dwarfs are indeed interacting but
their \hi\ appearance parameterised by Gini-\m20\ combined and their current star-formation do appear to be linked. 

%

\begin{figure*}
\begin{center}
     \begin{minipage}{0.65\linewidth}
	\includegraphics[width=\textwidth]{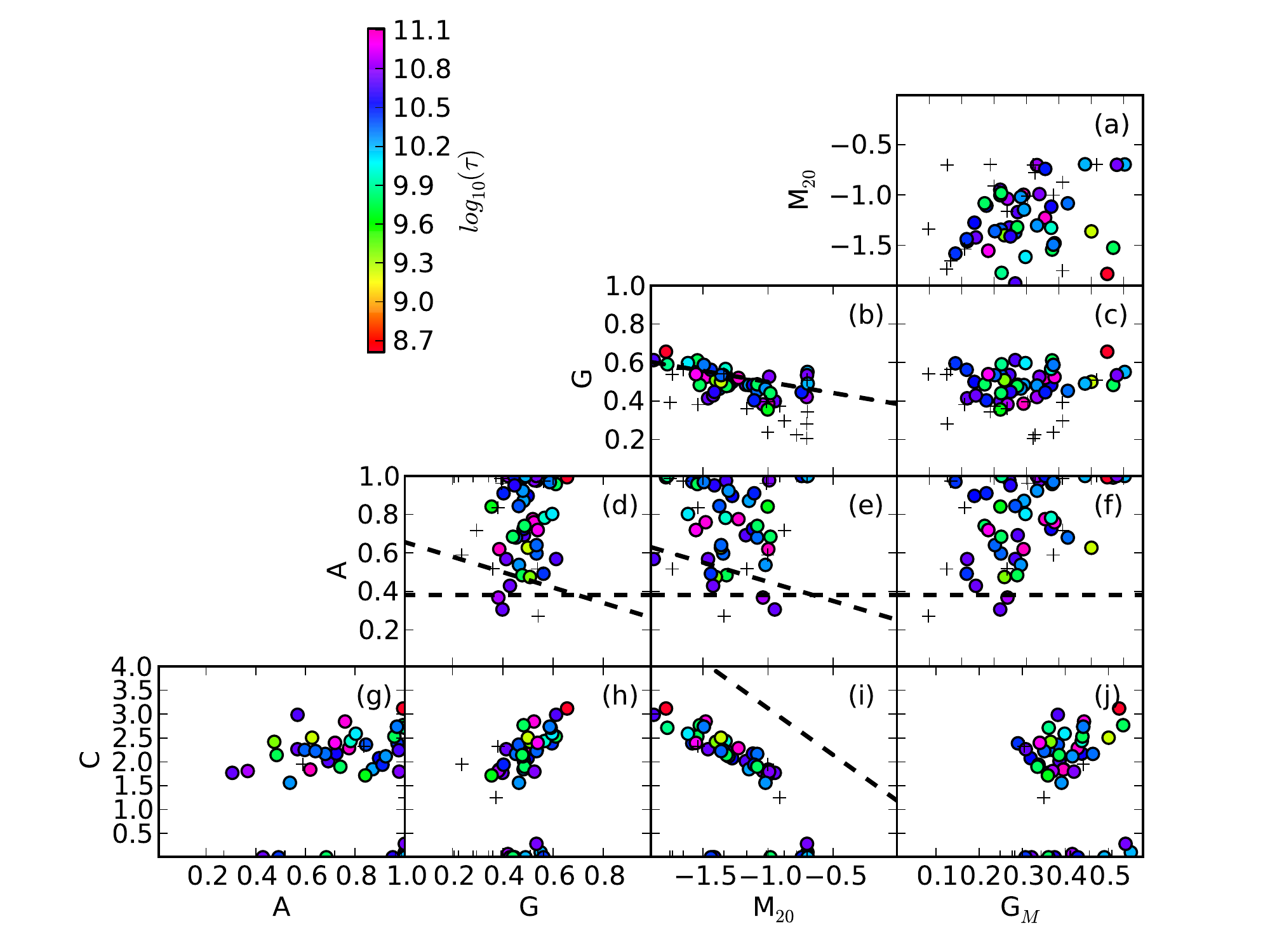}
    \end{minipage}\hfill
    \begin{minipage}{0.34\linewidth}
	\includegraphics[width=1.1\textwidth]{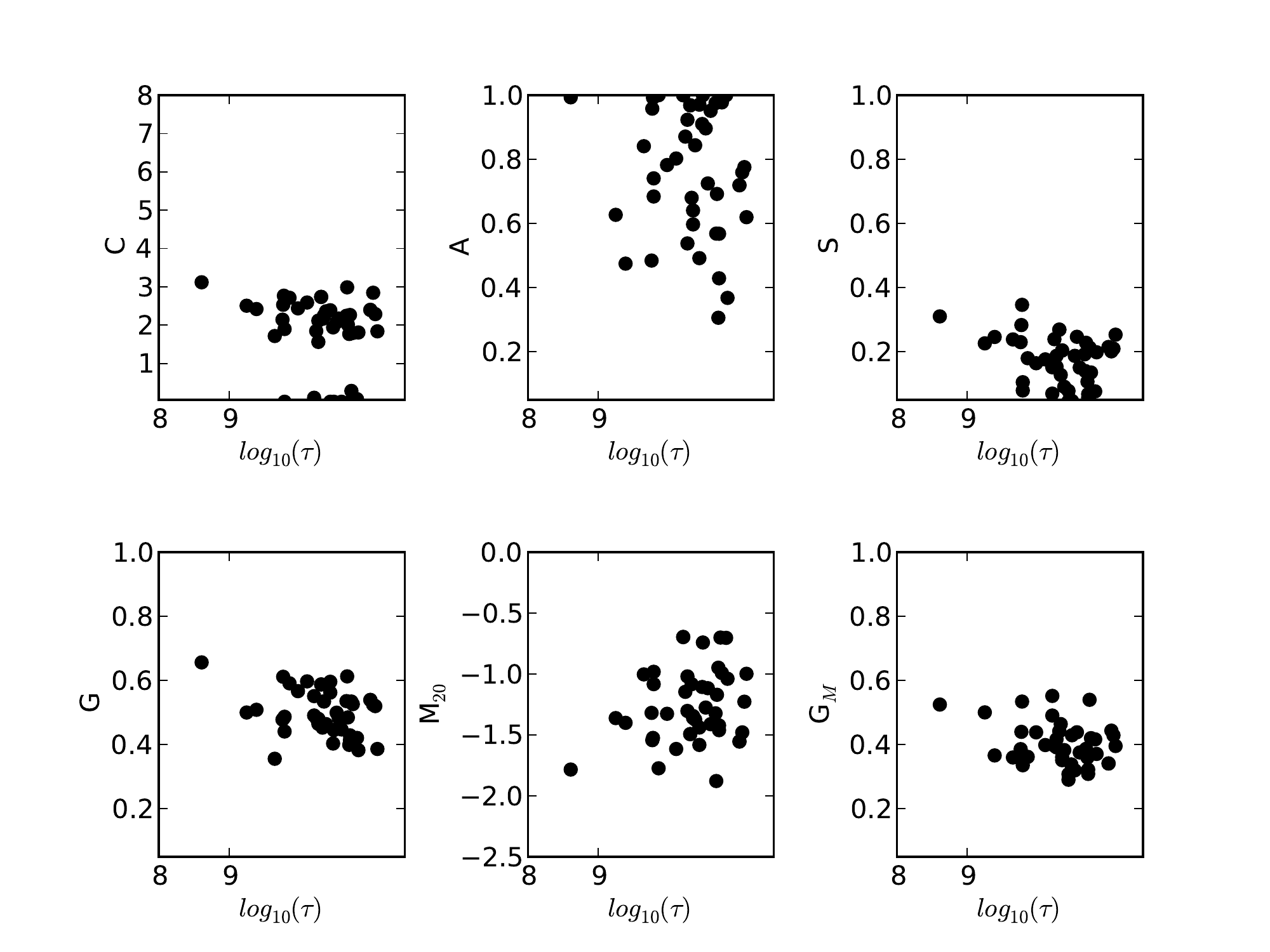}
	\caption{\label{f:tau:6par} The relation between gas exhaustion time estimate from \protect\cite{Hunter04} and the \hi\ morphological parameters. None of the \hi\ morphological parameters are closely related to the exhaustion time (Table \ref{t:spear}).}
    \end{minipage}
    \begin{minipage}{\linewidth}
	\caption{\label{f:tau} The distribution of the \hi\ morphological parameters colour coded by the gas exhaustion time estimate from \protect\cite{Hunter04}. Dashed lines are the criteria for major interaction from \protect\cite{Holwerda11c}, based on the WHISP sample or established morphological selections of mergers in optical data. None of the \hi\ morphological parameters are closely related to the exhaustion time (Table \ref{t:spear}).}
    \end{minipage}
\end{center}
\end{figure*}

\subsubsection{Gas Exhaustion Time $\tau_c$}
\label{s:tau}

\cite{Hunter04} supply an estimate of the time it will take each galaxy to exhaust its gas supply 
(estimated from single-dish observations) by their current star-formation. Figure \ref{f:tau} shows 
the morphology distribution colour-coded by the gas exhaustion time ($\tau$). The quickly exhausted 
galaxies are --unsurprisingly-- those with a high star-formation surface density. The exhaustion time 
estimate is a simple one, otherwise one could perhaps expect a relation with the concentration of fuel
or the Gini parameter (an indication of inequality). ISM in lumps close to the star-formation
would be consumed much faster than a smooth gas disk that would need to coalesce in star-forming clouds first.
However, the gas exhaustion time is not strongly related to any of the \hi\ morphology parameters (Figure \ref{f:tau:6par} and Table \ref{t:spear}).

\begin{figure*}
\begin{center}
     \begin{minipage}{0.65\linewidth}
	\includegraphics[width=\textwidth]{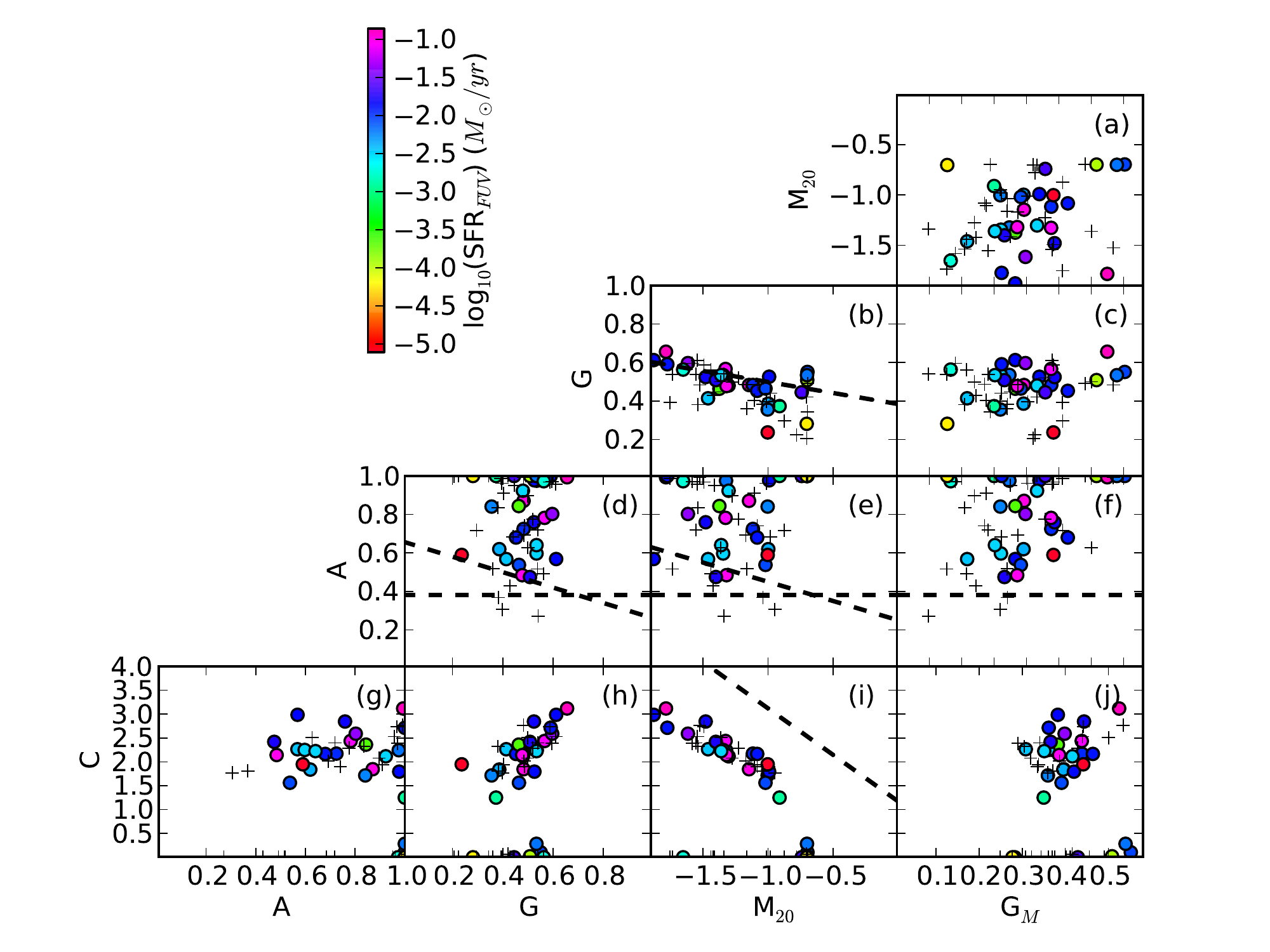}
    \end{minipage}\hfill
    \begin{minipage}{0.34\linewidth}
	\includegraphics[width=1.1\textwidth]{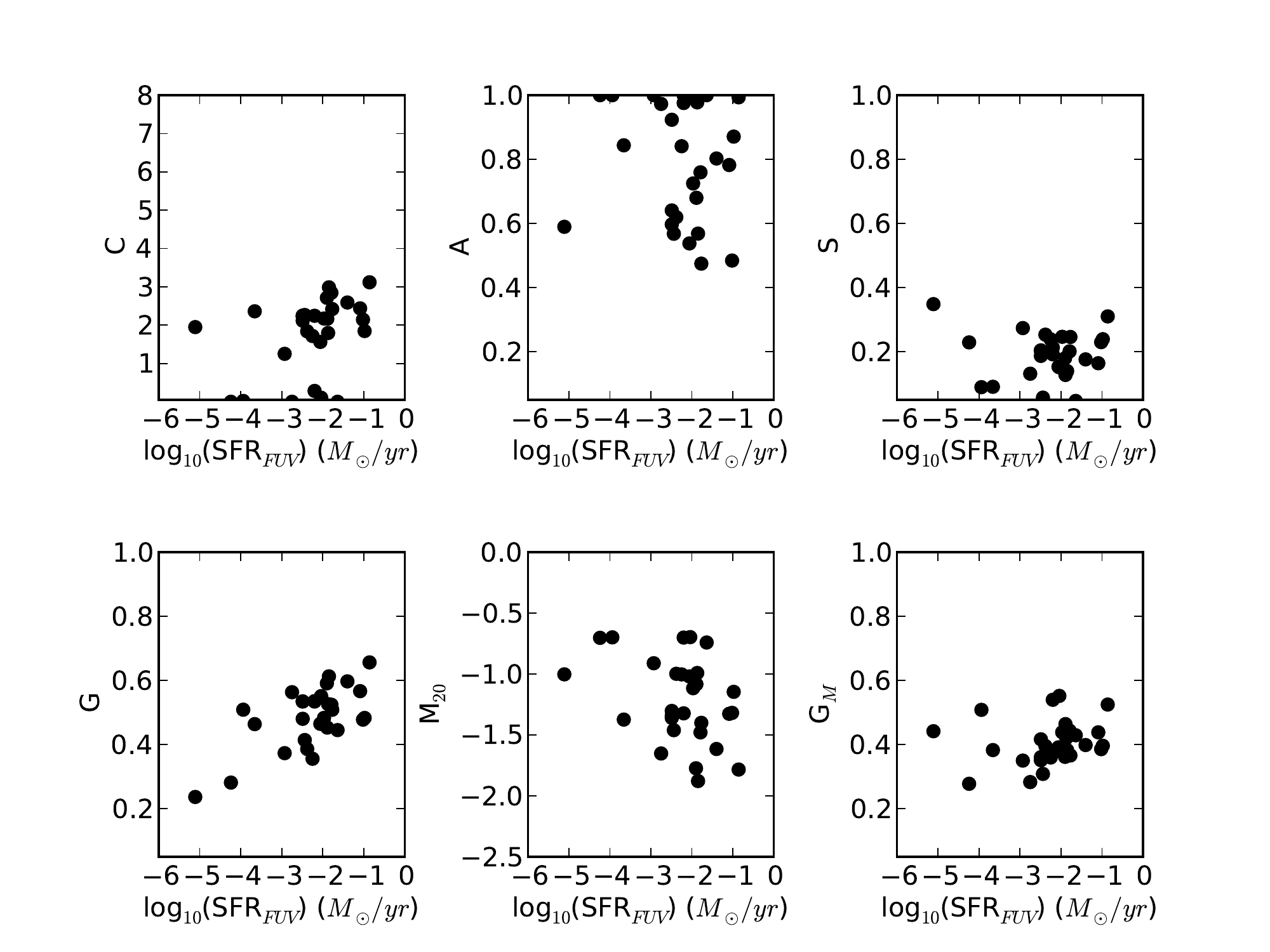}
	\caption{\label{f:sfr:fuv:6par}The relation between \fuv-derived SFR from \protect\cite{Hunter10} and the six morphology parameters. }
    \end{minipage}
    \begin{minipage}{\linewidth}
	\caption{\label{f:sfr:fuv}The distribution of the \hi\ morphological parameters colour coded by the star-formation rate based on \fuv\ flux ($log_{10} (SFR_{FUV})$) reported in \protect\cite{Hunter10}. Some weak trends visible between the SFR and the Gini and \m20\ values in \hi, but without clear delineations. The Spearman indices in Table \ref{t:spear} point to a strong correlation with both Concentration and Gini, a weaker one with \m20\ and \gm.}
    \end{minipage}
\end{center}
\end{figure*}

\begin{figure}
\begin{center}
\includegraphics[width=0.5\textwidth]{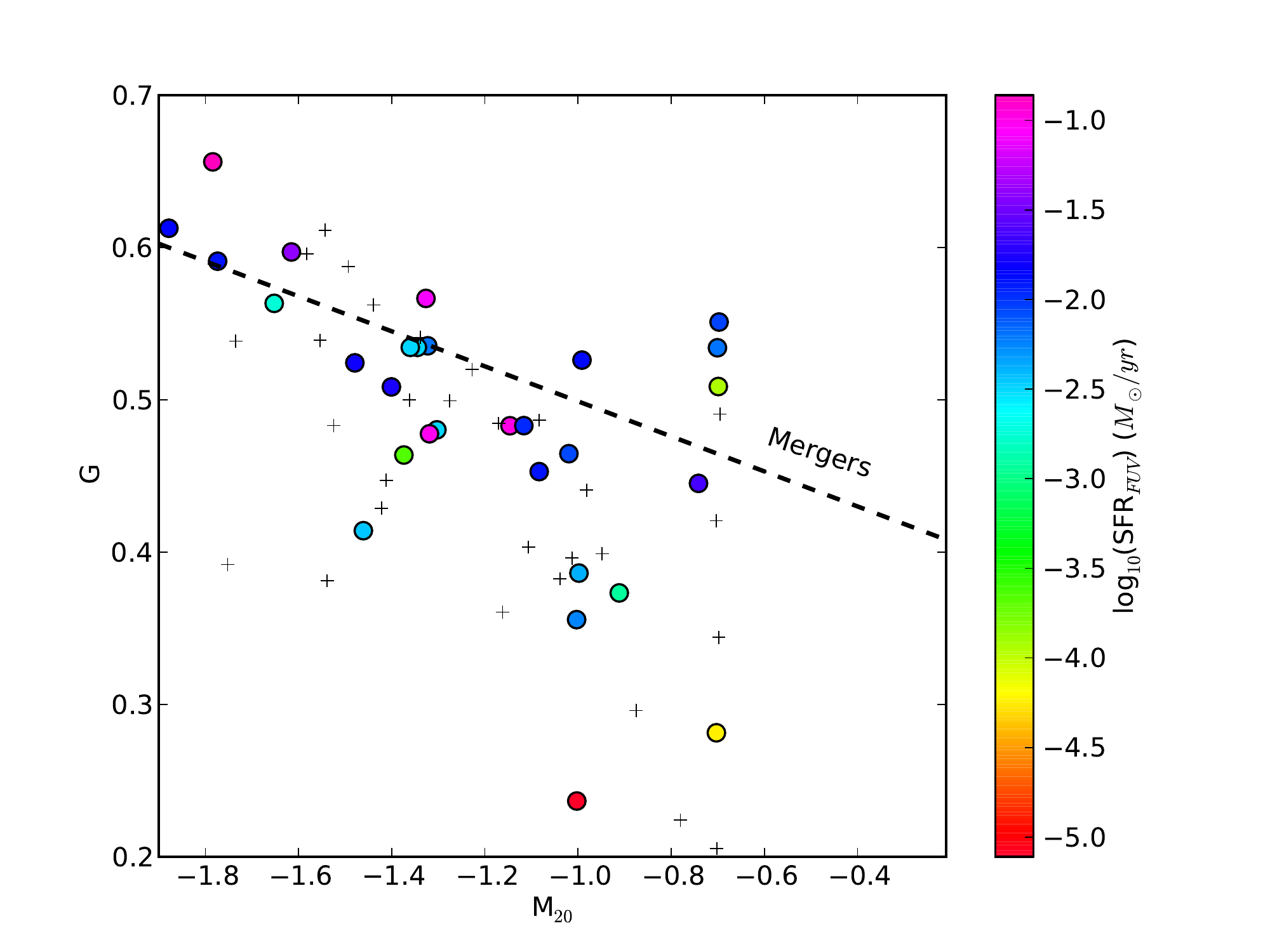}
\caption{The \m20\ and Gini parameters of the \hi\ maps colour coded by the star-formation rate based on \fuv\ flux ($log_{10} (SFR_{FUV})$) reported in \protect\cite{Hunter10}. The dashed line is a criterion for major interaction from \protect\cite{Holwerda11c, Lotz04}.  }
\label{f:GM20:sfr:fuv}
\end{center}
\end{figure}

\subsubsection{Recent Star-formation: $SFR_{FUV}$}
\label{s:sfr:fuv}

\cite{Hunter10} and McQuinn et al. {\em in prep} report \fuv\ fluxes and derived star-formation rates. 
The McQuinn results are corrected for dust extinction using an estimate of the total far-infrared flux 
based on Spitzer/MIPS 24, 70 and 160 $\mu$m maps. The Hunter et al. values are not corrected for 
dust extinction. The general agreement between the Hunter et al. and McQuinn et al. values in a couple 
of overlap cases is good enough for us to combine both \fuv\ star-formation values sets to mark the 
points in Figure \ref{f:sfr:fuv}.   
In Figure \ref{f:sfr:fuv}, there are some weak trends already evident between recent star-formation 
($SFR_{FUV}$) and the Gini and \m20\ values in \hi, but no clear delineations in parameter-space. 
Figure \ref{f:sfr:fuv:6par} confirms these trends: Gini increases with $SFR_{FUV}$ and \m20\ 
decreases, i.e., \hi\ disks become less smooth (higher G) but relatively fewer bright spots at higher 
radii (lower \m20). The Spearman indices in Table \ref{t:spear} corroborate this and also reveal (weak) 
relations with Concentration and \gm\ with recent star-formation.
Figure \ref{f:GM20:sfr:fuv} highlights the Gini-\m20 relation. The majority of galaxies with an $SFR_{FUV}$ measurement straddle the G-\m20\ interaction line. To better constrain the relation between \hi\ Gini and \m20 (and other \hi\ morphological parameters) the sample of $SFR_{FUV}$ measurements will need to be expanded to include all VLA-ANGST and LITTLE-THINGS galaxies. 

\cite{Hunter10} also provide comparisons to the str-formation rates inferred from H$\alpha$ fluxes from 
\cite{Hunter04} and a V-band photometry-based star-formation rate. These values could possibly provide 
a useful direct comparison to which time-scale of ongoing star-formation dominates the morphology for the 
same sample. The figures in Appendix \ref{s:h10} show the dependence of \hi\ morphology on these ratios, 
$SFR_{FUV}/SFR_{H\alpha}$ and $SFR_{FUV}/SFR_{V}$ in Figures \ref{f:ratHa} and \ref{f:ratV} respectively.
Apart from some suggestive clusterings of a few points, there is no real relation between the 
$SFR_{FUV}/SFR_{H\alpha}$ ratio and \hi\ morphology. There may be a some relation between the 
$SFR_{FUV}/SFR_{V}$ ratio and Gini (see Table \ref{t:spear}).

\begin{figure*}
\begin{center}
     \begin{minipage}{0.65\linewidth}
	\includegraphics[width=\textwidth]{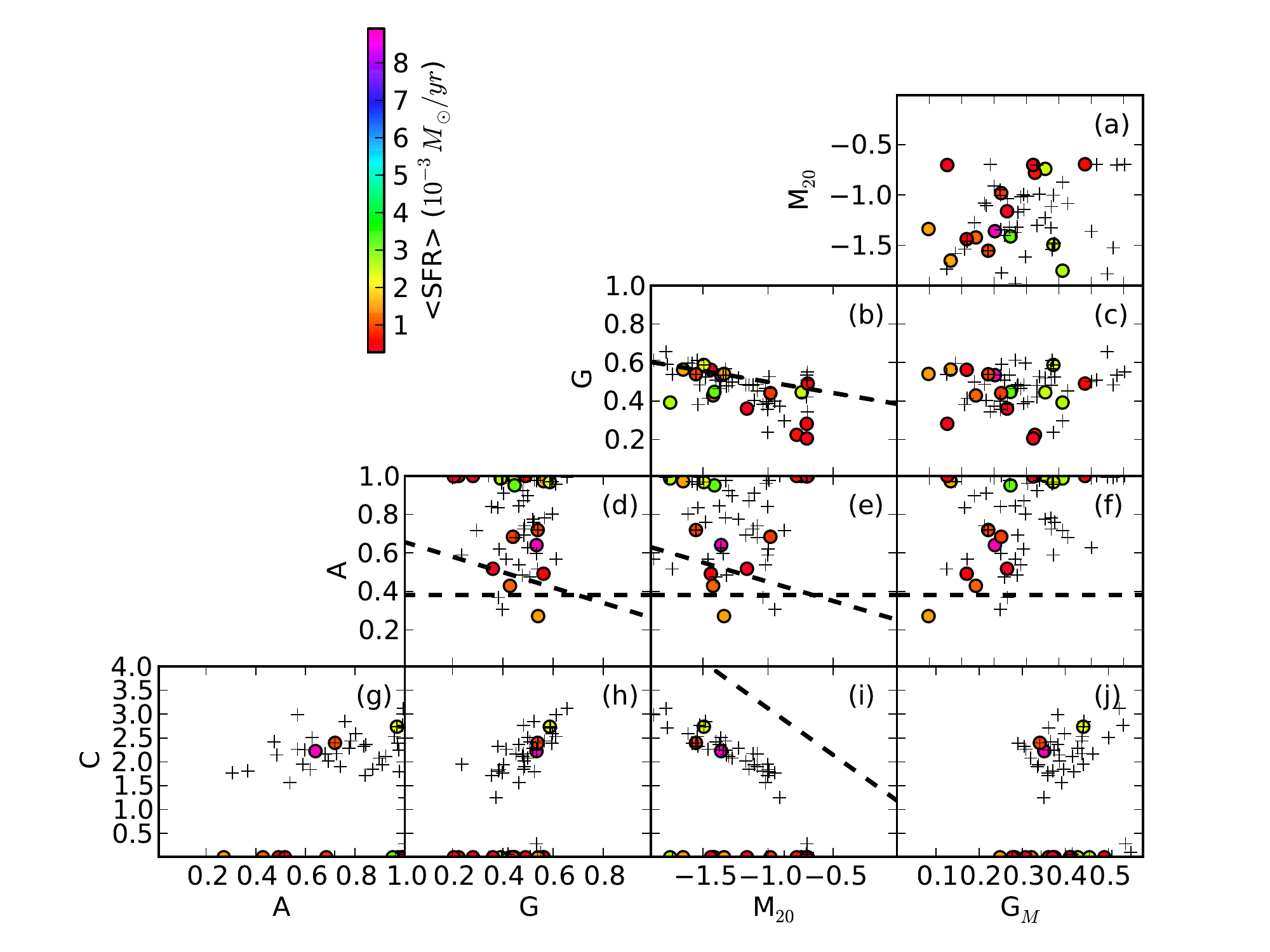}
    \end{minipage}\hfill
    \begin{minipage}{0.34\linewidth}
	\includegraphics[width=1.1\textwidth]{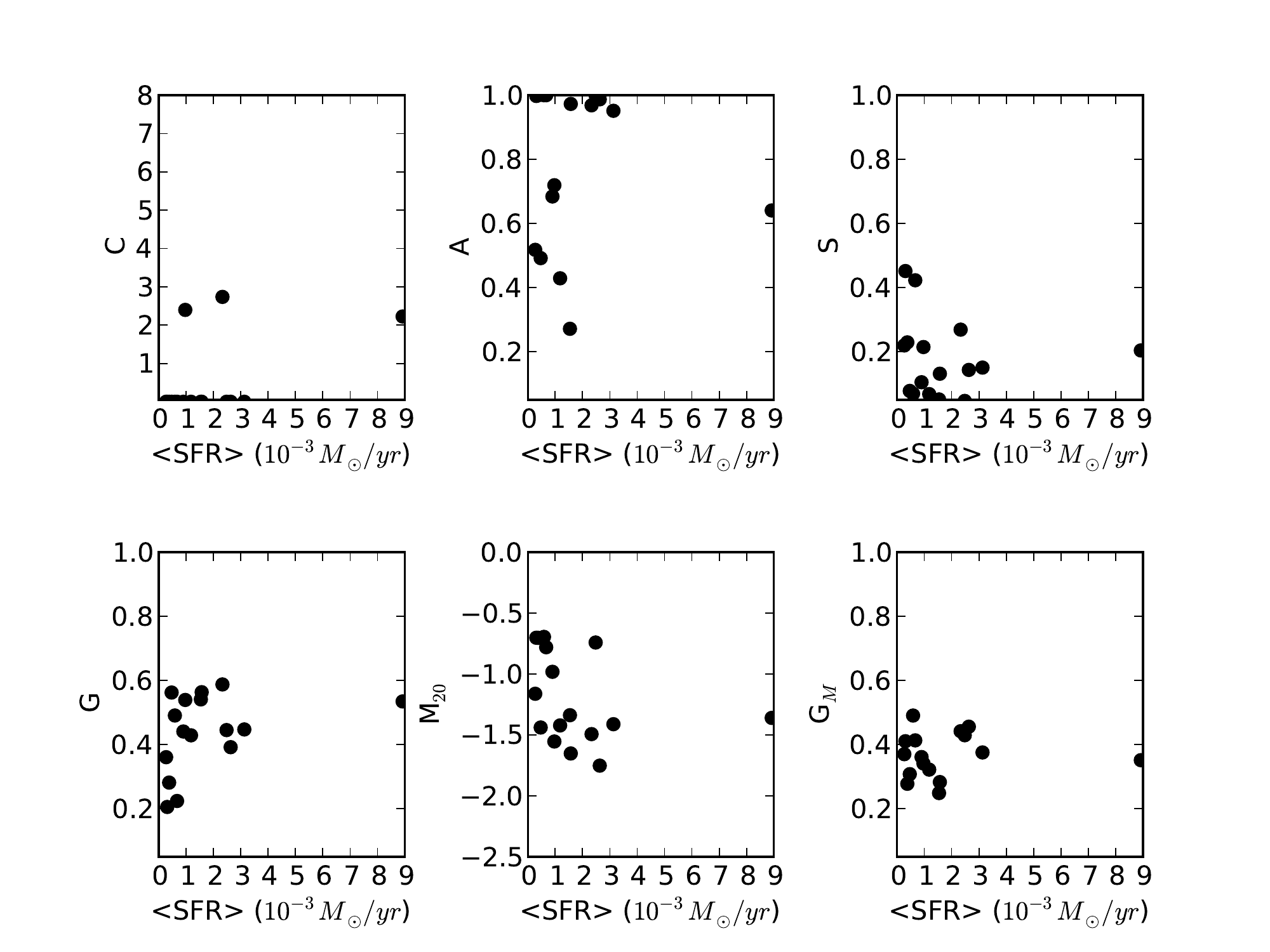}
	\caption{\label{f:SFtime:6par} The relation between $V$-band derived SFR from \protect\cite{Weisz11b} and the six morphology parameters.}
    \end{minipage}
    \begin{minipage}{\linewidth}
	\caption{\label{f:SFtime} The morphology of the \hi\ maps colour coded by the mean past star-formation rate from \protect\cite{Weisz11b} or \protect\cite{McQuinn12b}. There are relatively strong anti-correlations with \m20\ and related to the Gini parameter.}
    \end{minipage}
\end{center}
\end{figure*}

\subsection{Resolved stellar populations: Star-formation History}

One of the main science drivers behind the ANGST survey was to obtain an accurate star-formation 
history from the resolved stellar population as observed with the Hubble Space Telescope 
\citep[see][]{angst, Weisz11a, Weisz11b,Weisz11c, Weisz12a, Weisz12b, McQuinn10a,McQuinn10b, 
McQuinn12a,McQuinn12b}. The main values to compare to the \hi\ morphology are the mean 
star-formation time, the mass-to-light ratio and the mean age of the stars in each galaxy from \cite{Weisz11b}, 
their Tables 2 and 3, supplemented with a few average star-formation rates $<SFR>$ values from 
\cite{McQuinn12b}, their Table 1.  Based on the (weak) relations with ongoing star-formation tracers in 
the previous sections, one could expect some correlation between the shape of the atomic hydrogen 
distribution and the star-formation history, depending on the typical time-scale of the relation.
\begin{figure}
\begin{center}
\includegraphics[width=0.5\textwidth]{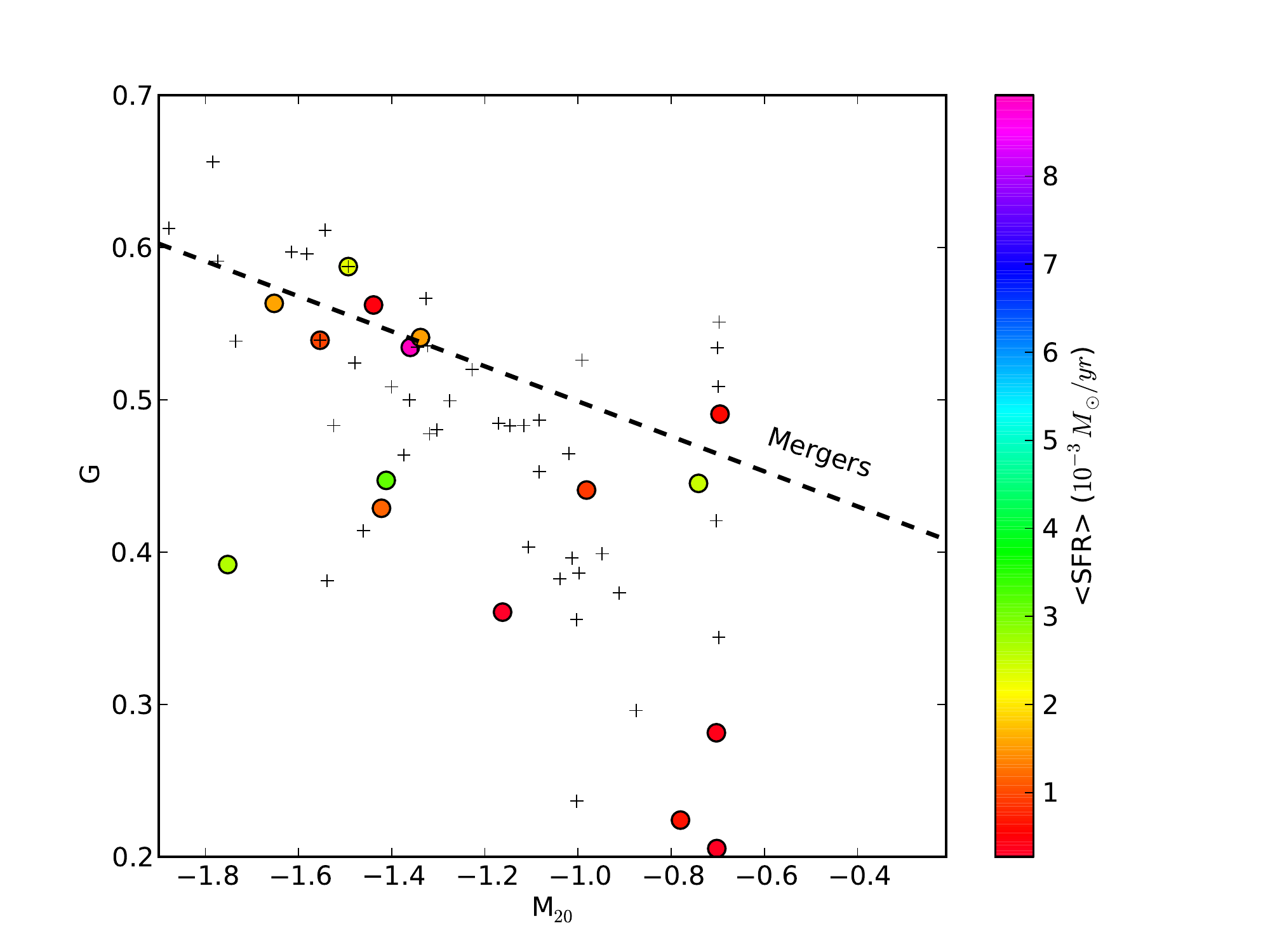}
\caption{The \m20\ and Gini parameters of the \hi\ maps colour coded by the mean past star-formation rate from \protect\cite{Weisz11b} or \protect\cite{McQuinn12b}. The dashed line is a criterion for major interaction from \protect\cite{Holwerda11c, Lotz04}.  }
\label{f:GM20:tau}
\end{center}
\end{figure}

Figure \ref{f:SFtime} shows the \hi\ morphology, colour-coded for the ANGST sample by the average 
star-formation rate ($<SFR>$) over the entire history of the galaxy, calculated over the past 10 Gyr 
\citep{Weisz11b} or 6 Gyr \citep{McQuinn12b}. Figure \ref{f:SFtime:6par} shows the direct relation between
the typical star-formation over these longer time-scales $<SFR>$ and the \hi\ morphology parameters.
Those galaxies with low star-formation rates are typically low in Gini, \gm, and Asymmetry, and high 
in \m20 values. The \gm\ and \m20\ values appear to be related for those galaxies with a low lifetime 
star-formation rate ($\sim10^{-3} M_\odot/yr$). The Spearman indices are high for \m20 and Gini: 
$<SFR>$ is anti-correlated to \m20, and correlated with Gini.

Thus, there is some relation between the mean star-formation rate in a dwarf galaxy and how 
the \hi\ is distributed. Galaxies that are not forming stars at a high rate right now and have not in the 
past ($<SFR> ~ \sim 1-2 \times 10^{-3} M_\odot/yr$) show lower values of Gini, \gm\ and higher values of \m20.

In comparison, appendix \ref{s:w11} shows that there is little or no relation with mass-to-light ratio 
or mean stellar age from \cite{Weisz11b}.

\begin{figure}
\begin{center}
\includegraphics[width=0.5\textwidth]{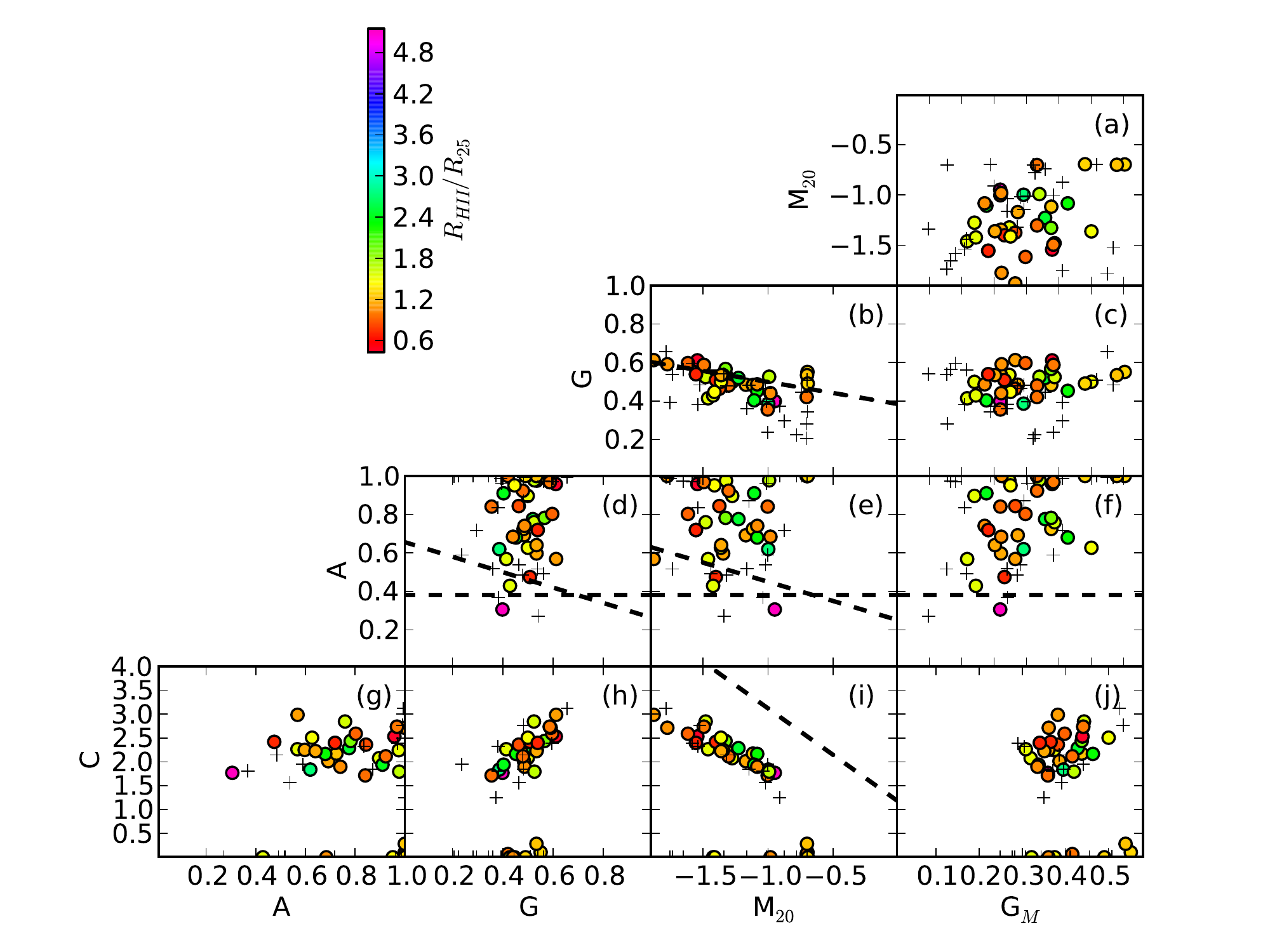}
\caption{The distribution of the \hi\ morphological parameters colour coded by the relative extent of the H$\alpha$ disk ($R_{H\alpha}/R_{R25}$) from \protect\cite{Hunter04}. Dashed lines are the criteria for major interaction from \protect\cite{Holwerda11c}, based on the WHISP sample or established morphological selections of mergers in optical data. The ratio is related with the \m20\ and anti-correlated to the Gini parameter (Table \ref{t:spear}), already identified in \protect\cite{Holwerda12c} as the right combination to identify extended star-formation disks.}
\label{f:xdisk}
\end{center}
\end{figure}
\begin{figure}
\begin{center}
\includegraphics[width=0.5\textwidth]{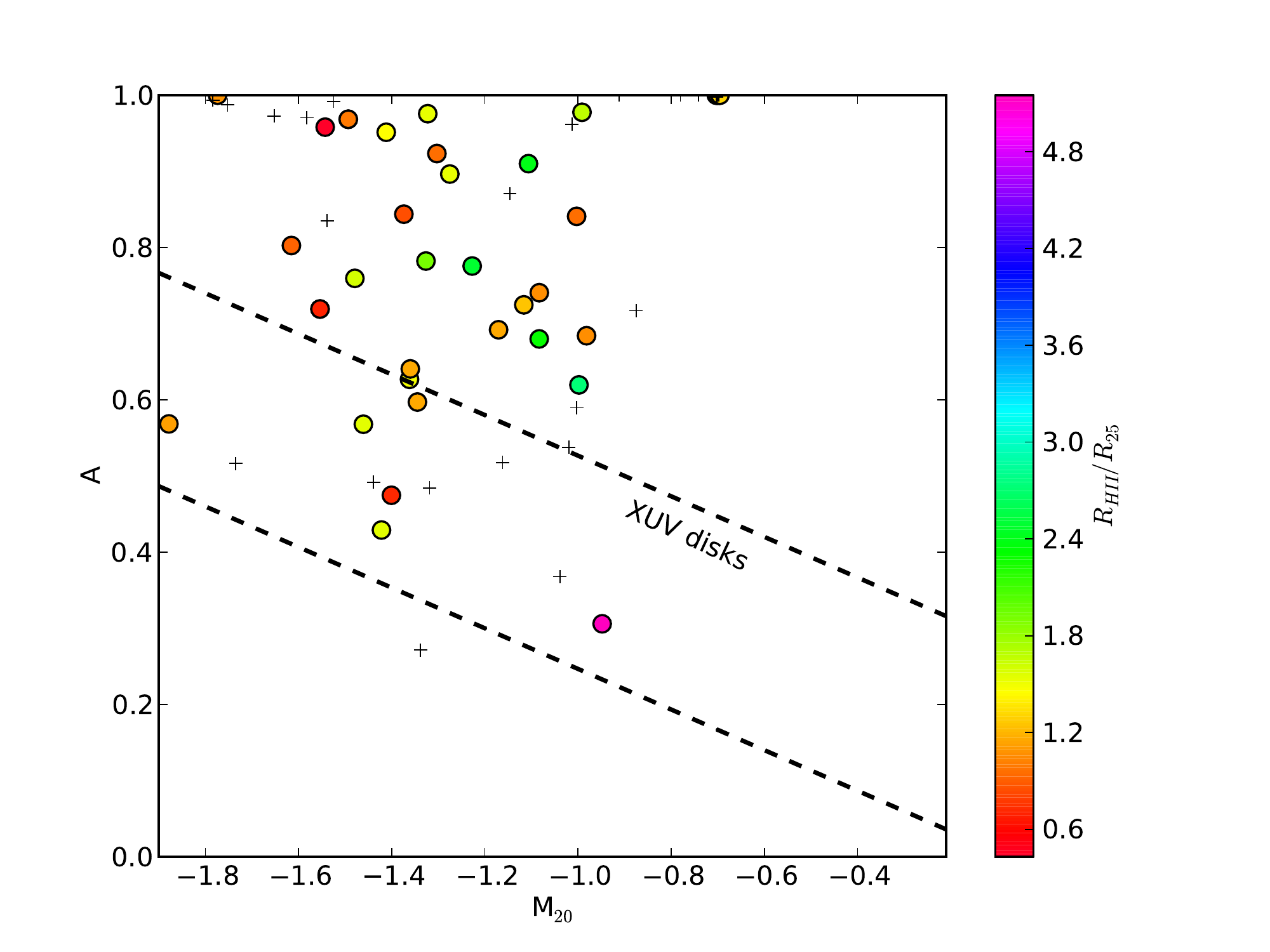}
\caption{The Asymmetry and \m20\ of the \hi\ maps colour coded by the relative extent of the H$\alpha$ disk ($R_{H\alpha}/R_D$) from \protect\cite{Hunter04}. The dashed lines are the \xuv\ disk criteria identified in \protect\cite{Holwerda12c} for \fuv\ images. }
\label{f:xuv}
\end{center}
\end{figure}

\subsection{Extended H$\alpha$ disks}
\label{s:RHII}

In \cite{Holwerda12c}, we used these parameters to identify extended \uv\ disks (\xuv) in the \whisp\ 
survey. \cite{Hunter04} note the relative scale of the HII regions to that of the optical disk ($R_{25}$), 
Holmberg radius ($R_H$), and disk scale length ($R_D$). Figure \ref{f:xdisk} shows the morphological 
parameters distribution, colour coded by the relative extent of the H$\alpha$ emission. The highest 
correlation between these axes ratios with the \hi\ parameters is with \m20\ and anti-correlated with 
Gini (Table \ref{t:spear}). 

We find that galaxies that have a high star-formation surface density, are also relatively compact. 
Figure \ref{f:xuv} shows the Asymmetry-\m20 for the \hi\ maps and the \xuv\ disk criterion for 6" \fuv\ 
data from \cite{Holwerda12c}. Eleven galaxies are in this selection. Five have extended HII radii 
($R_{HII}/R_{25} > 1$), and five of these do not have a HII radius from \cite{Hunter04} 
(Figure \ref{f:xdisk}). The one $R_{HII}/R_{25}$ extreme value in our sample is selected but the 
remainder is not exceptionally extended or concentrated. We conclude that extended HII disks are not related 
to extended \fuv\ disks. 



\begin{table*}
\caption{\label{t:spear}The Spearman correlation between various literature values and the HI  morphological parameters. Ranking is only computed for where both the literature values and HI morphological values are available. Notable rankings are marked in bold; these are parameters between which there is a (weak) linear relation. Positive values indicate correlation, negative ones anti-correlation.}
\begin{threeparttable}
\begin{tabular}{l l l l l l l l}
Name & C & A & S & \m20 & G & \gm & note\\
\hline
\hline
$v_{rot}$ 	 									& 0.23 	 & -0.06 	 & -0.04 	 & -0.02 	 & 0.05 	 & 0.33 		& \\ 
\hline
from \cite{Karachentsev04}: & & & & & & & \\
$\Theta$ 	 									& -0.03 	 & {\bf 0.42} & -0.02 	 & 0.14 	 & -0.03 	 & 0.21 		& (1)\\ 
\hline
from  \cite{Hunter06}: & & & & & & & \\
$M_{HI}/L$ 									& -0.13 	 & -0.13 	 & {\bf -0.46}  & -0.11 & 0.00 	 & -0.25 		& (2)\\ 
$log_{10}(SFR) M_\odot/yr$ 						& 0.17 	 & 0.15 	 & 0.16 	 & -0.06 	 & 0.13 	 & 0.32 		& (3)\\ 
$log_{10}(SFR_{M25})$ 	 						& 0.12 	 & -0.01 	 & 0.06 	 & 0.00 	 & 0.16 	 & 0.12 		& (4)\\ 
$log_{10}(SFR_D)$ 	 							& 0.32 	 & 0.17 	 & 0.23 	 & -0.24 	 & {\bf 0.44} 	 & 0.28 	& (5)\\ 
$log_{10}(\tau)$ 	 							& -0.20 	 & -0.14 	 & -0.23 	 & 0.13 	 & -0.21 	 & -0.22 		& (6)\\ 
$R_{HII}/R_{25}$ 	 							& -0.28 	 & -0.17 	 & -0.33 	 & {\bf 0.40} & -0.36 	 & 0.07 		& (7)\\ 
$R_{HII}/R_{H}$ 	 							& -0.29 	 & -0.11 	 & -0.21 	 & {\bf 0.61} & {\bf -0.40} 	 & 0.23 	& (8)\\ 
$R_{HII}/R_{D}$ 	 							& -0.09 	 & -0.17 	 & -0.15 	 & 0.34 	 & -0.10 	 & 0.23 		& (9)\\ 
\hline
from  \cite{Weisz11b} & & & & & & & \\
and \cite{McQuinn12a}: & & & & & & & \\
$f_{10}$ 	 									& -0.14 	 & -0.26 	 & -0.19 	 & 0.05 	 & -0.16 	 & 0.25 		& \\ 
$f_{06}$ 	 									& 0.03 	 & -0.21 	 & -0.14 	 & -0.16 	 & 0.16 	 & -0.02 		& \\ 
$f_{03}$ 	 									& -0.26 	 & 0.15 	 & {\bf 0.41} & 0.14 	 & -0.30 	 & 0.01 		& \\ 
$f_{02}$ 	 									& -0.28 	 & 0.28 	 & {\bf 0.48} & 0.27 	 & -0.42 	 & -0.13 		& \\ 
$f_{01}$ 	 									& -0.07 	 & -0.01 	 & 0.38 	 & 0.07 	 & -0.12 	 & -0.29 		& \\ 
Mean Stellar Age (Gyr) 	 						& -0.07 	 & -0.26 	 & -0.09 	 & -0.02 	 & -0.10 	 & 0.20 		& (10)\\ 
$<SFR>$ ($10^{-3} M_\odot/yr$) 	 				& 0.35 	 & -0.08 	 & -0.30 	 & {\bf -0.50} & {\bf 0.44} 	 & 0.15 	& (11)\\ 
Mass-to-light ratio 	 							& -0.11 	 & 0.08 	 & -0.11 	 & 0.11 	 & -0.07 	 & 0.05 		& (12)\\ 
\hline
from \cite{Hunter10}  & & & & & & & \\
and McQuinn et al. {\em in preparation}:  & & & & & & & \\
log$_{10}$(SFR$_{FUV}$) ($M_\odot/yr$) 		 	& {\bf 0.44} & -0.14 	 & 0.03 	 & -0.32 	 & {\bf 0.44} & 0.36 		& (13)\\ 
log$_{10}$(SFR$_{H\alpha}$) ($M_\odot/yr$) 	 		& 0.24 	 & 0.20 	 & 0.19 	 & -0.20 	 & {\bf 0.48} & 0.21 		& (14)\\ 
log$_{10}$(SFR$_{V}$) ($M_\odot/yr$) 	 			& 0.38 	 & -0.24 	 & 0.05 	 & -0.38 	 & 0.32 	 & 0.06 		& (15)\\ 
log$_{10}$(SFR$_{FUV}$/SFR$_{H\alpha}$) ($M_\odot/yr$) 	 & 0.15  & -0.38 	 & 0.18 	 & -0.15 	 & -0.21 	 & -0.04 	& (16)\\ 
log$_{10}$(SFR$_{FUV}$/SFR$_{V}$) ($M_\odot/yr$) 	 & {\bf -0.43}  & 0.06 	 & 0.06 	 & {\bf 0.40} 	 & {\bf -0.53} & -0.14 & (17)\\ 
Mass-to-light M/L$_V$ ($M_\odot/L_\odot$) 	 		& -0.01 	 & -0.25 	 & -0.04 	 & 0.04 	 & -0.22 	 & -0.18 		& (18)\\ 
\hline 
\hline
\end{tabular}
\begin{tablenotes}
\item [1] Tidal parameter from \cite{Karachentsev04}.
\item [2] The mass-to-light ratio inferred in \cite{Hunter06} from H$\alpha$ emission.
\item [3] The {\em total} star-formation rate inferred in \cite{Hunter06} from H$\alpha$ emission.
\item [4] The star-formation rate over the $R_{25}$ radius inferred in \cite{Hunter06} from H$\alpha$ emission.
\item [5] The star-formation rate over the $R_D$ radius inferred in \cite{Hunter06} from H$\alpha$ emission.
\item [6] The gas-depletion time inferred by \cite{Hunter06} from H$\alpha$ emission.
\item [7] The ratio between the radius containing the HII regions to the de Vaucouleur radius $R_{25}$.
\item [8] The ratio between the radius containing the HII regions to the Holmberg radius ($R_H$).
\item [9] The ratio between the radius containing the HII regions to the optical radius ($R_D$).

\item [10] The mean age of the stellar population computed from the fractions ($f_{1}-f_{10}$) reported in \protect\cite{Weisz11b}.
\item [11] The mean star-formation rate over the last 10 Gyr reported in \protect\cite{Weisz11b}.
\item [12] The mean stellar mass-to-light ratio inferred from the stellar population in \protect\cite{Weisz11b}.

\item [13] The star-formation rate inferred from \fuv\ flux reported in \protect\cite{Hunter10}.
\item [14] The star-formation rate inferred from $H\alpha$ flux reported in \protect\cite{Hunter10}.
\item [15] The star-formation rate inferred from V-band flux reported in \protect\cite{Hunter10}.
\item [16] The ratio between star-formation rates inferred from \fuv\ and $H\alpha$ flux reported in \protect\cite{Hunter10}.
\item [17] The ratio between star-formation rates inferred from \fuv\ and V-band flux reported in \protect\cite{Hunter10}.
\item [18] The inferred stellar mass-to-light ratio report reported in \protect\cite{Hunter10}.

\end{tablenotes}
\end{threeparttable}
\end{table*}

\section{The Star-formation and \hi\ Morphology Relations Spirals and Irregulars}
\label{s:sdss}

In the past few sections, we have tried to establish if there is a relation between the quantified 
\hi\ morphological parameters of dwarf galaxies and their star-formation measured over different time-scales by different tracers. To place this in a larger context, we will now compare the quantified \hi\ 
morphology parameters of {\em all} our samples, THINGS \citep{Holwerda11a}, WHISP \citep{Holwerda11b}, LITTLE-THINGS and VLA-ANGST (this paper) to a common star-formation and stellar mass estimate from the Sloan Digital Sky Survey from \cite{Brinchmann04}. 

Total stellar mass, total star-formation and specific star formation are available for the SDSS DR7 
sample of galaxies as described by \cite{Kauffmann03} and \cite{Brinchmann04} here:
\url{http://www.mpa-garching.mpg.de/SDSS/DR7/}. Cross-correlating with position, we obtain a common 
measurements for our full \hi\ sample. We note that these measures are based on a 3'' aperture typically 
centered on the galaxy nucleus.

The \hi\ samples divides into two sub-samples:  a high spatial resolution ($\sim$6"), the VLA-ANGST and 
THINGS surveys, and a lower resolution one ($\sim$10-12"), the WHISP and LITTLE-THINGS surveys. We will do the comparison with the SDSS parameters for the combined and the high- and low-resolution subsamples. To distinguish these plots from the previous comparisons, the markers are triangles for the low-resolution sample and squares for the high-resolution sample in the following plots.

We note two caveats in the comparison: first, the interferometric observations miss low-intensity, large-scale emission (see section \ref{s:samp}). This introduces a different sensitivity in the low and high-resolution maps to diffuse, large-scale tidal features for example. 
And secondly, the SDSS survey skipped very nearby galaxies which were resolved in individual HII regions for spectroscopic follow-up (e.g., M101 is not in the SDSS DR7 spectroscopic sample).

\subsection{Stellar Mass}
\label{s:sdss:m}

\begin{figure*}
\begin{center}
     \begin{minipage}{0.65\linewidth}
	\includegraphics[width=\textwidth]{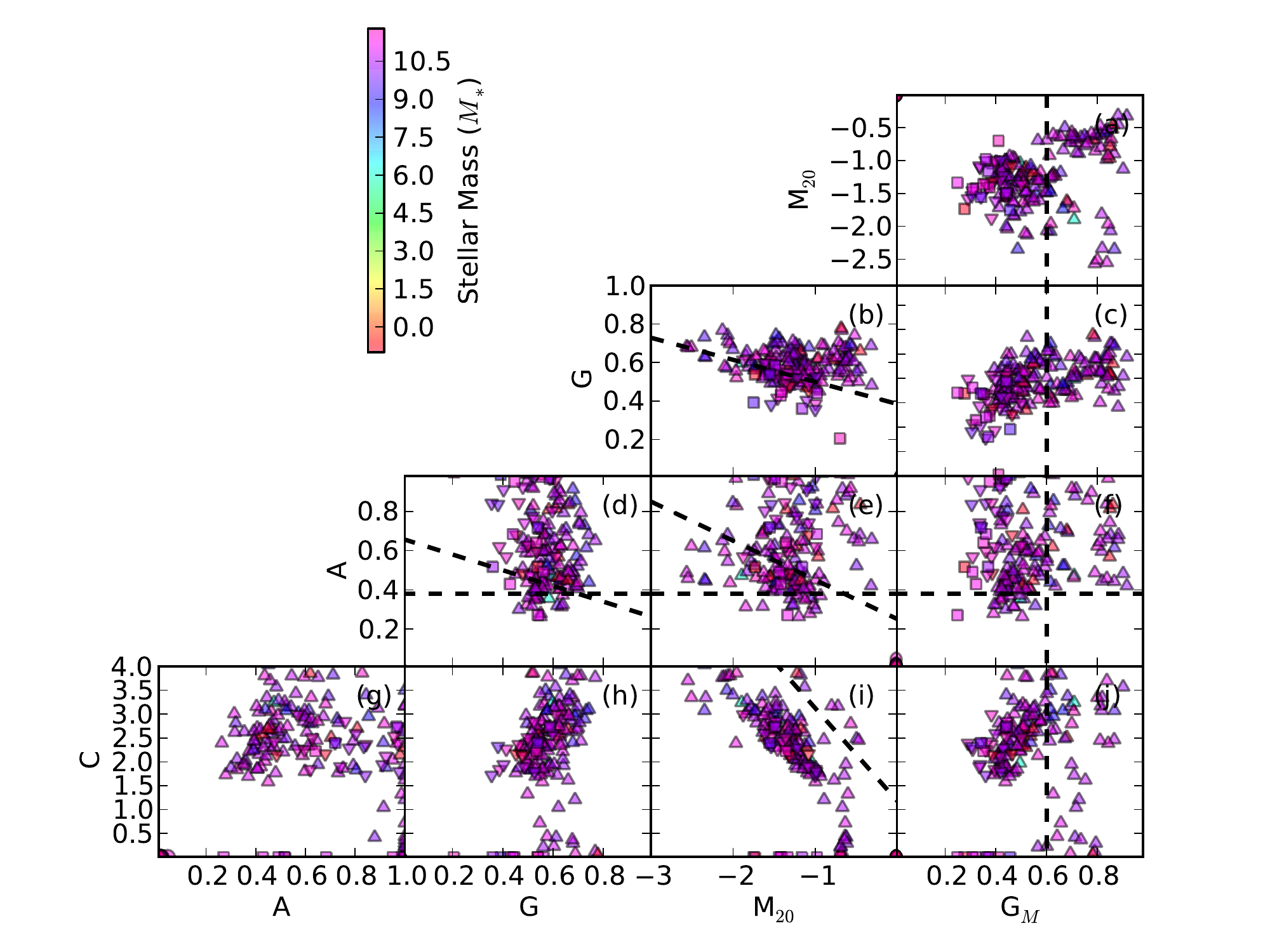}
    \end{minipage}\hfill
    \begin{minipage}{0.34\linewidth}
	\includegraphics[width=1.1\textwidth]{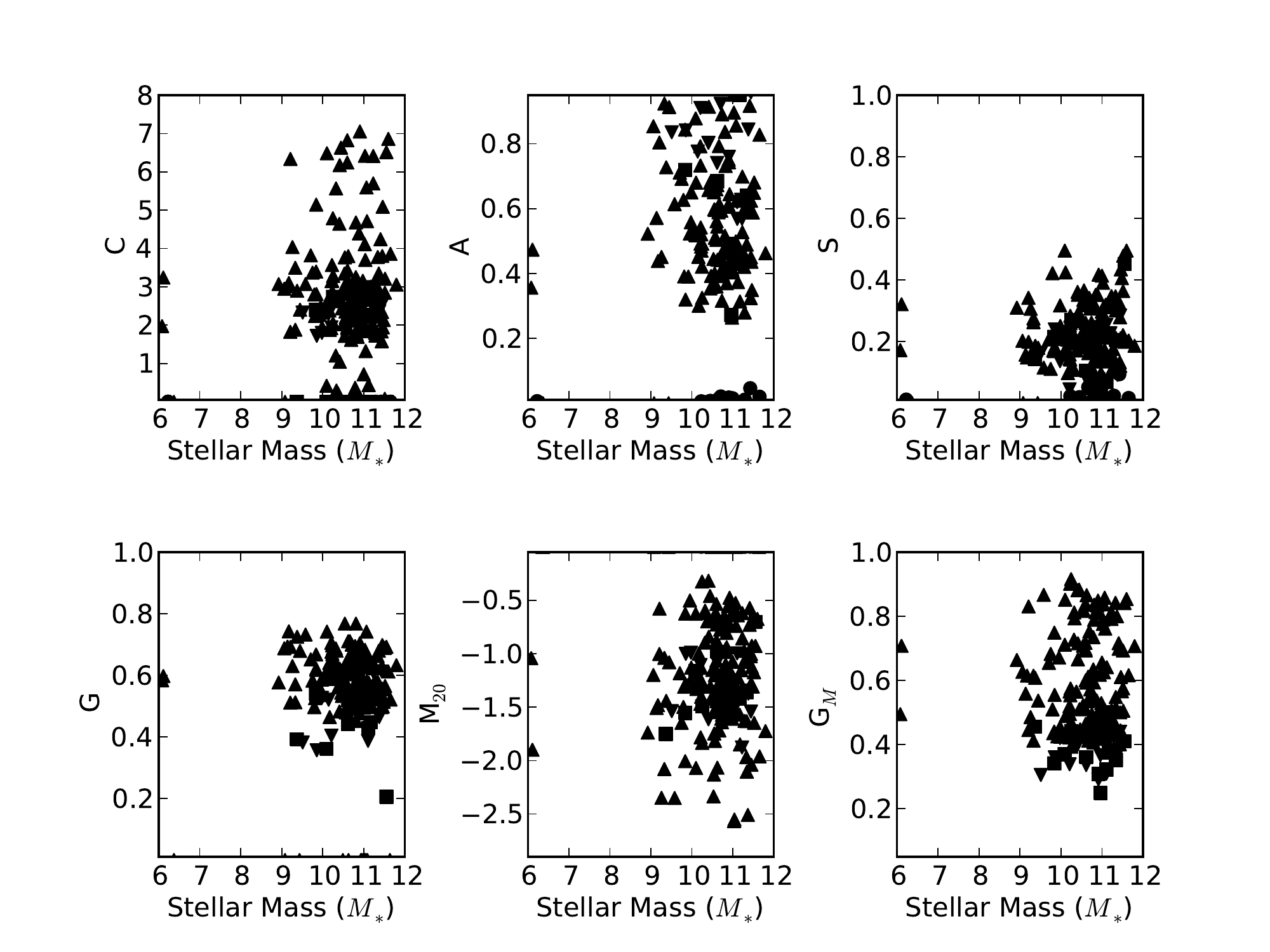}
	\caption{\label{f:sdss:sm:6par} The direct relation between \hi\ morphologies and their inferred stellar mass from \protect\cite{Brinchmann04}, updated to the SDSS DR7. }
    \end{minipage}
    \begin{minipage}{\linewidth}
	\caption{\label{f:sdss:sm} The distribution of \hi\ morphologies for the full sample, colour-coded by their inferred stellar mass from \protect\cite{Brinchmann04}, updated to the SDSS DR7. }
    \end{minipage}
\end{center}
\end{figure*}

The range in stellar masses is lower than one would expect (Figure \ref{f:sdss:sm}), given that two of the 
surveys target dwarf systems (VLA-ANGST and LITTLE-THINGS) but many of these galaxies do not 
feature in the SDSS DR7 spectroscopic catalog. Figure \ref{f:sdss:sm:6par} shows the relation (if any) 
between the six \hi\ morphological parameters and the stellar mass ($log_{10}(M_*/M_\odot)$). There 
appears to be little or no relation between a galaxy's stellar mass and the \hi\ morphology, in the full or 
either the high or lower-resolution samples (6" and 12" respectively, see Table \ref{t:sdss}). 

\subsection{Star-Formation}
\label{s:sdss:sfr}

\begin{figure*}
\begin{center}
     \begin{minipage}{0.65\linewidth}
	\includegraphics[width=\textwidth]{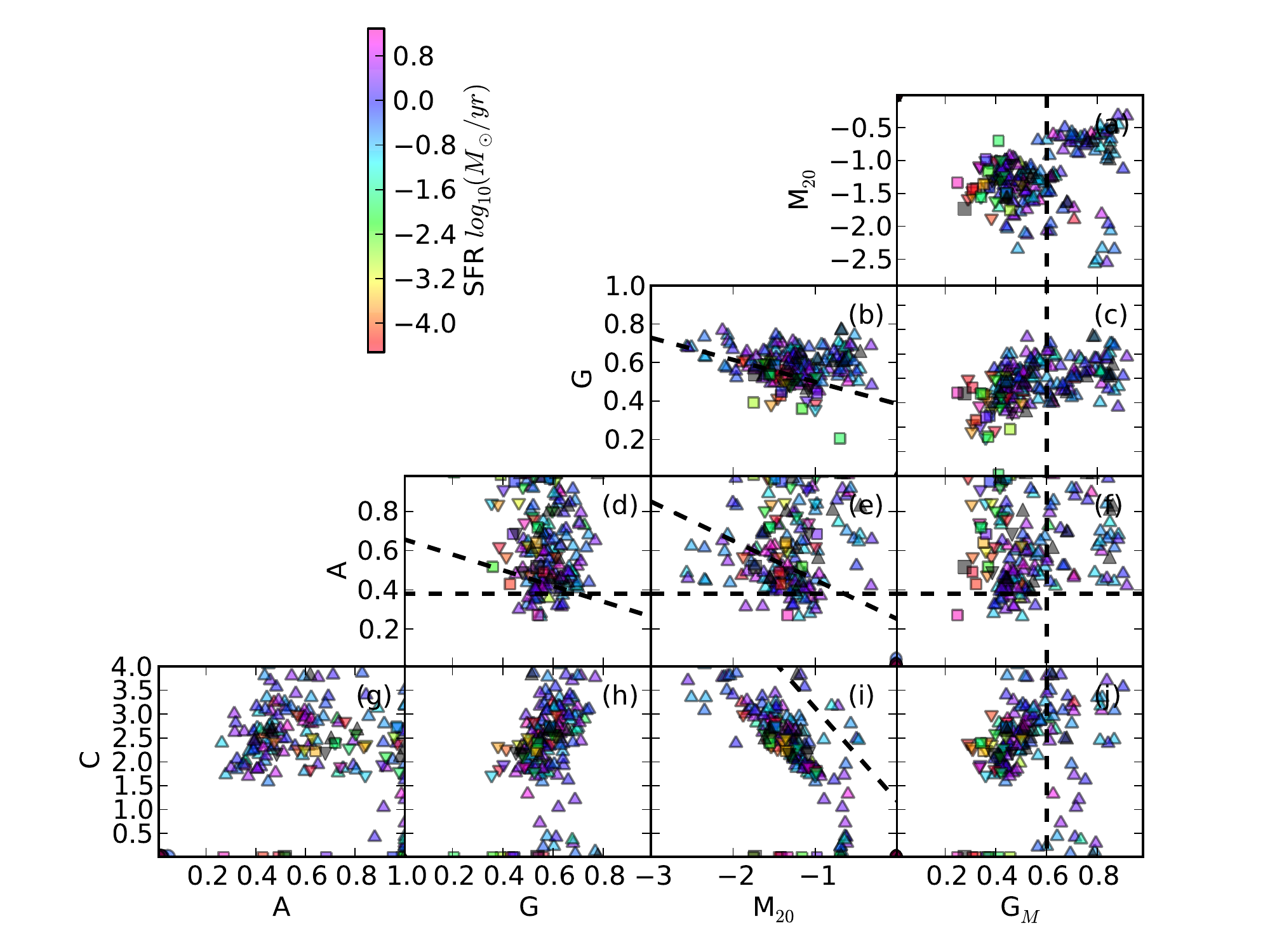}
    \end{minipage}\hfill
    \begin{minipage}{0.34\linewidth}
	\includegraphics[width=1.1\textwidth]{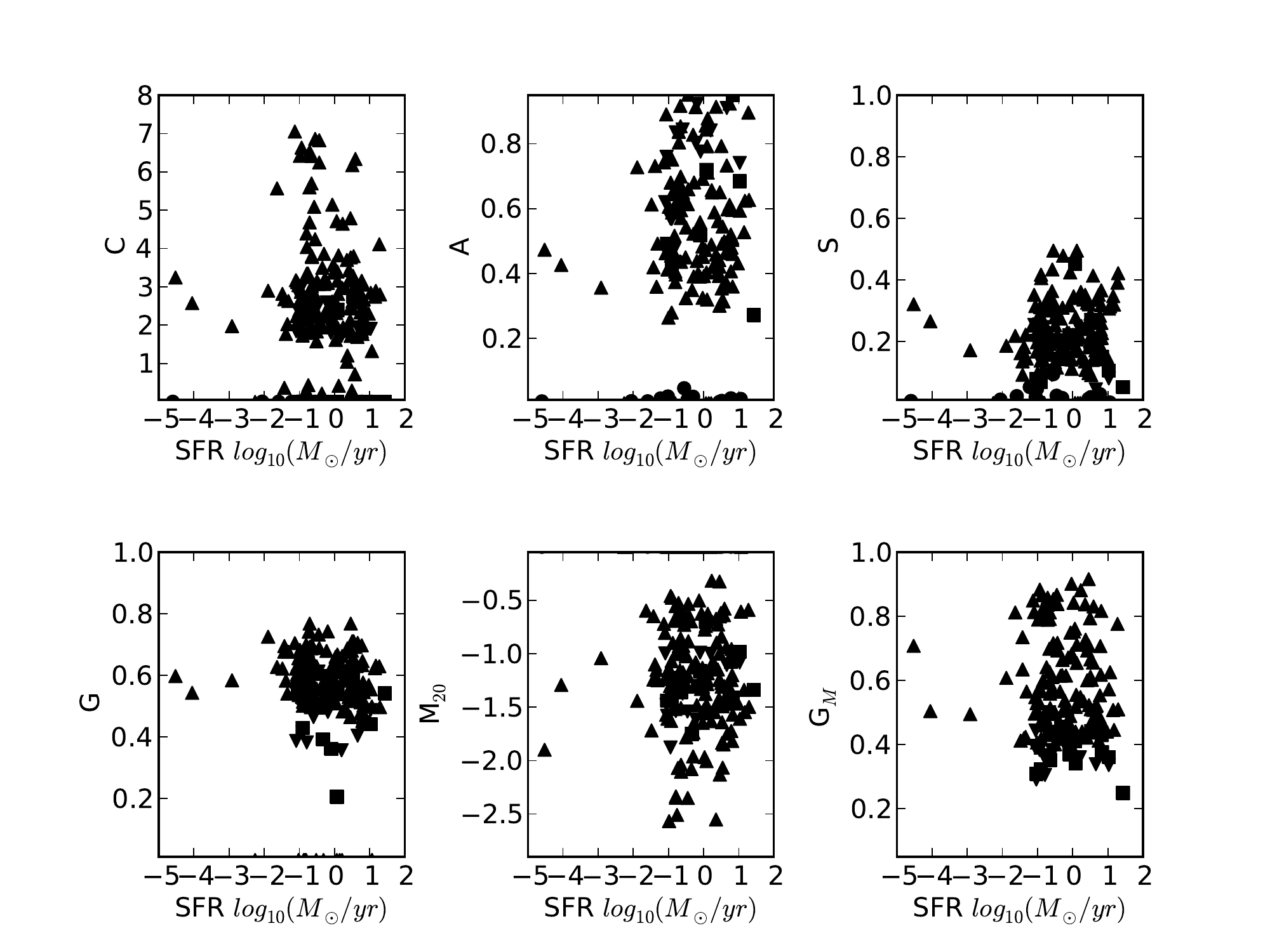}
	\caption{\label{f:sdss:sf:6par} The direct relation between \hi\ morphologies and their inferred total star-formation from \protect\cite{Brinchmann04}, updated to the SDSS DR7. }
    \end{minipage}
    \begin{minipage}{\linewidth}
	\caption{\label{f:sdss:sf} The distribution of \hi\ morphologies for the full sample, colour-coded by their inferred total star-formation from \protect\cite{Brinchmann04}, updated to the SDSS DR7. }
    \end{minipage}
\end{center}
\end{figure*}

Figures \ref{f:sdss:sf} shows the distribution of \hi\ morphology parameters but 
now colour-coded by total star-formation ($log(SFR)$). There are a few suggestive outliers, 
especially with respect to the Gini parameters (panel (b)), but a clean trend is impossible to distinguish. 
In the plot of star-formation rate agains the six \hi\ morphology parameters in Figure \ref{f:sdss:sf:6par}, 
clear trends are also absent, as reflected in the spearman rankings for the full sample (Table \ref{t:sdss}). 
However, there is an interesting difference between the low- and high-resolution samples: in there a 
much better correlations between the overall \hi\ morphology and the total star-formation in the 
high-resolution sample (6"). %

In the high-resolution sample, star-formation is weakly related to Asymmetric, Smoothness, \m20, 
Gini, and \gm. This is similar to what we found for the smaller systems (Table \ref{t:spear}), but 
some of the relations are inverted (e.g., SFR and Gini). 
The inversion of the relation between star-formation and those \hi\ parameters that measure the 
clumpiness of the ISM is intriguing: if one extends the mass range of the sample to high-mass 
galaxies (in fact the high-resolution sample is dominated by them), the Gini parameters {\em lowers} 
slightly with higher star-formation (Figure \ref{f:sdss:sf:6par}, and more clearly in Table \ref{t:sdss}). 
One could speculate that in high-mass, high-star-formation cases, the clumping is taken to extremes 
and the majority of hydrogen is in dense molecular clumps, leaving a relatively smooth \hi\ disk.

\begin{figure*}
\begin{center}
	\includegraphics[width=0.32\textwidth]{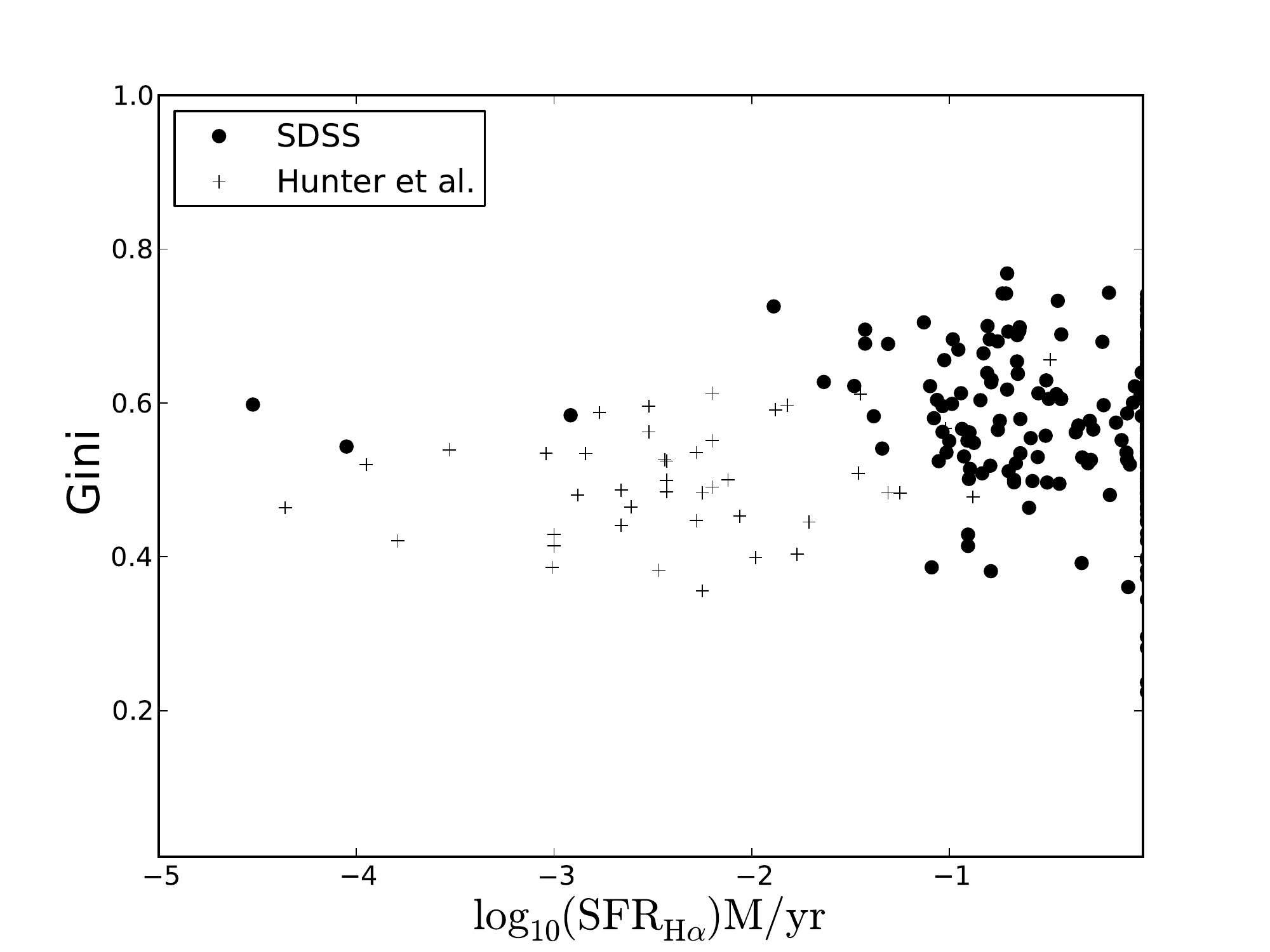}
	\includegraphics[width=0.32\textwidth]{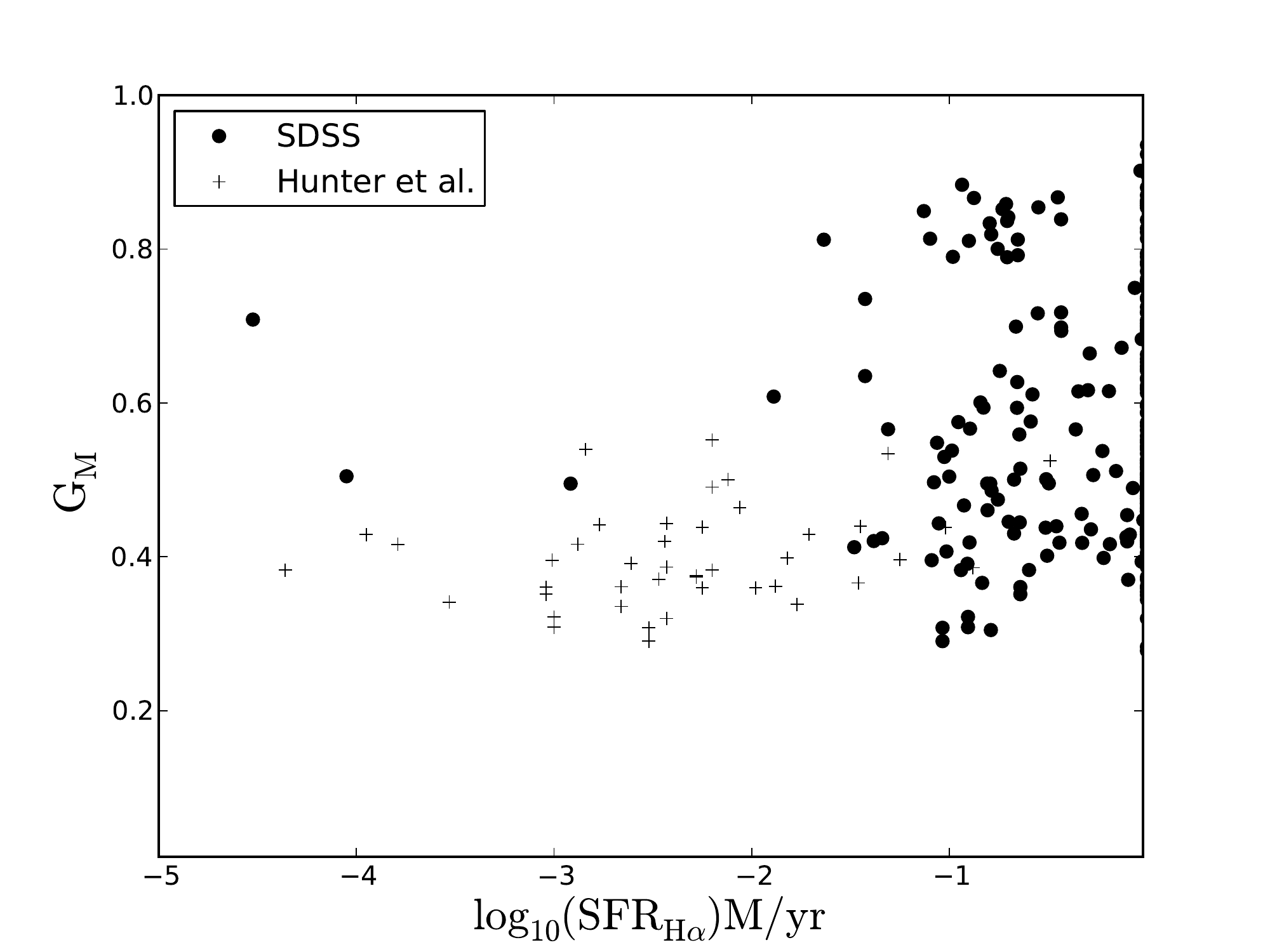}
	\includegraphics[width=0.32\textwidth]{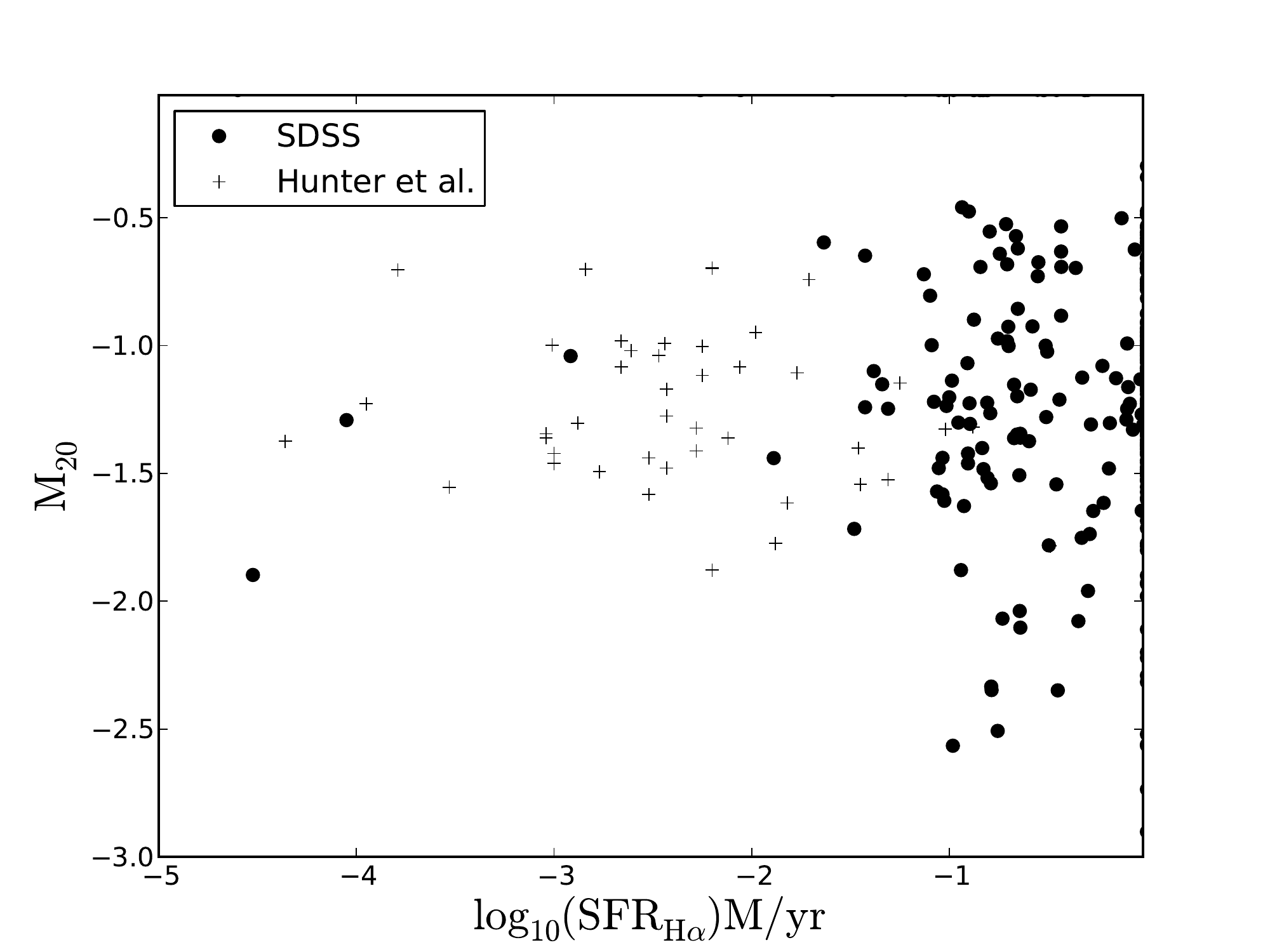}
	\caption{\label{f:sf:Ha} The relation between star-formation from H$\alpha$ and the Gini, \gm\ and \m20\ parameters. Crosses are the morphologies of the LITTLE-THINGS galaxies with H$\alpha$ star-formation from \protect\cite{Hunter06} and the dots are the other surveys with star-formation from \protect\cite{Brinchmann04}, based on the SDSS spectra. There is a weak relation with both Gini and \gm. We note that the spread in all three values goes up with star-formation.}
\end{center}
\end{figure*}

Because many of the dwarf systems do not have a reliable H$\alpha$ star-formation traces in SDSS, 
primarily because their stellar light is too diffuse, we plot the combination of the \cite{Hunter06} values
and the \cite{Brinchmann04} values in Figure \ref{f:sf:Ha} for Gini, \gm and \m20. We note that there 
is some relation, but most interestingly the range of \hi\ morphology values increases to higher star-formation (and the higher-mass
systems).

\subsection{Specific Star-Formation}

\begin{figure*}
\begin{center}
     \begin{minipage}{0.65\linewidth}
	\includegraphics[width=\textwidth]{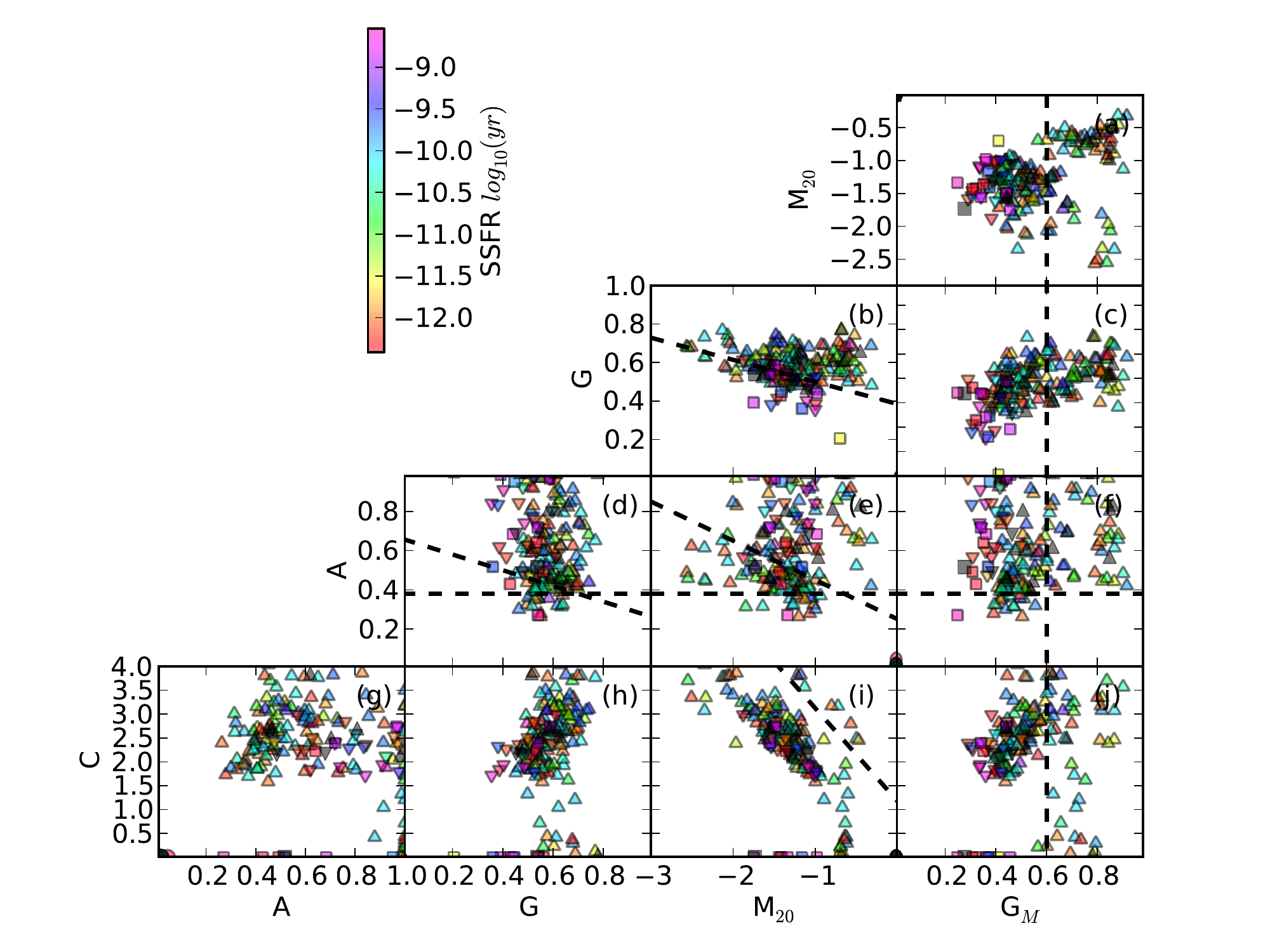}
    \end{minipage}\hfill
    \begin{minipage}{0.34\linewidth}
	\includegraphics[width=1.1\textwidth]{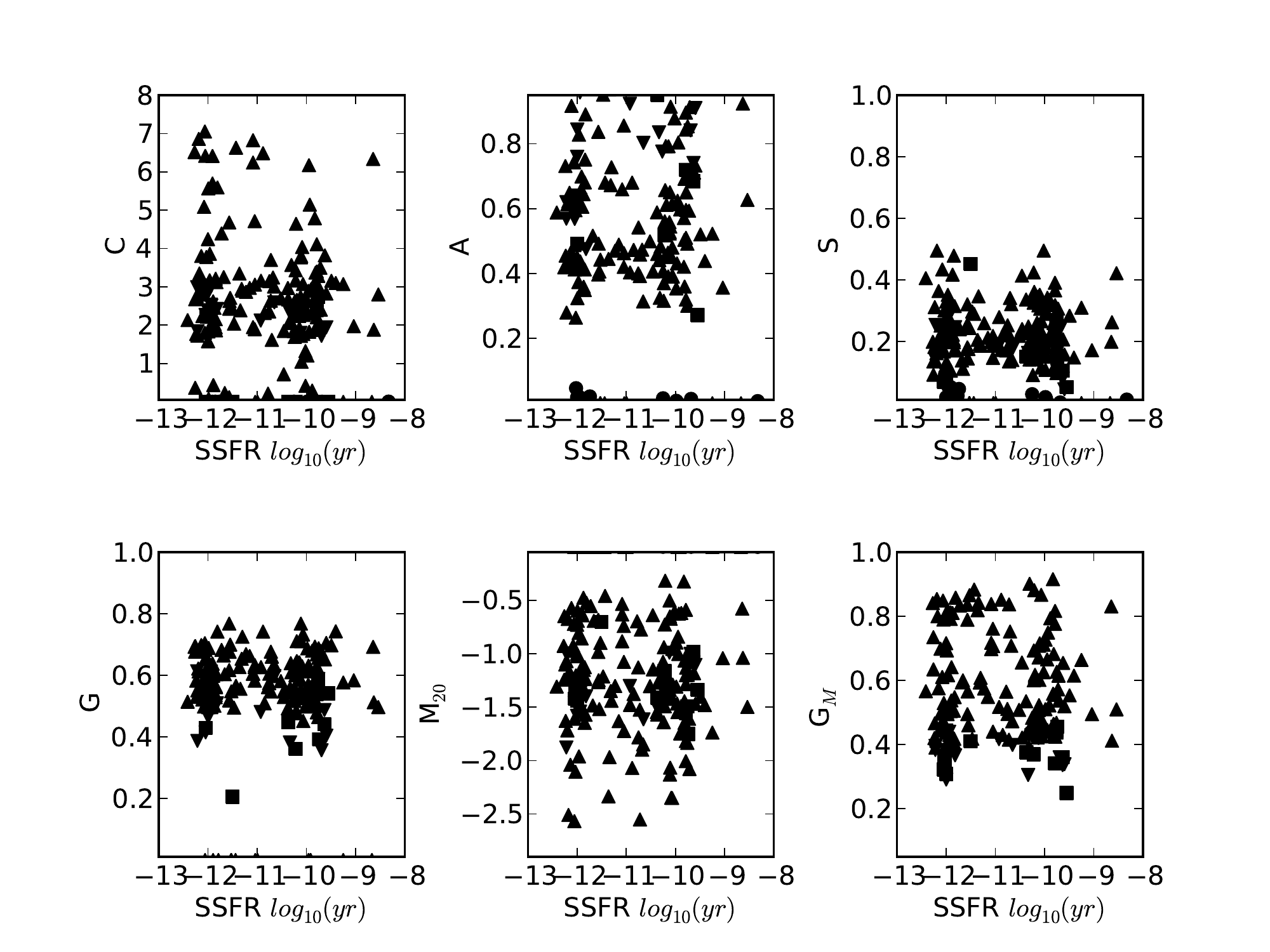}
	\caption{\label{f:sdss:ssf:6par} The direct relation between \hi\ morphologies and their inferred specific star-formation from \protect\cite{Brinchmann04}, updated to the SDSS DR7. }
    \end{minipage}
    \begin{minipage}{\linewidth}
	\caption{\label{f:sdss:ssf} The distribution of \hi\ morphologies for the full sample, colour-coded by their inferred specific star-formation from \protect\cite{Brinchmann04}, updated to the SDSS DR7. }
    \end{minipage}
\end{center}
\end{figure*}

Normalizing the total star-formation with the stellar mass, there is again little or no relation 
between the \hi\ morphology and the specific star-formation (Figures \ref{f:sdss:ssf} and 
\ref{f:sdss:ssf:6par} and table \ref{t:sdss}), either for the total or high- or low-resolution samples. 

\begin{table*}
\caption{The Spearman rank correlation of the \hi\ morphological parameters with stellar mass, total, and specific star-formation of the high- (6"), low-resolution (12") and full samples. There is very little relation between the \hi\ morphology and star-formation for data at 12" resolution. Star-formation's effects are more noticeable only in the high-resolution sample. }
\begin{center}
\begin{tabular}{l l l l l l l}

Sample \& parameter			& C & A & S & $M_{20}$ & G & $G_M$\\
\hline
\hline
6"	& & & & & & \\
Star Formation Rate (SFR)        & -0.05         & 0.44          & 0.34          & -0.35         & -0.36         & 0.36 \\ 
Specific Star Formation Rate (SSFR)      & -0.08         & -0.07         & -0.13         & -0.17         & -0.05         & 0.23 \\ 
Stellar Mass ($M_*$)     & -0.03         & 0.28          & 0.13          & 0.13          & -0.08         & -0.03 \\ 
\hline
12"	& & & & & & \\
Star Formation Rate (SFR)        & -0.04         & 0.01          & 0.15          & -0.05         & -0.07         & -0.02 \\ 
Specific Star Formation Rate (SSFR)      & -0.05         & -0.02         & 0.04          & -0.03         & -0.01         & -0.04 \\ 
Stellar Mass ($M_*$)     & -0.01         & -0.01         & 0.00          & -0.03         & -0.04         & -0.01 \\ 
\hline
All		& & & & & & \\
Star Formation Rate (SFR)        & 0.00          & 0.08          & 0.17          & -0.10         & -0.13         & 0.02 \\ 
Specific Star Formation Rate (SSFR)      & -0.05         & -0.00         & 0.02          & -0.04         & -0.02         & -0.03 \\ 
Stellar Mass ($M_*$)     & -0.01         & -0.00         & 0.01          & -0.01         & -0.05         & -0.01 \\ 
\hline
\end{tabular}
\end{center}
\label{t:sdss}
\end{table*}%

\section{Discussion}
\label{s:disc}

As part of the local volume of galaxies, many dwarf galaxies can be expected to be tidally 
disrupted by either a close dwarf companion or a nearby massive galaxy (the Milky Way, Andromeda or M81). 
However, the \hi\ morphology does not seem to be affected as much by tidal interaction as by star formation.
Since these galaxies are relatively shallow gravitational potentials, one would
expect the tidal forces to play a significant role in the general shape of the ISM. However, the short kinematic time
scales of the ISM may result in a quick relaxation of any tidal disturbance or tidal effects on the \hi\ in these galaxies 
are too subtle to detect in this parameter space. 

We compared the \hi\ morphology to a series of star-formation indicators, sensitive to different time-scales of 
star-formation. The measures of inequality in the \hi\ morphology, the Gini, \gm\ and \m20\ parameter are
weakly related to various tracers of star-formation (H$\alpha$, \fuv\ or based on resolved stellar population).
This result is largely in line with what is becoming the general picture of these galaxies: the local physics 
dominate over environment \citep{McQuinn12a}.

That is not to say that the star-formation is the shaping agent of the \hi\ morphology through, for example, 
supernova feedback and stellar winds. The reverse could well be true; the inequality distributed ISM more 
often generates the conditions for local star-formation. \cite{McQuinn12a} show that the star-formation is 
stochastic both in time and location within star-bursting galaxies, corroborating such a scenario.
And interestingly, the {\em mean} star-formation rate of the galaxies appear to be related to the \hi\ 
morphological parameters as well as the current star-formation indicator (Figure \ref{f:sfr}).
We propose a scenario where as soon as the ISM is transformed to an unequal or clumpy state, most 
likely by an external trigger --e.g., tidal or gravitational turbulence from the inflow of gas-- the star-formation 
rate is elevated. The combined effect of elevated star-formation and the external factor keep the ISM clumpy 
over longer time-scales than the current (ionizing, $H\alpha$) star-formation timescale.

We note a few things in the comparison of {\em all} the \hi\ morphological parameters available and a common 
stellar mass and star-formation estimates from SDSS (section \ref{s:sdss}). First, any relation manifests itself only in the 
{\em high-resolution} ($\sim$6") sample and not in the lower-resolution one. Because the sampling is sub-kpc 
typically in the high-resolution sample, this scale is one where the effects of feedback from star-formation (or 
the effect of \hi\ overdensities on star-formation) can be seen. Coarser observations simply wash any 
ISM-star-formation relation out. 

The importance of sampling on our study of the effects of star-formation on the ISM is noteworthy for the future all-sky 
\hi\ surveys (the WNSHS survey with the WSRT/APERTIF and the Wallaby survey with ASKAP, Koribalski et al. {\em in preparation}). The spatial resolution
of these surveys is expected to be $\sim$10", which means that the effects of star-formation on the \hi\ morphology will largely be smoothed out.
While these surveys will be indispensable to identify exceptional systems and characterize global gas characteristics in disks, follow-up observations at higher resolution will remain essential to characterize the interplay of star-formation and the ISM.

Secondly, the effects of star-formation reversed for some parameters when we compared for a low-mass 
sample (LITTLE-THINGS and VLA-ANGST) to the high-resolution sample (VLA-ANGST and THINGS). 
For example, the relation between total star-formation and the Gini parameter (or \m20 or \gm) reverses 
sign (Tables \ref{t:sdss} and \ref{t:spear}).
We put forward that in low-mass systems, it is predominantly the \hi\ that is related and regulates the 
star-formation \citep[as argued by ][]{Bigiel08} but in a higher-mass sample, the molecular component 
is the dominant ISM component in the relation with star-formation. 
That is to say, the clumping seen in the \hi\ of low-mass systems with increased star-formation happens 
in the molecular phase in more massive systems, leaving a relatively smooth \hi\ disk. 
Therefore the Gini parameter lowers with star-formation if one includes more massive galaxies. 

And the third point to make is that Asymmetry is strongly related to star-formation for the high-resolution 
subsample. This could be indirect evidence of gas accretion, especially for massive systems, as a strong relation
between any of the star-formation tracers and \hi\ Asymmetry is absent in the low-mass sample (Table \ref{t:spear}).
The relation between Asymmetry and \m20\ for \xuv\ disks reported in \cite{Holwerda12c} certainly appears to hint in that direction. 
However, this could also be taken as evidence of star-formation feedback on larger scales. 
Fact remains that the strongest and most consistent relationship in the high-resolution sample is between star-formation 
and \hi\ Asymmetry.

\section{Conclusion}
\label{s:concl}

We applied the quantified morphology parameterization to two \hi\ surveys of nearby dwarf galaxies, LITTLE-THINGS and VLA-ANGST, and compared these to indicators of interaction and star-formation. We find that:
\begin{enumerate}
\item The \hi\ morphological criteria for interaction, developed for use on massive galaxies do not apply to these smaller dwarf irregular galaxies  (Figure \ref{f:Theta}). 
\item A low value of the Asymmetry of the \hi\ map may point to an isolated dwarf (Figure \ref{f:Theta}). 
\item Current star-formation surface density (H$\alpha$) is related to the \hi\ morphology, specifically the Gini, \gm, and \m20\ parameters. These indicate a stronger inequality in the neutral ISM distribution  (Figure \ref{f:sfd} \& \ref{f:sfr}). 
\item Consequently, clumpy \hi\ is also the quickest depleted (Figure \ref{f:tau}).
\item Based on previous resolved stellar population results, the star-formation (current and history) is linked to the Gini, \gm, and \m20\ parameters of the \hi\ maps (Figure \ref{f:SFtime}): high star-formation and unequally distributed \hi\ are closely linked but not necessarily causal.
\item Over a large sample of galaxies, spanning a wider mass range, there is no relation between stellar mass, total or specific star-formation (Table \ref{t:sdss} and Figures \ref{f:sdss:sm}, \ref{f:sdss:sm:6par}, \ref{f:sdss:sf},\ref{f:sdss:sf:6par}, \ref{f:sdss:ssf}, and \ref{f:sdss:ssf:6par}). 
\item To detect any relation between star-formation tracers and \hi\ morphology, high-resolution ($\sim$6") \hi\ maps are critical (Table \ref{t:sdss}).
\item There is a relation between \hi\ Asymmetry and the ongoing {\em total} star-formation in massive galaxies (Figure \ref{f:sdss:sf:6par}, Table \ref{t:sdss}).
\end{enumerate}

Future applications of the quantified morphology parameters on \hi\ maps will be on the large catalogs of moderately resolved galaxies in the WNSHS (Jozsa et. al. {\em in preparation}) and WALLABY (Koribalski et al. {\em in preparation}) surveys. These will produce significantly improved statistics on \hi\ morphology which can then be combined with all-sky surveys of star-formation tracers (e.g., GALEX and WISE catalogs). 

\section*{Acknowledgements}

The authors would like to thank the anonymous referee whose comments and suggestions improved the manuscript significantly. The lead author thanks Dr McQuinn for the useful discussions on the topic of dwarfs and their star-formation and for making her \fuv\ star-formation estimates available to us and A. Leroy for the suggestion for the comparison against SDSS measures.
The authors thank C. Carignan for making the \hi\ moment-0 map of NGC 3109 from the Karoo Array Telescope available for a direct comparison with VLA data.

The lead author thanks the European Space Agency for the support of the Research Fellowship program.
We thank the National Radio Astronomy Observatory for their generous time allocation, observing, and data reduction support for these two Large Projects, the LITTLE-THINGS and VLA-ANGST. The National Radio Astronomy Observatory is a facility of the National Science Foundation operated under cooperative agreement by Associated Universities, Inc. We would like to thank the LITTLE-THINGS and VLA-ANGST teams for their effort on the calibration and imaging and making their products available to the astronomical community.
This research has made use of the NASA/IPAC Extragalactic Database (NED) which is operated by the Jet Propulsion Laboratory, California Institute of Technology, under contract with the National Aeronautics and Space Administration. 
This research has made use of NASA's Astrophysics Data System.

\newpage

\appendix

\begin{table*}
\section{The Morphological Parameters of the \hi\ column density maps based on the RO maps from LITTLE-THINGS and VLA-ANGST surveys.}

\begin{center}
\begin{tabular}{l l l l l l l l }

Name & Gini & \m20 & $C_{20/80}$ & A & S & E & \gm \\
\hline
\hline

CVnIdwA & 0.596 $\pm$ 0.015 	 & -1.582 $\pm$ 0.074 	 & 2.388 $\pm$ 0.135 	 & 1.941 $\pm$ 0.092 	 & 0.046 $\pm$ 0.049 	 & 0.276 $\pm$ 0.027 & 0.290 $\pm$ 0.039  \\
DDO43 & 0.499 $\pm$ 0.013 	 & -1.275 $\pm$ 0.060 	 & 2.076 $\pm$ 0.114 	 & 1.793 $\pm$ 0.049 	 & 0.187 $\pm$ 0.043 	 & 0.201 $\pm$ 0.013 & 0.320 $\pm$ 0.015  \\
DDO46 & 0.485 $\pm$ 0.012 	 & -1.171 $\pm$ 0.038 	 & 2.014 $\pm$ 0.088 	 & 1.384 $\pm$ 0.050 	 & 0.228 $\pm$ 0.038 	 & 0.079 $\pm$ 0.018 & 0.387 $\pm$ 0.017  \\
DDO47 & 0.399 $\pm$ 0.008 	 & -0.948 $\pm$ 0.012 	 & 1.769 $\pm$ 0.036 	 & 0.612 $\pm$ 0.058 	 & 0.107 $\pm$ 0.022 	 & 0.118 $\pm$ 0.009 & 0.360 $\pm$ 0.010  \\
DDO50 & 0.483 $\pm$ 0.007 	 & -1.146 $\pm$ 0.015 	 & 1.846 $\pm$ 0.043 	 & 1.742 $\pm$ 0.014 	 & 0.239 $\pm$ 0.020 	 & 0.118 $\pm$ 0.011 & 0.396 $\pm$ 0.007  \\
DDO52 & 0.386 $\pm$ 0.012 	 & -0.998 $\pm$ 0.056 	 & 1.838 $\pm$ 0.083 	 & 1.239 $\pm$ 0.159 	 & 0.253 $\pm$ 0.025 	 & 0.337 $\pm$ 0.017 & 0.396 $\pm$ 0.014  \\
DDO53 & 0.535 $\pm$ 0.018 	 & -1.323 $\pm$ 0.053 	 & 2.244 $\pm$ 0.119 	 & 1.951 $\pm$ 0.052 	 & 0.192 $\pm$ 0.043 	 & 0.086 $\pm$ 0.030 & 0.374 $\pm$ 0.025  \\
DDO63 & 0.483 $\pm$ 0.013 	 & -1.116 $\pm$ 0.029 	 & 2.172 $\pm$ 0.059 	 & 1.449 $\pm$ 0.041 	 & 0.247 $\pm$ 0.030 	 & 0.059 $\pm$ 0.021 & 0.438 $\pm$ 0.011  \\
DDO69 & 0.520 $\pm$ 0.012 	 & -1.227 $\pm$ 0.041 	 & 2.287 $\pm$ 0.102 	 & 1.551 $\pm$ 0.154 	 & 0.210 $\pm$ 0.024 	 & 0.477 $\pm$ 0.016 & 0.429 $\pm$ 0.011  \\
DDO70 & 0.414 $\pm$ 0.011 	 & -1.461 $\pm$ 0.025 	 & 2.264 $\pm$ 0.055 	 & 1.136 $\pm$ 0.037 	 & 0.057 $\pm$ 0.018 	 & 0.101 $\pm$ 0.015 & 0.308 $\pm$ 0.010  \\
DDO75 & 0.551 $\pm$ 0.006 	 & -0.697 $\pm$ 0.001 	 & 0.104 $\pm$ 0.012 	 & 2.000 $\pm$ 0.000 	 & 0.152 $\pm$ 0.025 	 & 0.187 $\pm$ 0.012 & 0.552 $\pm$ 0.006  \\
DDO87 & 0.383 $\pm$ 0.014 	 & -1.039 $\pm$ 0.028 	 & 1.808 $\pm$ 0.059 	 & 0.736 $\pm$ 0.066 	 & 0.198 $\pm$ 0.026 	 & 0.158 $\pm$ 0.018 & 0.371 $\pm$ 0.016  \\
DDO101 & 0.356 $\pm$ 0.022 	 & -1.003 $\pm$ 0.025 	 & 1.716 $\pm$ 0.144 	 & 1.682 $\pm$ 0.162 	 & 0.238 $\pm$ 0.046 	 & 0.298 $\pm$ 0.030 & 0.360 $\pm$ 0.031  \\
DDO126 & 0.453 $\pm$ 0.018 	 & -1.083 $\pm$ 0.061 	 & 2.166 $\pm$ 0.124 	 & 1.360 $\pm$ 0.266 	 & 0.128 $\pm$ 0.045 	 & 0.530 $\pm$ 0.017 & 0.464 $\pm$ 0.018  \\
DDO133 & 0.403 $\pm$ 0.022 	 & -1.107 $\pm$ 0.031 	 & 1.940 $\pm$ 0.059 	 & 1.820 $\pm$ 0.091 	 & 0.044 $\pm$ 0.054 	 & 0.196 $\pm$ 0.019 & 0.338 $\pm$ 0.020  \\
DDO154 & 0.524 $\pm$ 0.006 	 & -1.479 $\pm$ 0.031 	 & 2.845 $\pm$ 0.110 	 & 1.519 $\pm$ 0.058 	 & 0.201 $\pm$ 0.031 	 & 0.523 $\pm$ 0.013 & 0.443 $\pm$ 0.007  \\
DDO155 & 0.487 $\pm$ 0.017 	 & -1.083 $\pm$ 0.057 	 & 1.898 $\pm$ 0.151 	 & 1.481 $\pm$ 0.138 	 & 0.079 $\pm$ 0.040 	 & 0.153 $\pm$ 0.023 & 0.335 $\pm$ 0.028  \\
DDO165 & 0.526 $\pm$ 0.012 	 & -0.992 $\pm$ 0.045 	 & 1.795 $\pm$ 0.103 	 & 1.955 $\pm$ 0.055 	 & 0.135 $\pm$ 0.052 	 & 0.360 $\pm$ 0.031 & 0.420 $\pm$ 0.018  \\
DDO167 & 0.480 $\pm$ 0.014 	 & -1.303 $\pm$ 0.129 	 & 2.116 $\pm$ 0.212 	 & 1.847 $\pm$ 0.167 	 & 0.187 $\pm$ 0.038 	 & 0.473 $\pm$ 0.018 & 0.416 $\pm$ 0.020  \\
DDO168 & 0.613 $\pm$ 0.009 	 & -1.878 $\pm$ 0.057 	 & 2.986 $\pm$ 0.136 	 & 1.137 $\pm$ 0.043 	 & 0.140 $\pm$ 0.052 	 & 0.393 $\pm$ 0.013 & 0.383 $\pm$ 0.012  \\
DDO187 & 0.539 $\pm$ 0.013 	 & -1.554 $\pm$ 0.093 	 & 2.397 $\pm$ 0.212 	 & 1.438 $\pm$ 0.238 	 & 0.215 $\pm$ 0.058 	 & 0.248 $\pm$ 0.028 & 0.341 $\pm$ 0.022  \\
DDO210 & 0.509 $\pm$ 0.013 	 & -0.699 $\pm$ 0.001 	 & 0.021 $\pm$ 0.003 	 & 2.000 $\pm$ 0.000 	 & 0.089 $\pm$ 0.037 	 & 0.161 $\pm$ 0.055 & 0.508 $\pm$ 0.013  \\
DDO216 & 0.464 $\pm$ 0.019 	 & -1.374 $\pm$ 0.113 	 & 2.360 $\pm$ 0.256 	 & 1.688 $\pm$ 0.101 	 & 0.091 $\pm$ 0.073 	 & 0.480 $\pm$ 0.020 & 0.383 $\pm$ 0.036  \\
F564-V3 & 0.381 $\pm$ 0.013 	 & -1.539 $\pm$ 0.113 	 & 2.324 $\pm$ 0.155 	 & 1.670 $\pm$ 0.204 	 & 0.162 $\pm$ 0.024 	 & 0.325 $\pm$ 0.022 & 0.305 $\pm$ 0.023  \\
IC10 & 0.483 $\pm$ 0.004 	 & -1.525 $\pm$ 0.024 	 & 2.767 $\pm$ 0.033 	 & 1.984 $\pm$ 0.003 	 & 0.346 $\pm$ 0.004 	 & 0.310 $\pm$ 0.005 & 0.534 $\pm$ 0.002  \\
IC1613 & 0.465 $\pm$ 0.008 	 & -1.020 $\pm$ 0.024 	 & 1.562 $\pm$ 0.044 	 & 1.076 $\pm$ 0.029 	 & 0.153 $\pm$ 0.021 	 & 0.289 $\pm$ 0.015 & 0.391 $\pm$ 0.008  \\
LGS3 & 0.237 $\pm$ 0.021 	 & -1.003 $\pm$ 0.053 	 & 1.948 $\pm$ 0.176 	 & 1.179 $\pm$ 0.092 	 & 0.348 $\pm$ 0.051 	 & 0.317 $\pm$ 0.017 & 0.442 $\pm$ 0.015  \\
M81dwA & 0.373 $\pm$ 0.015 	 & -0.912 $\pm$ 0.029 	 & 1.251 $\pm$ 0.097 	 & 2.000 $\pm$ 0.000 	 & 0.274 $\pm$ 0.031 	 & 0.116 $\pm$ 0.023 & 0.350 $\pm$ 0.019  \\
NGC1569 & 0.656 $\pm$ 0.003 	 & -1.784 $\pm$ 0.073 	 & 3.118 $\pm$ 0.273 	 & 1.987 $\pm$ 0.002 	 & 0.310 $\pm$ 0.036 	 & 0.294 $\pm$ 0.030 & 0.525 $\pm$ 0.005  \\
NGC2366 & 0.567 $\pm$ 0.007 	 & -1.326 $\pm$ 0.031 	 & 2.437 $\pm$ 0.052 	 & 1.565 $\pm$ 0.084 	 & 0.164 $\pm$ 0.017 	 & 0.533 $\pm$ 0.006 & 0.438 $\pm$ 0.005  \\
NGC3738 & 0.611 $\pm$ 0.008 	 & -1.543 $\pm$ 0.063 	 & 2.531 $\pm$ 0.226 	 & 1.916 $\pm$ 0.019 	 & 0.283 $\pm$ 0.034 	 & 0.317 $\pm$ 0.015 & 0.440 $\pm$ 0.007  \\
NGC4163 & 0.534 $\pm$ 0.017 	 & -1.345 $\pm$ 0.097 	 & 2.249 $\pm$ 0.188 	 & 1.194 $\pm$ 0.224 	 & 0.204 $\pm$ 0.049 	 & 0.091 $\pm$ 0.038 & 0.361 $\pm$ 0.020  \\
NGC4214 & 0.478 $\pm$ 0.006 	 & -1.319 $\pm$ 0.023 	 & 2.143 $\pm$ 0.044 	 & 0.969 $\pm$ 0.040 	 & 0.230 $\pm$ 0.017 	 & 0.167 $\pm$ 0.014 & 0.386 $\pm$ 0.008  \\
SagDIG & 0.421 $\pm$ 0.021 	 & -0.703 $\pm$ 0.002 	 & 0.068 $\pm$ 0.004 	 & 2.000 $\pm$ 0.000 	 & 0.076 $\pm$ 0.049 	 & 0.175 $\pm$ 0.016 & 0.416 $\pm$ 0.021  \\
UGC8508 & 0.587 $\pm$ 0.011 	 & -1.493 $\pm$ 0.073 	 & 2.736 $\pm$ 0.224 	 & 1.937 $\pm$ 0.073 	 & 0.269 $\pm$ 0.054 	 & 0.584 $\pm$ 0.044 & 0.442 $\pm$ 0.014  \\
WLM & 0.534 $\pm$ 0.005 	 & -0.701 $\pm$ 0.003 	 & 0.282 $\pm$ 0.014 	 & 2.000 $\pm$ 0.000 	 & 0.213 $\pm$ 0.018 	 & 0.553 $\pm$ 0.004 & 0.540 $\pm$ 0.005  \\
Haro29 & 0.508 $\pm$ 0.016 	 & -1.401 $\pm$ 0.149 	 & 2.419 $\pm$ 0.241 	 & 0.949 $\pm$ 0.088 	 & 0.246 $\pm$ 0.030 	 & 0.199 $\pm$ 0.019 & 0.366 $\pm$ 0.018  \\
Haro36 & 0.597 $\pm$ 0.016 	 & -1.615 $\pm$ 0.146 	 & 2.590 $\pm$ 0.177 	 & 1.605 $\pm$ 0.076 	 & 0.176 $\pm$ 0.041 	 & 0.316 $\pm$ 0.018 & 0.398 $\pm$ 0.028  \\
Mrk178 & 0.500 $\pm$ 0.020 	 & -1.362 $\pm$ 0.072 	 & 2.506 $\pm$ 0.292 	 & 1.254 $\pm$ 0.142 	 & 0.226 $\pm$ 0.049 	 & 0.429 $\pm$ 0.020 & 0.500 $\pm$ 0.019  \\
VIIZw403 & 0.591 $\pm$ 0.012 	 & -1.774 $\pm$ 0.178 	 & 2.714 $\pm$ 0.378 	 & 2.000 $\pm$ 0.000 	 & 0.180 $\pm$ 0.089 	 & 0.464 $\pm$ 0.025 & 0.362 $\pm$ 0.018  \\
NGC247 & 0.344 $\pm$ 0.006 	 & -0.698 $\pm$ 0.002 	 & 0.000 $\pm$ 0.004 	 & 2.000 $\pm$ 0.000 	 & 0.061 $\pm$ 0.009 	 &   \dots $\pm$ 0.002 & 0.344 $\pm$ 0.005  \\
NGC404 & 0.296 $\pm$ 0.005 	 & -0.875 $\pm$ 0.011 	 & 0.000 $\pm$ 0.033 	 & 1.434 $\pm$ 0.019 	 & 0.356 $\pm$ 0.013 	 &   \dots $\pm$ 0.008 & 0.456 $\pm$ 0.004  \\
UGC4483 & 0.563 $\pm$ 0.014 	 & -1.652 $\pm$ 0.128 	 & 0.000 $\pm$ 0.169 	 & 1.946 $\pm$ 0.075 	 & 0.131 $\pm$ 0.039 	 &   \dots $\pm$ 0.029 & 0.283 $\pm$ 0.024  \\
SextansB & 0.429 $\pm$ 0.011 	 & -1.422 $\pm$ 0.032 	 & 0.000 $\pm$ 0.044 	 & 0.858 $\pm$ 0.071 	 & 0.068 $\pm$ 0.016 	 &   \dots $\pm$ 0.013 & 0.322 $\pm$ 0.013  \\
NGC3109 & 0.445 $\pm$ 0.010 	 & -0.741 $\pm$ 0.002 	 & 0.000 $\pm$ 0.010 	 & 2.000 $\pm$ 0.000 	 & 0.047 $\pm$ 0.031 	 &   \dots $\pm$ 0.006 & 0.429 $\pm$ 0.011  \\
Antlia & 0.281 $\pm$ 0.015 	 & -0.703 $\pm$ 0.001 	 & 0.000 $\pm$ 0.004 	 & 2.000 $\pm$ 0.000 	 & 0.229 $\pm$ 0.036 	 &   \dots $\pm$ 0.019 & 0.278 $\pm$ 0.015  \\
SextansA & 0.491 $\pm$ 0.009 	 & -0.695 $\pm$ 0.002 	 & 0.000 $\pm$ 0.002 	 & 2.000 $\pm$ 0.000 	 & 0.069 $\pm$ 0.021 	 &   \dots $\pm$ 0.011 & 0.490 $\pm$ 0.009  \\
DDO82 & 0.396 $\pm$ 0.020 	 & -1.013 $\pm$ 0.089 	 & 0.000 $\pm$ 0.169 	 & 1.924 $\pm$ 0.053 	 & 0.278 $\pm$ 0.038 	 &   \dots $\pm$ 0.022 & 0.402 $\pm$ 0.031  \\
KDG73 & 0.224 $\pm$ 0.034 	 & -0.780 $\pm$ 0.048 	 & 0.000 $\pm$ 0.377 	 & 2.000 $\pm$ 0.000 	 & 0.423 $\pm$ 0.058 	 &   \dots $\pm$ 0.014 & 0.413 $\pm$ 0.050  \\
NGC3741 & 0.392 $\pm$ 0.010 	 & -1.752 $\pm$ 0.059 	 & 0.000 $\pm$ 0.079 	 & 1.974 $\pm$ 0.010 	 & 0.143 $\pm$ 0.014 	 &   \dots $\pm$ 0.006 & 0.456 $\pm$ 0.009  \\
DDO99 & 0.447 $\pm$ 0.010 	 & -1.412 $\pm$ 0.035 	 & 0.000 $\pm$ 0.097 	 & 1.903 $\pm$ 0.028 	 & 0.151 $\pm$ 0.028 	 &   \dots $\pm$ 0.010 & 0.375 $\pm$ 0.011  \\
NGC4163 & 0.534 $\pm$ 0.013 	 & -1.360 $\pm$ 0.083 	 & 2.225 $\pm$ 0.159 	 & 1.281 $\pm$ 0.136 	 & 0.204 $\pm$ 0.055 	 & 0.091 $\pm$ 0.038 & 0.351 $\pm$ 0.021  \\
MCG9-20-131 & 0.539 $\pm$ 0.016 	 & -1.735 $\pm$ 0.146 	 & 0.000 $\pm$ 0.205 	 & 1.033 $\pm$ 0.357 	 & 0.131 $\pm$ 0.057 	 &   \dots $\pm$ 0.021 & 0.277 $\pm$ 0.049  \\
UGCA292 & 0.562 $\pm$ 0.019 	 & -1.439 $\pm$ 0.104 	 & 0.000 $\pm$ 0.137 	 & 0.984 $\pm$ 0.474 	 & 0.078 $\pm$ 0.051 	 &   \dots $\pm$ 0.037 & 0.308 $\pm$ 0.040  \\
GR8 & 0.441 $\pm$ 0.017 	 & -0.982 $\pm$ 0.055 	 & 0.000 $\pm$ 0.108 	 & 1.368 $\pm$ 0.222 	 & 0.105 $\pm$ 0.038 	 &   \dots $\pm$ 0.015 & 0.361 $\pm$ 0.018  \\
UGC8508 & 0.587 $\pm$ 0.012 	 & -1.493 $\pm$ 0.106 	 & 2.736 $\pm$ 0.185 	 & 1.937 $\pm$ 0.086 	 & 0.269 $\pm$ 0.063 	 & 0.584 $\pm$ 0.033 & 0.442 $\pm$ 0.017  \\
KKH86 & 0.205 $\pm$ 0.032 	 & -0.702 $\pm$ 0.058 	 & 0.000 $\pm$ 0.267 	 & 1.995 $\pm$ 0.173 	 & 0.452 $\pm$ 0.079 	 &   \dots $\pm$ 0.037 & 0.410 $\pm$ 0.047  \\
UGC8833 & 0.541 $\pm$ 0.022 	 & -1.338 $\pm$ 0.113 	 & 0.000 $\pm$ 0.218 	 & 0.543 $\pm$ 0.268 	 & 0.051 $\pm$ 0.045 	 &   \dots $\pm$ 0.040 & 0.249 $\pm$ 0.035  \\
KK230 & 0.361 $\pm$ 0.022 	 & -1.162 $\pm$ 0.079 	 & 0.000 $\pm$ 0.195 	 & 1.036 $\pm$ 0.368 	 & 0.220 $\pm$ 0.034 	 &   \dots $\pm$ 0.015 & 0.370 $\pm$ 0.050  \\
DDO187 & 0.539 $\pm$ 0.011 	 & -1.554 $\pm$ 0.128 	 & 2.397 $\pm$ 0.137 	 & 1.438 $\pm$ 0.252 	 & 0.215 $\pm$ 0.044 	 & 0.248 $\pm$ 0.037 & 0.341 $\pm$ 0.027  \\
\hline 

\end{tabular}
\caption{The morphological parameters of the \hi\ column density maps for the combined LITTLE-THINGS and VLA-ANGST surveys. }
\end{center}
\label{t:qm}
\end{table*}%
\newpage

\section{Star-formation ratios from Hunter et al 2010}
\label{s:h10}

\begin{figure*}
\begin{center}
     \begin{minipage}{0.65\linewidth}
	\includegraphics[width=\textwidth]{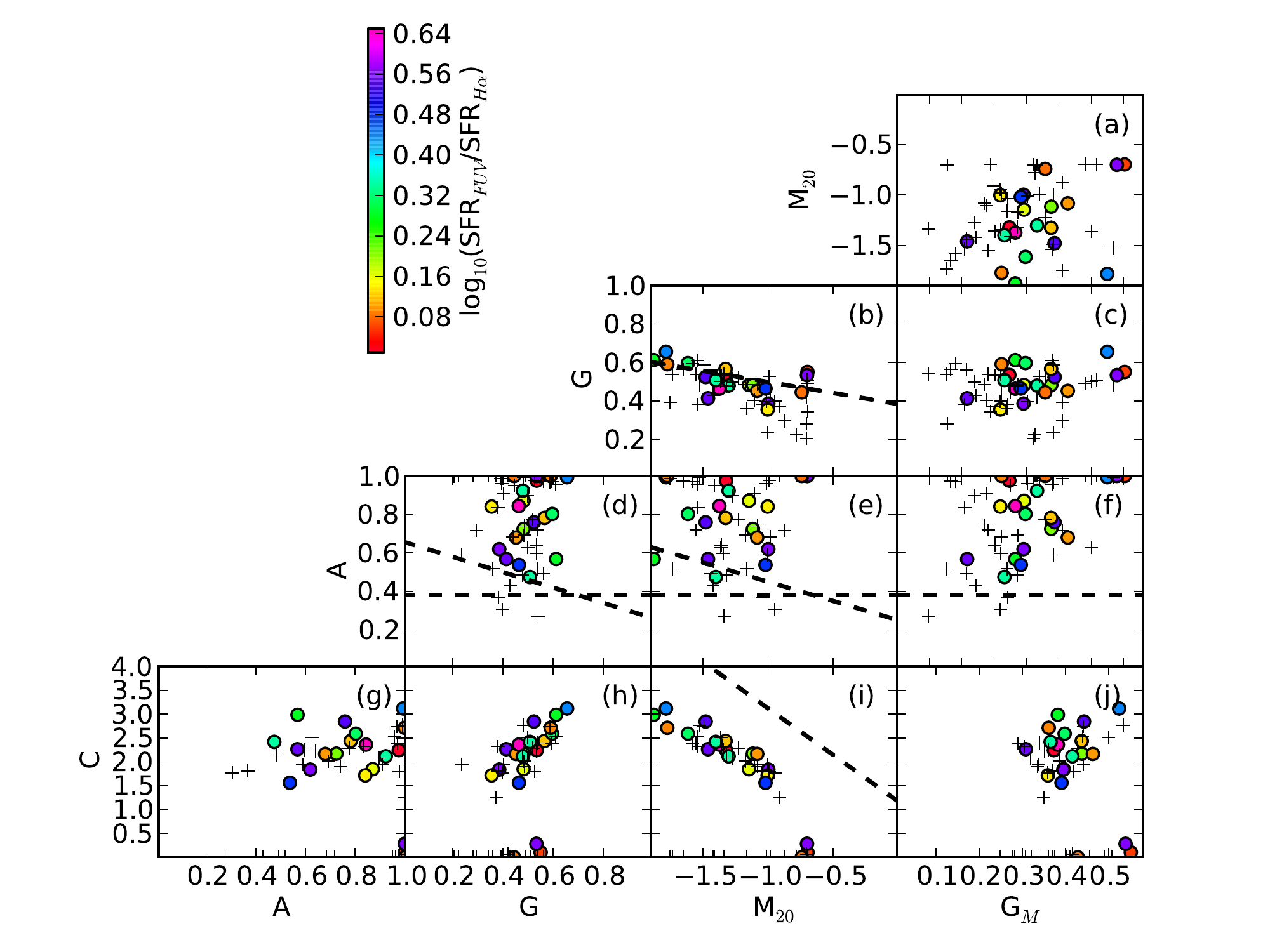}
    \end{minipage}\hfill
    \begin{minipage}{0.34\linewidth}
	\includegraphics[width=1.1\textwidth]{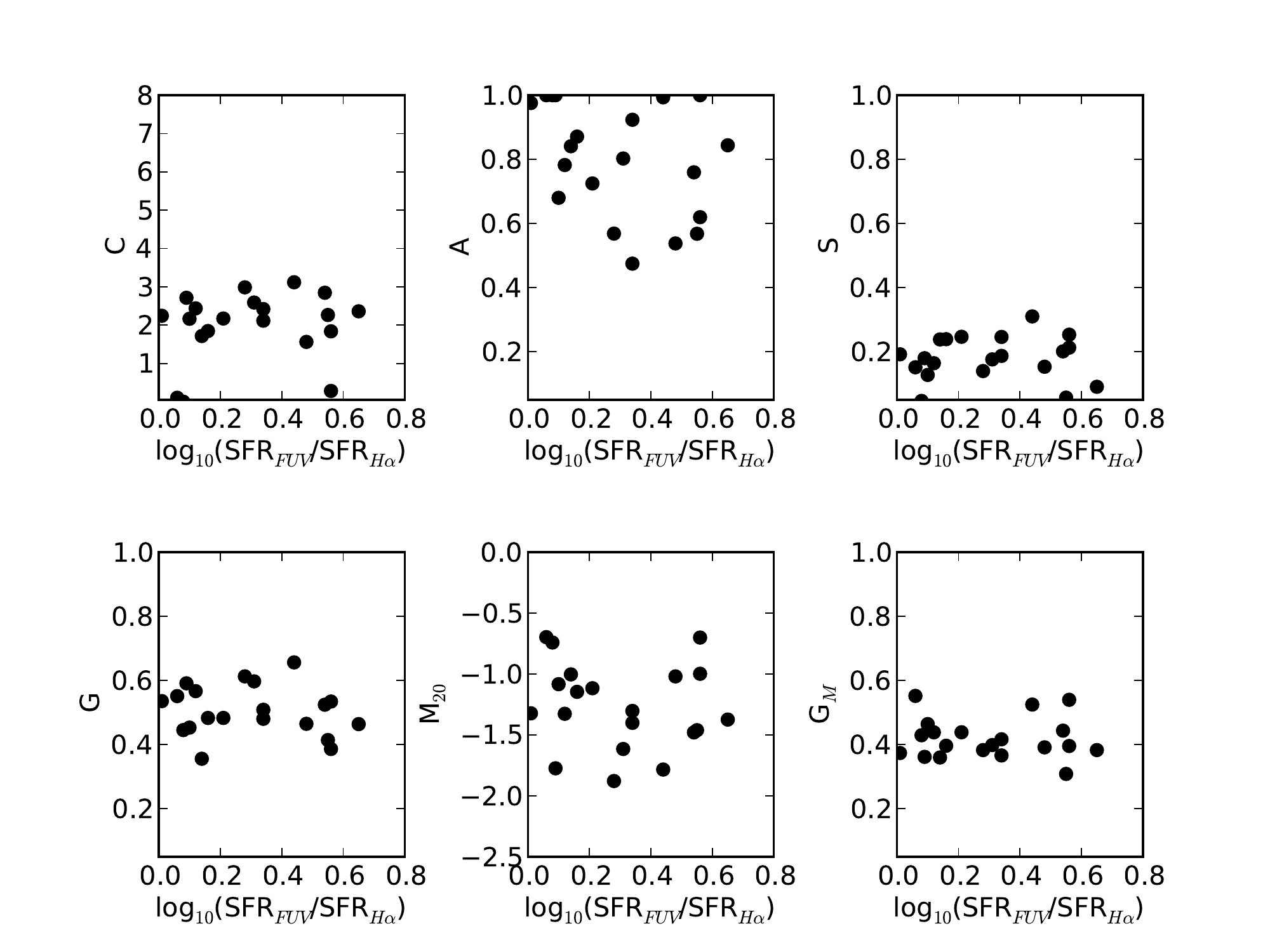}
	\caption{\label{f:ratHa:6par}The distribution of the \hi\ morphological parameters versus the ratio of star-formation rate based on \fuv/H$\alpha$ from \protect\cite{Hunter10}. }
    \end{minipage}
    \begin{minipage}{\linewidth}
	\caption{\label{f:ratHa}The distribution of the \hi\ morphological parameters colour coded by the ratio of star-formation rate based on \fuv/H$\alpha$ from \protect\cite{Hunter10}. }
    \end{minipage}
\end{center}
\end{figure*}

\begin{figure*}
\begin{center}
     \begin{minipage}{0.65\linewidth}
	\includegraphics[width=\textwidth]{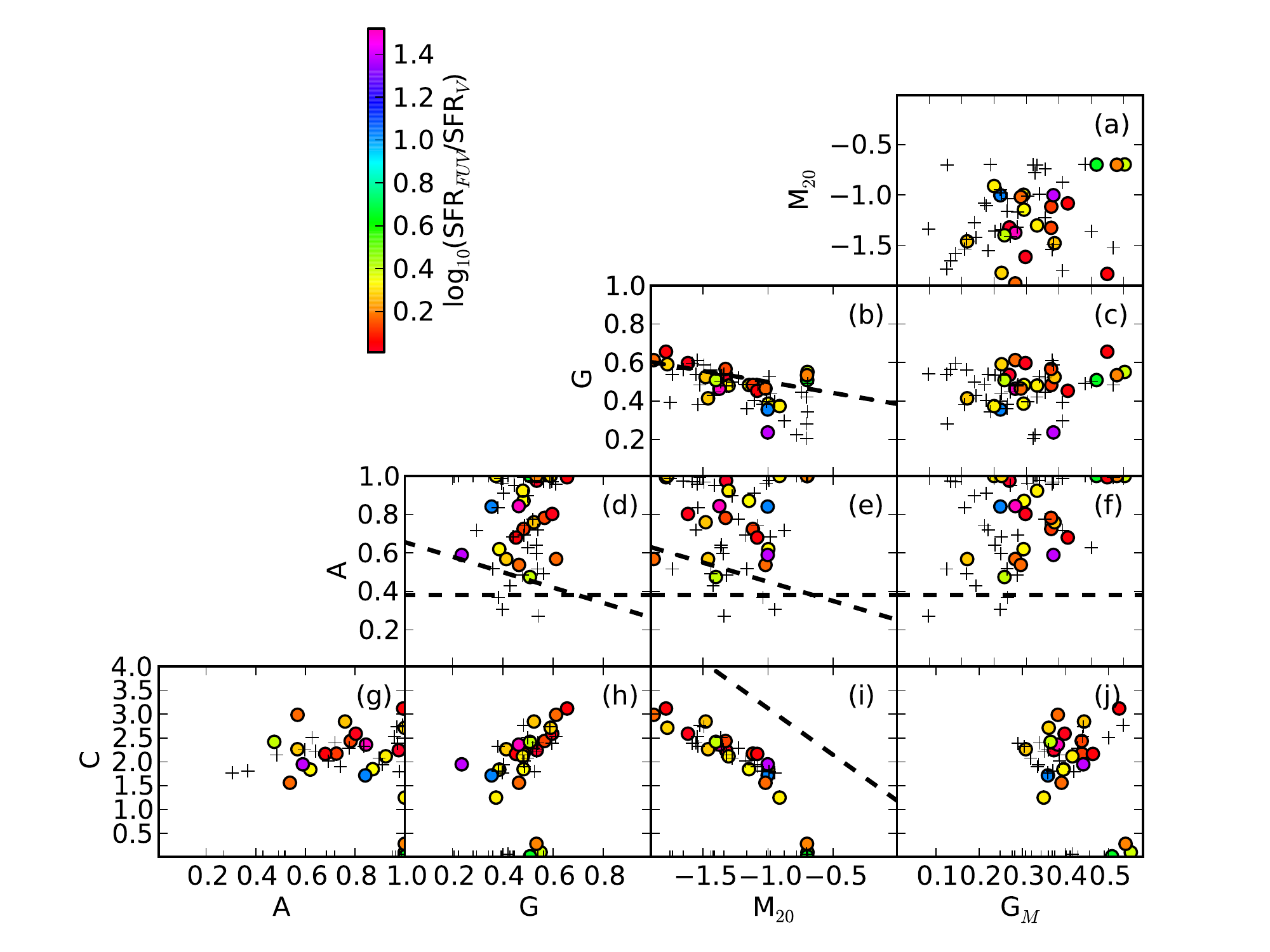}
    \end{minipage}\hfill
    \begin{minipage}{0.34\linewidth}
	\includegraphics[width=1.1\textwidth]{./holwerda_f19a.pdf}
	\caption{\label{f:ratV:6par}The distribution of the \hi\ morphological parameters versus the ratio of star-formation rate based on \fuv/V-band from \protect\cite{Hunter10}. }
    \end{minipage}
    \begin{minipage}{\linewidth}
	\caption{\label{f:ratV}The distribution of the \hi\ morphological parameters colour coded by the ratio of star-formation rate based on \fuv/V-band from \protect\cite{Hunter10}. }
    \end{minipage}
\end{center}
\end{figure*}

\begin{figure*}
\begin{center}
     \begin{minipage}{0.65\linewidth}
	\includegraphics[width=\textwidth]{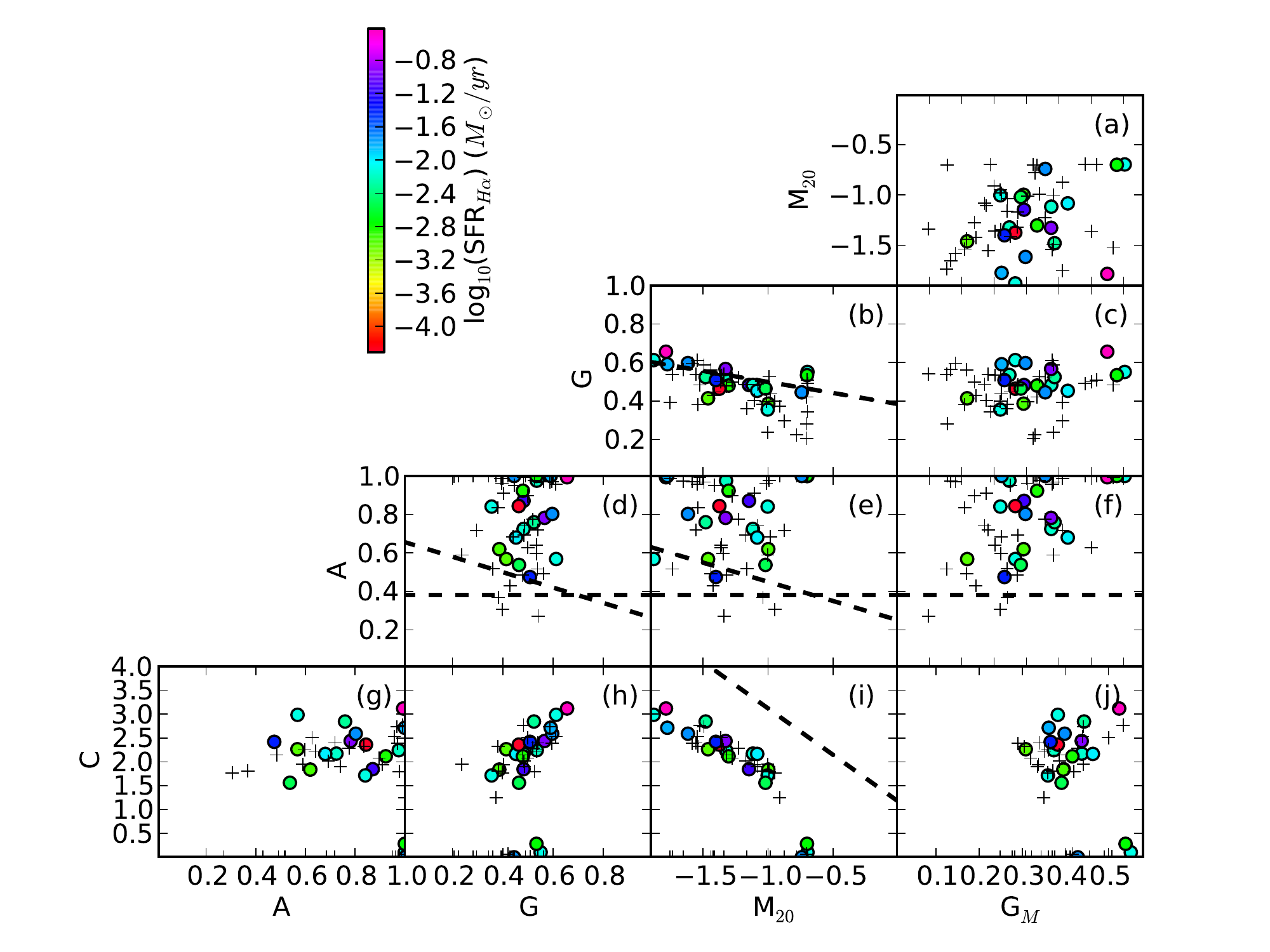}
    \end{minipage}\hfill
    \begin{minipage}{0.34\linewidth}
	\includegraphics[width=1.1\textwidth]{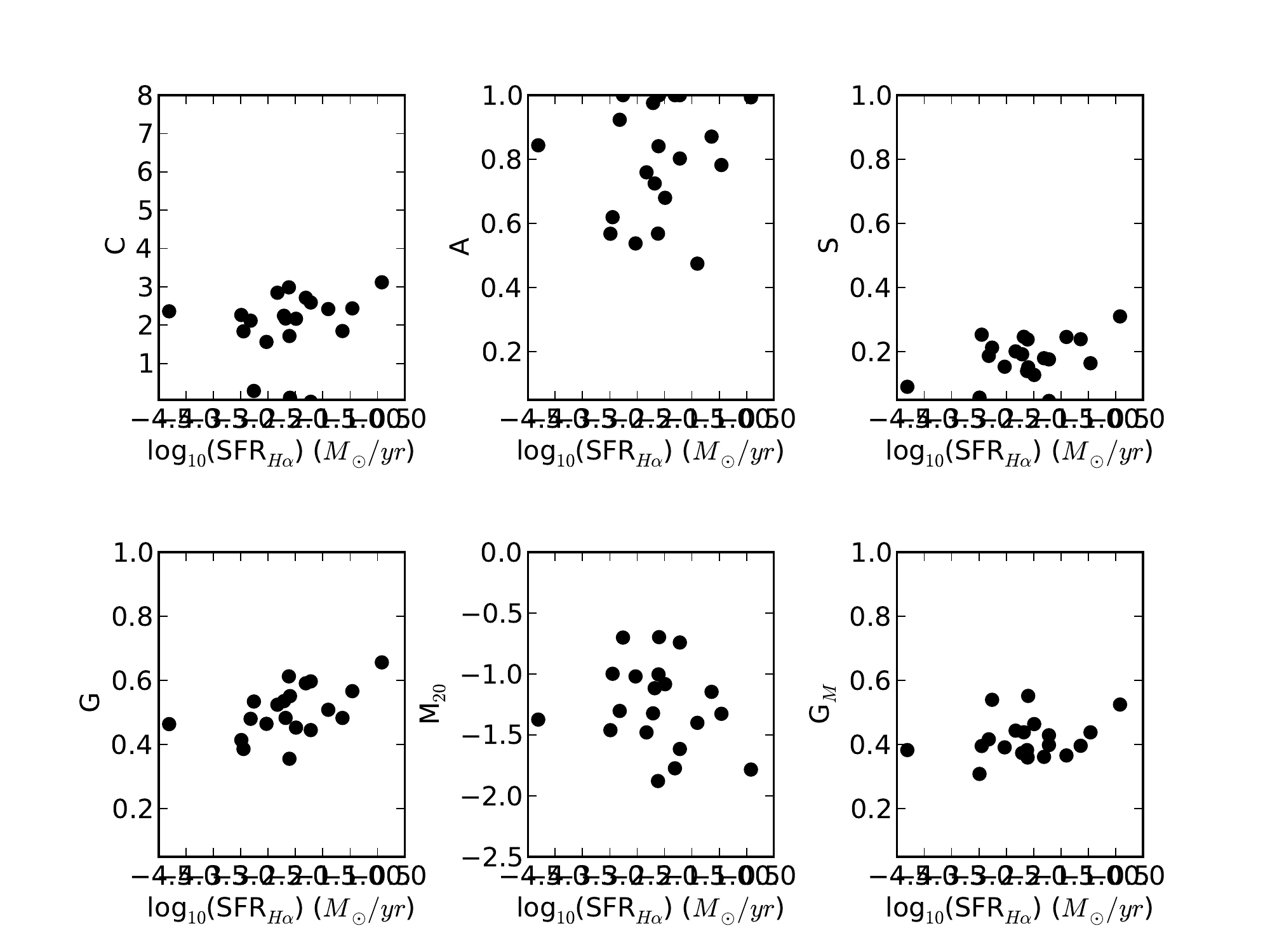}
	\caption{\label{f:sfr:Ha:6par}The relation between H$\alpha$-derived SFR from \protect\cite{Hunter10} and the six morphology parameters. }
    \end{minipage}
    \begin{minipage}{\linewidth}
	\caption{\label{f:sfr:Ha}The distribution of the \hi\ morphological parameters colour coded by the star-formation rate based on H$\alpha$ flux reported in \protect\cite{Hunter10}. }
    \end{minipage}
\end{center}
\end{figure*}

\cite{Hunter10} provide ratios between the star-formation derived from \fuv\ and those from H$\alpha$ and broadband V.
Figures \ref{f:ratHa} and \ref{f:ratHa:6par} show the relations with the $SFR_{FUV}/SFR_{H\alpha}$ and Figures \ref{f:ratV} and \ref{f:ratV:6par} show the relations with the $SFR_{FUV}/SFR_{V}$. Only very weak trends are visible (see also Table \ref{t:spear}).
Figures \ref{f:sfr:Ha}, \ref{f:sfr:Ha:6par}, \ref{f:sfr:Ha} and \ref{f:sfr:V:6par} show the direct relations with the H$\alpha$ and V-band 
star-formation estimates from \cite{Hunter10} with the \hi\ morphology parameters.

\begin{figure*}
\begin{center}
     \begin{minipage}{0.65\linewidth}
	\includegraphics[width=\textwidth]{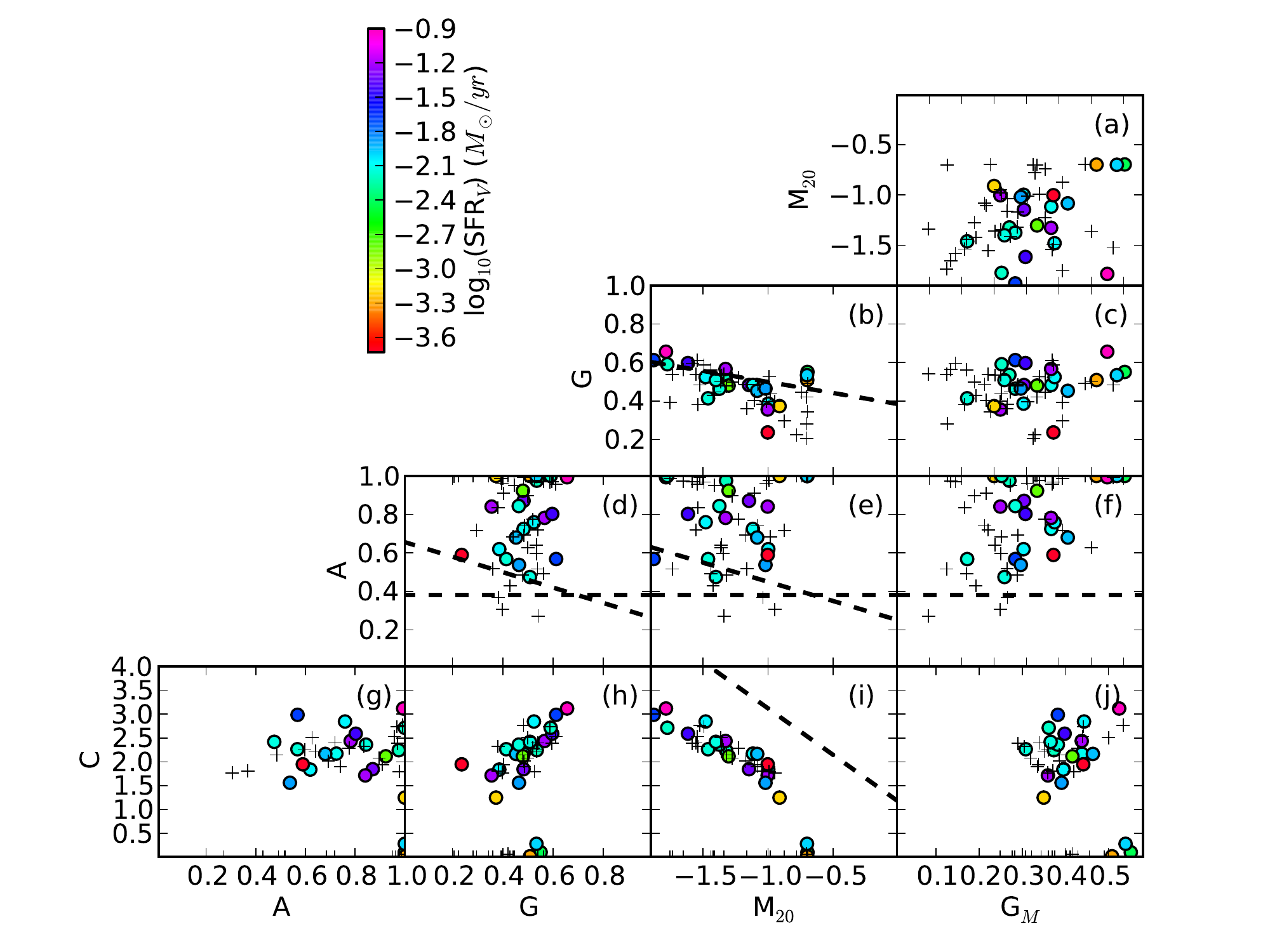}
    \end{minipage}\hfill
    \begin{minipage}{0.34\linewidth}
	\includegraphics[width=1.1\textwidth]{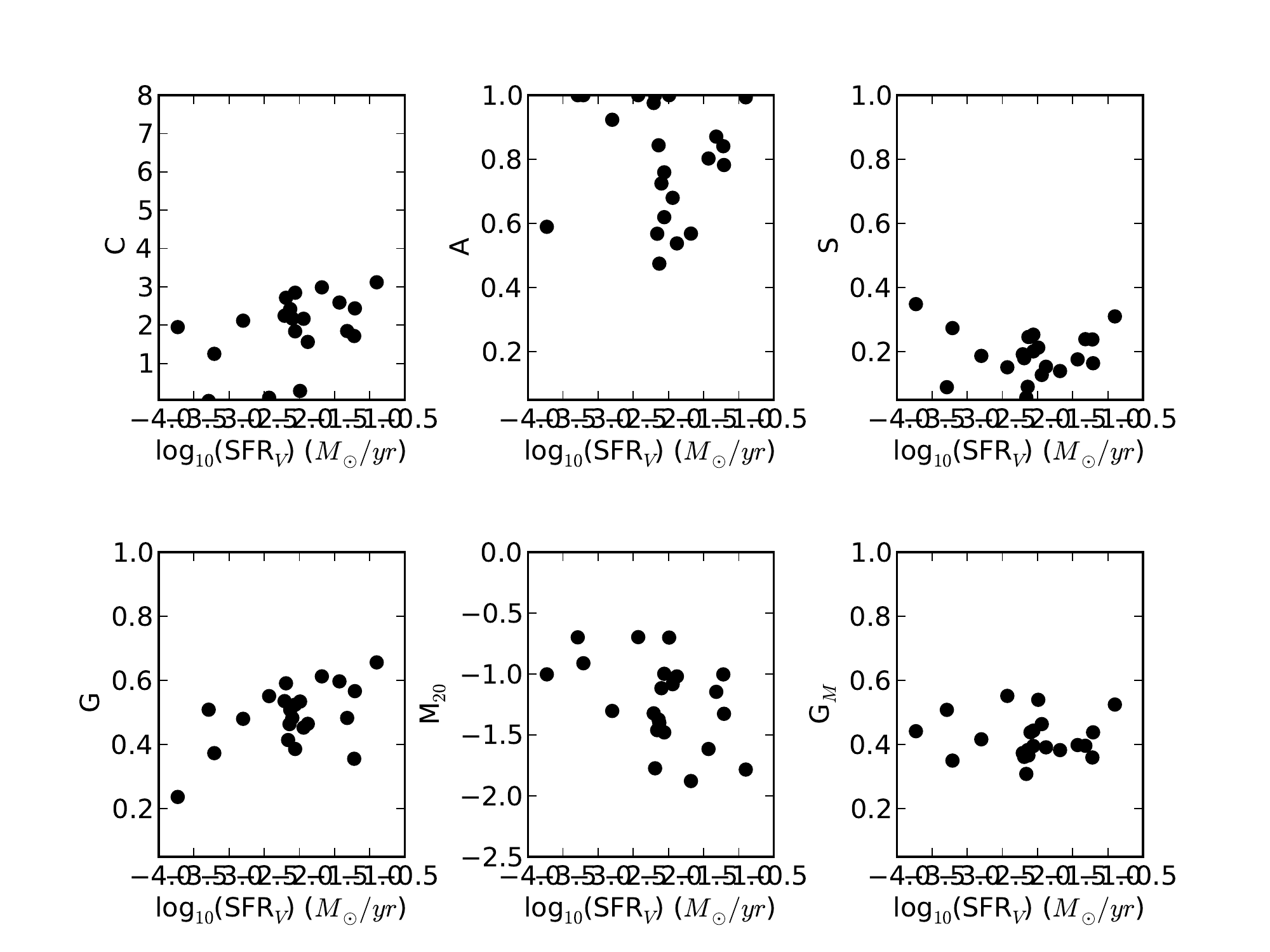}
	\caption{\label{f:sfr:v:6par} The relation between $V$-band derived SFR from \protect\cite{Hunter10} and the six morphology parameters. }
    \end{minipage}
    \begin{minipage}{\linewidth}
	\caption{\label{f:sfr:v} The distribution of the \hi\ morphological parameters colour coded by the star-formation rate based on V-band flux reported in \protect\cite{Hunter10}. }
    \end{minipage}
\end{center}
\end{figure*}

\section{Values from Weisz et al 2011}
\label{s:w11}

\begin{figure*}
\begin{center}
     \begin{minipage}{0.65\linewidth}
	\includegraphics[width=\textwidth]{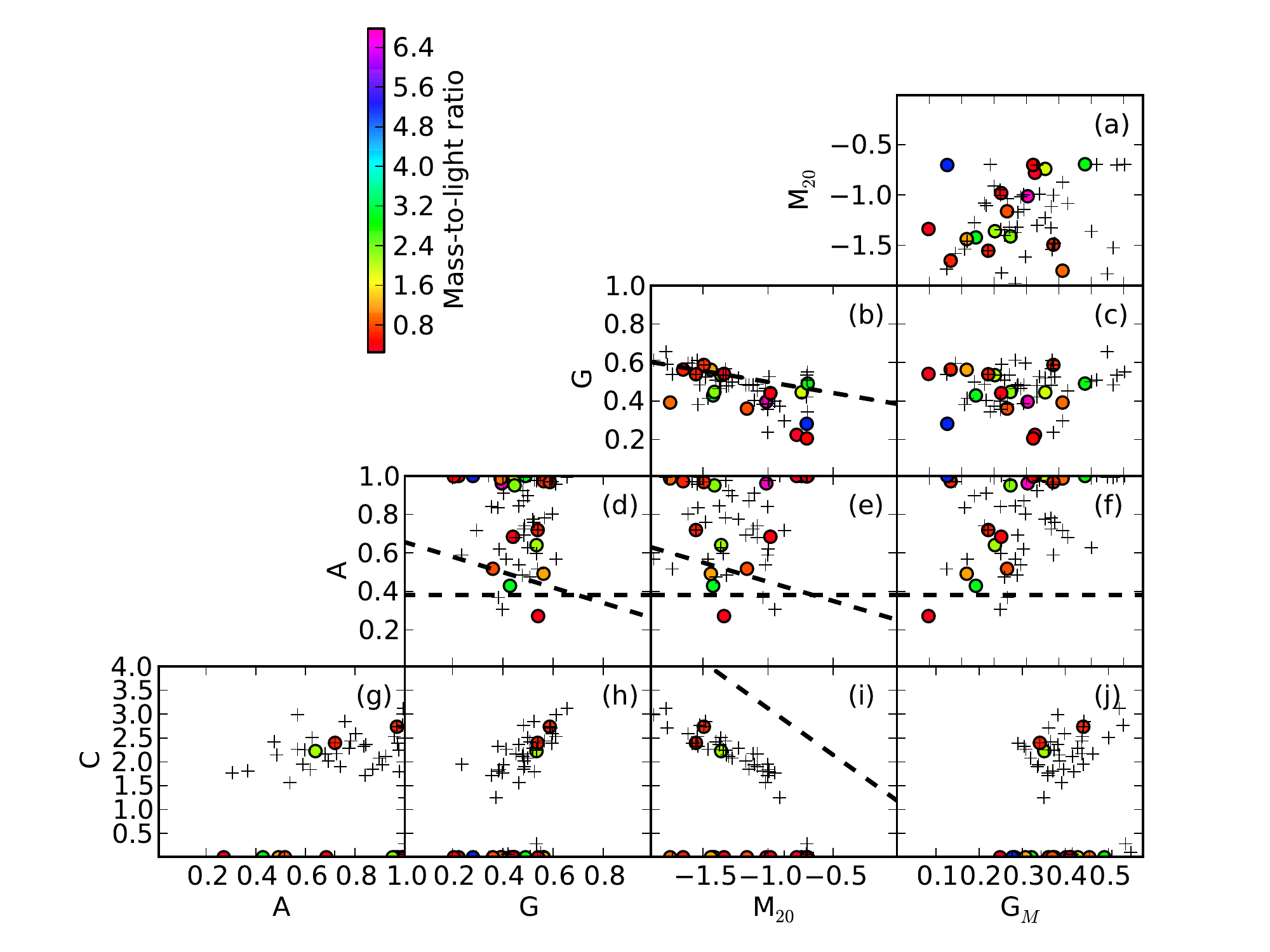}
    \end{minipage}\hfill
    \begin{minipage}{0.34\linewidth}
	\includegraphics[width=1.1\textwidth]{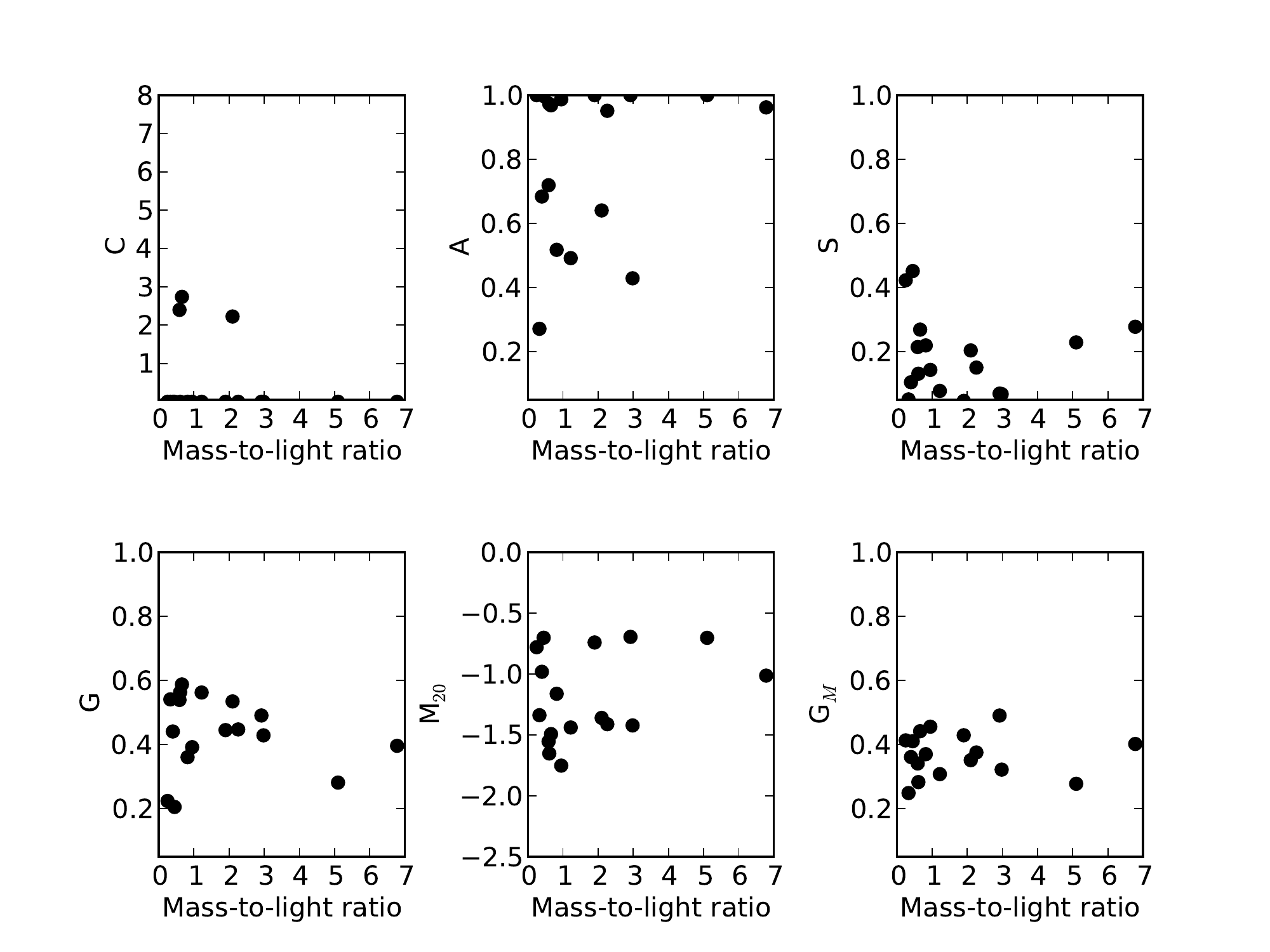}
	\caption{\label{f:ML:6par} The morphology of the \hi\ maps colour versus the mass-to-light ratio from \protect\cite{Weisz11b}. }
    \end{minipage}
    \begin{minipage}{\linewidth}
	\caption{\label{f:ML} The morphology of the \hi\ maps colour coded by the mass-to-light ratio from \protect\cite{Weisz11b}. }
    \end{minipage}
\end{center}
\end{figure*}

\begin{figure*}
\begin{center}
     \begin{minipage}{0.65\linewidth}
	\includegraphics[width=\textwidth]{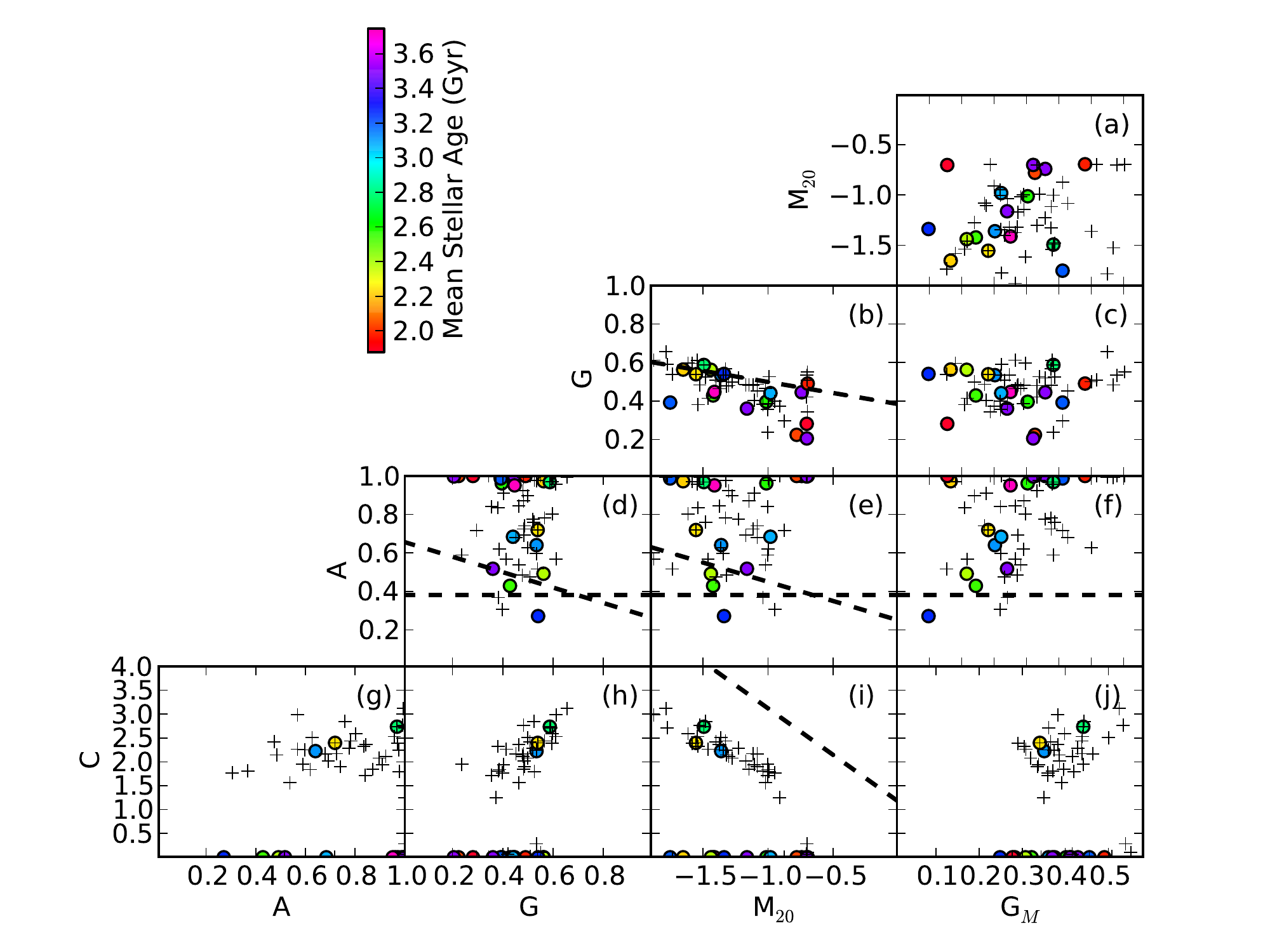}
    \end{minipage}\hfill
    \begin{minipage}{0.34\linewidth}
	\includegraphics[width=1.1\textwidth]{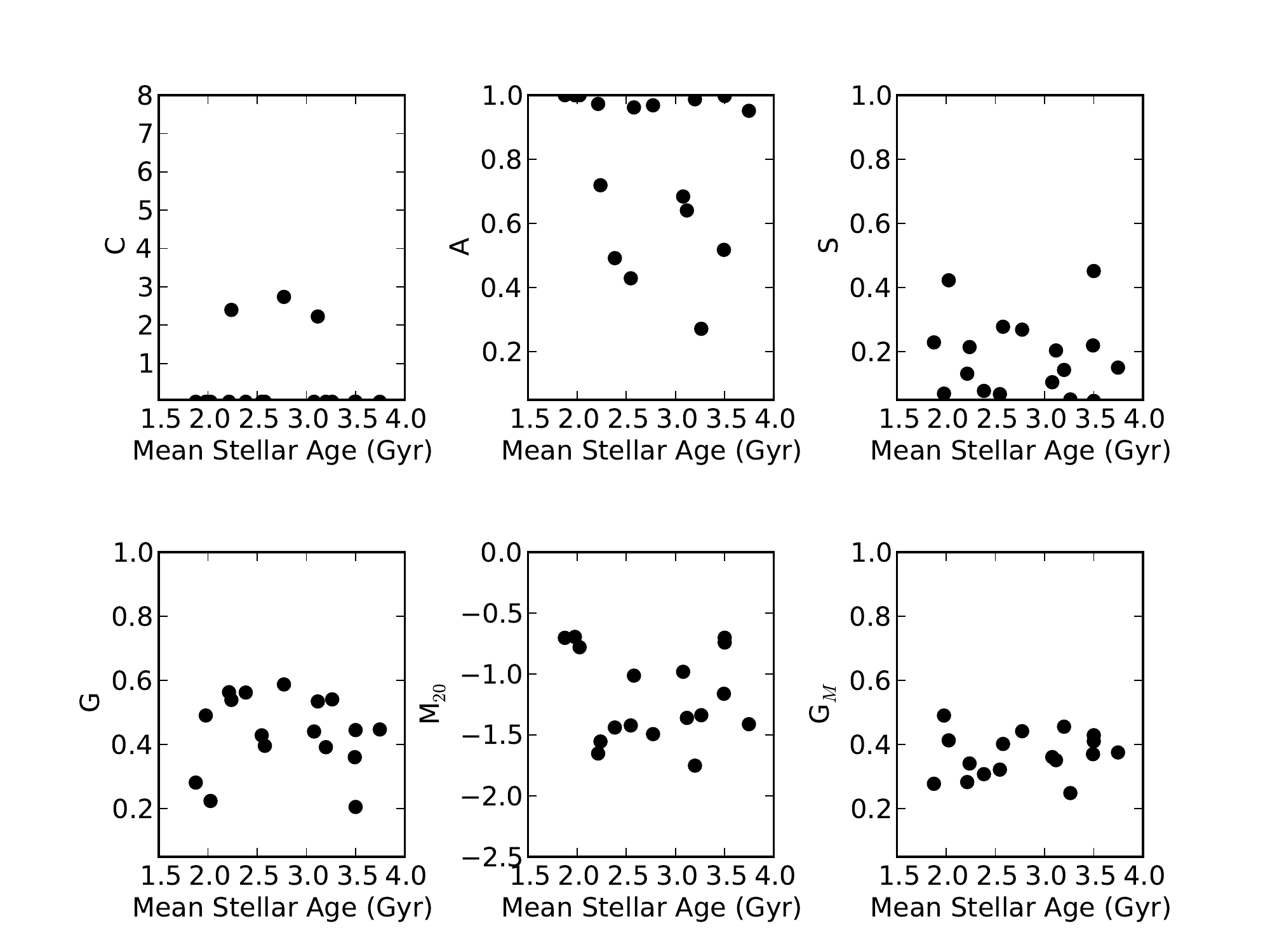}
	\caption{\label{f:meanage} The morphology of the \hi\ maps colour coded by the mean stellar age from \protect\cite{Weisz11b}. }
    \end{minipage}
    \begin{minipage}{\linewidth}
	\caption{\label{f:meanage:6par} The morphology of the \hi\ maps versus the mean stellar age from \protect\cite{Weisz11b}. }
    \end{minipage}
\end{center}
\end{figure*}

In addition to a mean star-formation rate, \cite{Weisz11b} provide a mass-to-light ratio from their stellar populations. Figure \ref{f:ML} 
shows the ANGST galaxies colour-coded by their M/L ratio and Figure \ref{f:ML:6par} the direct relations 
but no trend emerges with any of the \hi\ morphological parameters (low values in Table \ref{t:spear}).
%
\cite{Weisz11b} also give the fraction of the stars formed at a given epoch (1, 2, 3, 6, or 10 Gyr ago). 
Converting these fractions to a mean age, we colour-code the ANGST galaxies with this mean age in 
Figure \ref{f:meanage}. There is no relation between the mean stellar age and the distribution of the 
\hi\ morphology parameters from the VLA-ANGST (Table \ref{t:spear}).

\end{document}